\documentclass[12pt]{article}

\usepackage{amsmath}
\usepackage{latexsym}
\usepackage{float}
\usepackage[caption = false]{subfig}
\usepackage{graphicx,psfrag,epsf}
\graphicspath{{}}
\usepackage{enumerate}
\usepackage{natbib}
\usepackage{bbm}

\usepackage{chngcntr}
\usepackage[colorlinks,citecolor=blue,urlcolor=blue,filecolor=blue,backref=page]{hyperref}

\newcommand{\blind}{0}

\begin{document}

\bibliographystyle{plainnat}

\def\spacingset#1{\renewcommand{\baselinestretch}%
{#1}\small\normalsize} \spacingset{1}


\if0\blind
{
  \title{\bf Variational Bayes for Functional Data Registration, Smoothing, and Prediction }
  \author{Cecilia Earls\\
   Cornell University\\
    and \\
    Giles Hooker \\
    Cornell University}
  \maketitle
} \fi

\if1\blind
{
  \bigskip
  \bigskip
  \bigskip
  \begin{center}
    {\LARGE\bf Title}
\end{center}
  \medskip
} \fi

\bigskip
\begin{abstract}

We propose a model for functional data registration that extends current inferential capabilities for unregistered data by providing a flexible probabilistic framework that 1) allows for functional prediction in the context of registration and 2) can be adapted to include smoothing and registration in one model.  The proposed inferential framework is a Bayesian hierarchical model where the registered functions are modeled as Gaussian processes.  To address the computational demands of inference in high-dimensional Bayesian models, we propose an adapted form of the variational Bayes algorithm for approximate inference that performs similarly to MCMC sampling methods for well-defined problems.  The efficiency of the adapted variational Bayes (AVB) algorithm allows variability in a predicted registered, warping, and unregistered function to be depicted separately via bootstrapping. Temperature data related to the El-Ni\~no phenomenon is used to demonstrate the unique inferential capabilities for prediction provided by this model.

\end{abstract}

\noindent%

\spacingset{1.45}

\section{INTRODUCTION}
\label{sec:intro}

This paper introduces a novel approach to functional data registration within a Bayesian hierarchical model.  In this model, both registered and warping functions are modeled in terms of Gaussian processes with registration described by restrictions on the covariance of the registered functions.  This enables both MCMC and variational Bayes methods to be employed for estimation and prediction.  Our model for registration extends current inferential procedures to include both 1) functional prediction in the context of registration and 2) registration and smoothing in one model. 

The primary advantage to our proposed registration model in comparison to current registration methods is that it provides a probabilistic framework in which new observations can be considered.  Assuming a new unregistered function has been partially recorded, with this framework, we can obtain estimates of the registered partial function and also the corresponding partial warping function.  Using these estimates, the complete registered function, the complete warping function, and the complete unregistered function can be predicted.  Details of the prediction model can be found in Section \ref{sec:pred}.  To the authors' knowledge this is the first time functional prediction is considered in the context of registration.

Additionally, our model can be extended to allow for noisy observations.  Most current registration methods consider functional regularization as a pre-processing step with the exception of \cite{rak:14}.  However, in the paper by \citet{rak:14}, the authors' maximum likelihood approach to registering latent functions does not provide an easy way to quantify variability in the estimates of the registered functions.  In Appendix C.3, we provide an illustration of how smoothing functions in a pre-processing step can significantly underestimate the variability in the estimates of the registered functions.  

The simplest definition of functional data registration is any algorithm that aligns functions in a way that eliminates all phase variability between functions ( \citet{ram:05}).  Without registration, basic summary statistics such as the sample mean and covariance are less interpretable as time variation between significant features in the functions tends to dampen the amplitude variation in these features.  Furthermore, the average timing of significant features may also be of interest and is difficult to obtain under traditional methods of analyzing functional data.  There has been much recent interest in proper ways to define and measure registration as well as in developing registration methods with desirable statistical properties.   

The evolution of registration dates back to \citet{sak:78} where the authors use a dynamic programming algorithm for landmark registration.  Landmark registration was again considered by \citet{kneip:92} and \citet{kneip:95}.  In 1997, a new cost function for functional registration was introduced by \citet{wang:97}.  A significant advancement in registration literature can be traced to \citet{sil:95} and \citet{ram:98} where the authors introduce global registration procedures, and Ramsey considers the use of a flexible family of monotone warping functions.  Parametric and B-spline base warping functions are considered by \citet{brum:04} and \citet{gerv:04}, respectively.  Nonparametric maximum likelihood approaches to registration are considered by both \citet{ronn:01} and \citet{gerv:05}.  A moments based approach to registration is introduced by \citet{jam:07}.  \citet{tang:08} propose pairwise curve synchronization.  The first Bayesian approach to registration can be found in \citet{tel:07}.  Registration to principal components is considered by \citet{kneip:08}. Finally, with regard to improvements in registration, the recent work by \citet{sriv:11} offers the most comprehensive framework for registration to date.

Much of the focus in combining registration with other types of inference in one model has been in the area of functional data clustering and registration.  Current work in this area can be found in \citet{liu:09}, \citet{san:10}, and also a Bayesian approach in \citet{zhang:14}.  Recent work by \citet{rak:14} includes a model for functional smoothing and registration.  While these extensions to registration procedures offer additional tools for functional data analysis, they tend to focus less on high-quality registration.

In this paper, we develop Bayesian hierarchical models that address both areas of development in registration procedures. First, a model is proposed to register functional data that gives estimates that compare favorably with those from the best current registration methods available, notably, \citet{sriv:11}.  Then, we demonstrate how this model can be extended to incorporate other inferential procedures.  The two examples provided in this paper are extensions for both a functional prediction model and a model for simultaneous registration and smoothing. 

This paper also addresses the computational issues associated with high-dimensional Bayesian hierarchical models.  To this end, we propose an alternative algorithm to variational Bayes approximation that can be used for models in which the full conditional distributions of a subset of the parameters are not from a known parametric family.  To distinguish our algorithm from pure variational Bayes, in this paper we will refer to this approximate inference procedure as  Adapted Variational Bayes (AVB).  

This paper is organized as follows.  Section \ref{sec:meth} presents our basic registration model.   The Adapted Variational Bayes algorithm is discussed in detail in Section \ref{sec:VB}.   A comparison of results from our model to current methods can be found in Section \ref{sec:OM}.  Additionally, a comparison of results obtained using AVB and those given by MCMC can be found in \ref{sec:MCMC}.  The prediction model is presented in Section \ref{sec:pred}.  In Section \ref{sec:elnino},  the prediction model is used to forecast the future trajectory of sea-surface temperatures that are associated  with the El-Ni\~no phenomenon.  An adaptation to our model that allows for noisy data is found in Appendix C.  Finally, a discussion is found in Section \ref{sec:conc}.

\section{GAUSSIAN PROCESS MODELS FOR REGISTRATION}
\label{sec:meth}

The functional registration models proposed in this paper are foremost designed to extend and improve on the minimum eigenvalue registration criterion for continuous registration first introduced by \citet{ram:98} .  Accordingly, we will consider two functions perfectly registered if the variation between the two functions can be described entirely in terms of one functional direction - the \textit{target function}.  Our method of registration improves on Ramsay and Li's Procrustes method, \citet{ram:98}, by implicitly accounting for vertical shifts between registered functions and by allowing the target curve to evolve throughout the registration procedure. In Section \ref{sec:comp}, we will demonstrate how using the minimum eigenvalue criterion under these conditions provides a more complete curve registration.  Our results are comparable to those of \citet{sriv:11}.  

The theoretical basis for modeling functional data as Gaussian processes in a hierarchical Bayesian environment is established in \citet{earls:14}. In our registration model, each registered function, $X_i(h_i(t))$, $i = 1, \hdots, N$, is the composition of an observed unregistered function, $X_i(t)$, with an unknown warping function, $h_i(t)$, over some fixed time domain $\mathcal T = [t_1,t_p]$.  The function $h_i(t)$ is represented as  $h_i(t) = t_1 + \int_{t_1}^{t} \exp{({w_i(s)}})ds$ to enforce its monotonicity where $w_i(t)$ is specified as having a Gaussian process distribution, resulting in a functional random effect. In this paper, we will refer to $w_i(t)$ as the \textit{base function} associated with warping function $h_i(t)$.  
The base functions are non-parametrically specified for optimal registration.  We, however, impose the following restrictions on the warping functions:

\begin{enumerate}
\item{$h(t_1) = t_1$}, 
\item{$h(t_p) = t_p$}, and 
\item{if $t_k>t_j$, then $h(t_k)>h(t_j)$ for all $t_k, t_j \in \mathcal{T}$}.
\end{enumerate}

Restrictions (1) and (3) are built into the definition of $h_i(t)$.  Restriction (2) is imposed through the characteristic function in the expression for the prior defined for each base function, $w_i(t)$.  Note that $w_i(t)=0$ corresponds to the identity warping, $h_i(t) = t$.  An important feature of our definition of the warping functions is that it defines an identifiable relationship between $h_i(t_j)$ and $w_i(t)$ which is necessary for predicting future outcomes of curves that are only partially observed. In Section \ref{sec:pred}  is a more thorough discussion of the prediction model.

We also model each registered function, $X_i(h_i(t))$, $i = 1\hdots N$, as a Gaussian process such that  
\begin{eqnarray}
X_i(h_i(t))\mid z_{0i},z_{1i}, f(t)  \sim GP(z_{0i}+z_{1i}f(t),\gamma_R^{-1}\Sigma(s,t)),\quad s,t\in \mathcal{T}.
\end{eqnarray}
Here, $f(t)$ is the \textit{target function}. The target function serves as the primary functional direction in which the registered functions vary.  Accordingly, the above covariance function, $\gamma_R^{-1}\Sigma(s,t)$, is defined to penalize all variation in the registered functions beyond a scaling and vertical shifting of the target function (the function-specific mean). In these models we will define $\gamma_R$ as a registration parameter that determines the severity of this penalty.  

Given a sample of unregistered functions, $X_i(t)$, $i=1\hdots N$, defined over the interval $\mathcal T = [t_1,t_p]$, we are interested in estimating the warping functions, $h_i(t)$, the shifting and scaling parameters, $z_{0i}$ and $z_{1i}$, the target curve, $f(t)$, and the registered functions. 

For now, we will assume the functions are recorded without noise.  If the functions are recorded with noise, it is common practice in the current literature to first perform a pre-processing smoothing step.  An undesirable result of this pre-processing step is that the subsequent inference procedure is unable to capture the extra variability associated with the smoothing process.  This phenomenon is illustrated using the Berkeley boys growth velocity data in Appendix C.3.  Appendices C.1 and C.2 detail how our basic registration model can be modified to both smooth and register functions.

Inference is accomplished through a Bayesian hierarchical model.  The distributional assumptions and prior specifications for this model are

\begin{eqnarray}
X_i(h_i(t))\mid z_{0i},z_{1i}, f(t) &\sim& GP(z_{0i}+z_{1i}f(t),\gamma_R^{-1}\Sigma(s,t)), \quad s,t\in \mathcal {T},\quad i=1, \hdots, N\label{eq:reg}, \\
\Sigma(s,t) &=& P_1(s,t) + P_2(s,t), \label{eq:regcov}\\
h_i(t) &=& t_1 + \int_{t_1}^{t} \exp({w_i(s)})ds,  \quad t \in \mathcal T, \quad i=1, \hdots, N,\nonumber \\
w_i(t) &\propto& GP(0,\gamma_w^{-1}\Sigma(s,t)+\lambda_w^{-1}P_w(s,t))\mathbf{1}\{ t_1 + \int_{t_1}^{t_p} \exp({w_i(s)})ds=t_p\}, \nonumber \\
&& s,t\in \mathcal {T}, \quad i=1, \hdots, N, \label{eq:base}\\
P_w(s,t) &=& P_2(s,t), \label{eq:basecov} \\
z_{0i}\mid \sigma_{z0}^2 &\sim& N(0,\sigma_{z0}^2), \quad i=1, \hdots, (N-1), \quad z_{0N}=-\sum_{i=1}^{N-1} z_{0i},  \nonumber\\
\sigma_{z0}^2 &\sim& IG(a,b), \nonumber \\
z_{1i}\mid \sigma_{z1}^2 &\sim& N(1,\sigma_{z1}^2),  \quad i=1, \hdots, N, \label{eq:z1}\\
\sigma_{z1}^2 &\sim& IG(a,b), \label{eq:sigz1}\\
f(t) \mid \eta_f, \lambda_f &\sim& GP(0,\Sigma_f(s,t)), \quad s,t\in \mathcal {T}, \label{eq:target}\\
\Sigma_f(s,t) &=& \eta_f^{-1}P_1(s,t) + \lambda_f^{-1}P_2(s,t), \label{eq:targcov} \\
\eta_f &\sim& G(c,d), \text{ and} \nonumber\\
\lambda_f &\sim& G(c,d).  \nonumber
\end{eqnarray}

For this model, the parameters, $z_{0i}, i = 1, \hdots N$, allow the registered functions to vary by vertical shifts from a scaling of the target function, $f(t)$.  The constraint, $z_{0N}=-\sum_{i=1}^{N-1} z_{0i}$, ensures the average vertical shift is estimated to be 0.  The parameters, $z_{1i}, i = 1, \hdots N$ , quantify amplitude variation in the registered functions. Note, the Gaussian distribution on $\mathbf z_1= (z_{11}\hdots z_{1N})'$ can be replaced by a Dirichlet distribution on $\mathbf z_1/N$.  The result is a slightly more complicated model that has the nice effect of scaling the target function to the empirical mean of the estimated registered functions. Priors \eqref{eq:z1} and \eqref{eq:sigz1} would then be omitted.

In the above model specifications, all covariance functions are composed of a linear combinations of two bi-variate functions, $P_1(s,t)$ and $P_2(s,t)$.  $P_1(s,t)$ penalizes variation in constant and linear functions and $P_2(s,t)$ penalizes function variability in all other directions. Together they define a proper covariance function.   For each covariance function above, the specification of the registration and smoothing parameters indicate the extent the two different types of variability should be penalized for each function.  For example, for both the registered functions and the  base functions, we want to penalize variation in \textit{any} direction other than that of the mean function.  The covariance specifications of  $\gamma_R^{-1}\Sigma(s,t)$ and $\gamma_w^{-1}\Sigma(s,t)$ reflect these penalties, where the magnitude of the penalty is controlled by registration parameters, $\gamma_R$ and $\gamma_w$, (distributional assumptions  \ref{eq:reg},\ref{eq:regcov}, and \ref{eq:base}).  We can use $P_2(s,t)$ to penalize roughness in a given function.  Here we would like both the target function and the base functions to be smooth.  This is achieved by the inclusion of  $\lambda_f^{-1}P_2(s,t)$ and $\lambda_w^{-1}P_2(s,t)$ in the priors for these functions (distributional assumptions \ref{eq:target}, \ref{eq:targcov}, \ref{eq:base}, and \ref{eq:basecov}) where the level of the penalty is controlled by the smoothing parameters $\lambda_f$ and $\lambda_w$.  For the exact definitions of $P_1(s,t)$ and $P_2(s,t)$, see \citet{earls:14}.  

Given the above model, in practice we will proceed by using finite approximations to each function and functional distribution.  In \citet{earls:14} we establish some theoretical properties of these types of approximations.  The following finite-dimensional distributions are used in the final model in lieu of their infinite dimensional counterparts above:

 \begin{eqnarray}
\mathbf{X}_i(\mathbf{h}_i)\mid z_{0i},z_{1i}, \mathbf f &\sim& N_p(z_{0i}\mathbf{1}+z_{1i}\mathbf {f},\gamma_R^{-1}\boldsymbol{\Sigma}), \quad i=1\hdots N, \label{eq:RegPrior}\\
\mathbf{h}_i(t_j) &=& t_1 + \sum_{k=2}^{j} (t_k-t_{k-1})\exp({w_i(t_{k-1})}),  \quad  i=1\hdots N, \quad j = 1 \hdots p, \nonumber \\
\mathbf w_i &\propto& N_{p-1}(\mathbf 0,\gamma_w^{-1}\boldsymbol\Sigma+\lambda_w^{-1}\mathbf P_w)\mathbf{1}\{  t_1 + \sum_{k=2}^{p} (t_k-t_{k-1})\exp({w_i(t_{k-1})})=t_p\}, \nonumber\\
&& i=1\hdots N,  \label{eq:Prior}  \\
\mathbf f \mid \eta_f, \lambda_f &\sim& N_p(0,\boldsymbol\Sigma_f), \text{ and} \nonumber\\
\boldsymbol\Sigma_f &=& \eta_f^{-1}\mathbf{P}_1+\lambda_f^{-1}\mathbf{P}_2.\nonumber\label{eq:P1}
\end{eqnarray}

Section \ref{sec:comp} provides several examples that illustrate how allowing the target function to be estimated within the model results in a more complete functional registration in comparison to the Procrustes method, \citet{ram:98}.  However, the Gaussian process model does not constrain the timing of a feature in the target function to occur at the average time of the corresponding unregistered features.  Although the model for the $w_i(t)$ is centered on zero, it is still possible that the average of the estimated warped time points, $\overline{h_\mathbf{{\cdot}}(t_1)}$,$\ldots$,$\overline{h_\mathbf{{\cdot}}(t_p)}$, does not correspond to the original time points. Shifting these by an additional registration so that the warped times average to the original time does not affect our prediction model, but it will then allow an explicit comparison of $h_i(t_j)$ to $t_j$ to tell us whether the process is running ahead or behind ``standard'' time.  \citet{sriv:11} use a similar ``correction" to determine their target function.  Details on how to perform this final registration can be found in Appendix A.3.

\subsection{Registration and Warping Parameter Selection}

Registration in this model is controlled by three parameters, $\gamma_R$, $\gamma_w$, and $\lambda_w$.  The parameter $\gamma_R$ determines the extent the registered functions will be penalized for varying from a shifting and scaling of the target function.  This penalty for lack of registration is tempered by penalties for roughness in the warping functions.  The parameter $\gamma_w$ determines how far the warping functions can deviate from the identity warping while $\lambda_w$ controls the smoothness of the warping functions.  This model can also be adapted to allow for function specific warping penalties.  In Section \ref{sec:elnino}, we will give an example where function specific penalties for the base functions have been utilized to preserve significant covariance relationships in the estimated registered functions.

For a given statistical analysis, the registration parameters are chosen by the user.  This is because variation in registered functions outside of shifts and rescaling $f(t)$ (controlled by $\gamma_R$) is nearly non-identifiable from variation in warping functions, as regulated by $\gamma_w$ and $\lambda_w$. That is, there are linear directions of variation in the $w(t)$ that result in nearly linear directions of variation in the resulting $X_i(t)$ (a simplified example is the identity $\sin(t + \delta) = \cos(\delta) \sin(t) + \sin(\delta) \cos(t)$ which confounds horizontal shifts with vertical variation in periodic functions).   An exact description of this form of confounding is beyond the scope of this paper, but it yields unstable estimates when these parameters are not fixed and we therefore choose these by hand. In this model a large registration penalty, $\gamma_R$, in comparison to the penalty on warping, $\gamma_w$, will result in registered functions that no longer retain significant features in the data.  Alternatively, a registration parameter that is too small will not properly align features.  Desirable values of these parameters can be determined using short runs of the adapted variational Bayes algorithm described in Section \ref{sec:AVB}.  In practice, we have found these penalties should be adjusted by powers of ten to see a significant change in estimates of the registered functions.  Once determined, $\gamma_R$, $\gamma_w$, and $\lambda_w$ are fixed and can be used with the adapted variational Bayes estimates to initialize an MCMC sampler.  

\section{VARIATIONAL APPROXIMATION}
\label{sec:VB}

\subsection{Variational Bayes}
\label{sec:AVB}

For our registration model, it is appropriate to use Markov Chain Monte Carlo (MCMC) methods to sample from the joint posterior distribution of all unknown parameters.  However, for most applications, the dimensionality of the parameter space will require exceptionally long chains that are impractical and expensive to obtain.  Here, we suggest a variational Bayes alternative to MCMC sampling to at the very least obtain good starting values for a MCMC sampler.  Alternatively, we will show in Section \ref{sec:MCMC} that differences in the estimated parameters obtained through adapted variational Bayes and MCMC sampling tend to be small, and estimation via adapted variational Bayes alone is likely sufficient for many inferential procedures.

The variational Bayes procedure described here is based on the variational methods proposed by \citet{omer:10} and \citet{bish:06}.  Their proposed method optimizes a lower bound of the marginal likelihood which results in finding an approximate joint posterior density that has the smallest Kullback-Leibler (KL) distance from the true joint posterior density.  Both fixed form and nonparametric forms of variational Bayes algorithms are currently available.  The variational Bayes algorithm that we propose is most closely related to fixed form variational Bayes.  A clear explanation of fixed form variational Bayes can be found in \citet{gold:11} where the authors utilize variational Bayes for a functional regression model.

Suppose $q(\boldsymbol\theta)$ is the approximated posterior joint distribution.  The fixed form variational Bayes algorithm assumes for some partition of $\boldsymbol\theta = \{\boldsymbol\theta_1, \hdots ,\boldsymbol\theta_d\}$, $q(\boldsymbol\theta) = 
\prod_{k=1}^{d}q_k(\boldsymbol\theta_k)$, where each distribution $q_k$ is of a known parametric form.  Traditionally this requirement is satisfied through the use of conditionally conjugate priors.

In our model, the Gaussian process priors for the base functions, $w_i(t)$, $i=1, \hdots, N$, are not conditionally conjugate to the likelihood function.  Therefore, the fixed form variational Bayes optimization method does not apply directly since $q_k(\mathbf w_i)$, $i=1, \hdots, N$ are not known parametric distributions. 

\subsection{Adapted Variational Bayes (AVB)}

Suppose we order the parameter vector, $\boldsymbol\theta$, so that, $\boldsymbol\theta = \{\mathbf w_1, \hdots , \mathbf w_N, \boldsymbol\theta_{N+1}, \hdots, \boldsymbol\theta_d\}$, for  $k= \{(N+1), \hdots, d\}$.  While the $q$ distributions on the approximated base functions, $\mathbf w_i$, $i=1, \hdots, N$ are not of known parametric forms, each $q_k(\boldsymbol\theta_k)$, $k= \{(N+1), \hdots, d\}$, is a known parametric distribution that can be estimated using the standard fixed form variational Bayes algorithm. The following variational Bayes algorithm is adapted to include estimation for parameters without conditionally conjugate priors in addition to all other parameters that typically can be estimated using fixed form variational Bayes.  This adaptation is similar to the variational approximation to the EM algorithm described by \citet{tzikas:08} which performs approximate inference based on the EM algorithm for models where the posterior distribution of the latent variables is of an unknown form.

\textbf{The Adapted Variational Bayes Algorithm}  

Define $f(\mathbf X,\mathbf w, \boldsymbol\theta)$ as the joint distribution of the data, $\mathbf X$,  and parameters of a Bayesian hierarchical model.  Suppose an approximate joint posterior distribution of the parameters, $\boldsymbol\theta = \{\mathbf w_1, \hdots , \mathbf w_N, \boldsymbol\theta_{N+1}, \hdots, \boldsymbol\theta_d\}$, is of the form $\prod_{j=1}^{N}\prod_{k=N+1}^d q_j(\mathbf{w_j})q_k(\boldsymbol{\theta_k})$ where each $q_j(\mathbf{w_j}), j = 1, \hdots, N$, is known only up to a constant of proportionality and the distributions, $q_k(\boldsymbol\theta_k), k= N+1,\hdots, d$, are of known parametric forms.  We will define the following estimation procedure for $\boldsymbol\theta$ as the adapted variational Bayes algorithm. 

\begin{enumerate}
\item Initialize $\boldsymbol\theta$.
\item For each iteration, $m$, and each $k$, $k = 1, \hdots, N$, update the estimate for \\ $\mathbf w_k$ so that $\mathbf w_k^{(m)} = \sup_{\mathbf w_k}$$q_k(\mathbf w_k\mid \boldsymbol\theta_j^{(m-1)}, j = (N+1), \hdots, d$).  This is equivalent to setting $\mathbf w^{(m)} = \sup_{\mathbf w}$$f(\mathbf X,\mathbf w \mid \boldsymbol\theta_j^{(m-1)}, j = (N+1), \hdots, d$).
\item For each iteration, $m$, and each $k$, $k = (N+1), \hdots, d$, update $q_k$ so that $q_k^{(m)}$ $\propto$ $\exp[E_{(\boldsymbol\theta_{-k})}(\log f(\boldsymbol\theta_k\mid rest)]$, where the expectation is taken with respect to the distributions $q_j^{(m-1)}(\boldsymbol\theta_j)$, $j=1,\hdots, d$, $j\neq k$.
\item Repeat steps (2) and (3) until the desired convergence criterion is met.
\end{enumerate}

Here the notation, $E_{(\mathbf\theta_{-k})}$, denotes the expected value over all parameters except $\theta_k$.  In the next section, we will drop the subscript $k$, and $E_{(\mathbf\theta_{-\theta_k})}$ will represent the expectation over all parameters except for $\theta_k$ (e.g. $E_{(\mathbf\theta_{-\eta_f})}$ will represent the expectation taken over all parameters except for $\eta_f$).  \\

\textbf{Theorem} \textit{The adapted variational Bayes algorithm converges to estimated parameters, $\hat{\boldsymbol{\theta}}$, that minimize the Kullback-Leibler distance between the approximate posterior distribution, $q(\boldsymbol{\theta_{-\mathbf w}})$, and the posterior distribution, $f(\boldsymbol{\theta_{-\mathbf w}}\mid \mathbf X, \mathbf w)$, for a local optimization of $f(\mathbf X,\mathbf w \mid \boldsymbol{ \theta_{-\mathbf w}})$ in $\mathbf w$.}\\

\textbf{\textit{Proof}:} Assume $\mathbf w$ is known.  \citet{gold:11} demonstrate that minimizing the K-L distance between $q(\boldsymbol{\theta_{-\mathbf w}})$ and  $f(\boldsymbol{\theta_{-\mathbf w}}\mid \mathbf X, \mathbf w)$ is equivalent to maximizing the following log of a q-specific lower bound of the joint marginal distribution of $\mathbf X$ and $\mathbf w$ in $q$.
 
\begin{eqnarray}
    \log (\mathbf X,\mathbf w; q) &=& \int q(\boldsymbol\theta_{-\mathbf w}) \log \left \{\frac{f(\mathbf X,\mathbf w,\boldsymbol\theta_{-\mathbf w})}{q(\boldsymbol\theta_{-\mathbf w})}\right \}d\boldsymbol\theta_{-\mathbf w} \nonumber \\
    &=& E_{q(\boldsymbol\theta_{-\mathbf w})}[\log[f(\mathbf X,\mathbf w,\boldsymbol\theta_{-\mathbf w})]-\log[q(\boldsymbol\theta_{-\mathbf w})]] \nonumber
\end{eqnarray}

The adapted variational Bayes algorithm alternates between: 1) maximizing $f(\mathbf X,\mathbf w \mid \boldsymbol{ \theta_{-\mathbf w}})$ in $\mathbf w$ (possibly locally), and 2) fixing $\mathbf w$ at the value determined by the previous step and using traditional variational Bayes to maximize $f(\mathbf X,\mathbf w; q)$.  Here we demonstrate this process results in a monotonic increasing sequence in $\log f(\mathbf X,\mathbf w; q)$ which guarantees the convergence of this algorithm.

For each iteration, $m$ of our adapted variational Bayes algorithm,
\begin{eqnarray}
    \log f(\mathbf X,\mathbf w^{(m)}; q^{(m)}) &=&  E_{q(\boldsymbol\theta_{-\mathbf w})}[\log[f(\mathbf X,\mathbf w^{(m)},\boldsymbol\theta^{(m)}_{-\mathbf w})]-\log[q(\boldsymbol\theta^{(m)}_{-\mathbf w})]] \nonumber\\
    &\leq&  E_{q(\boldsymbol\theta_{-\mathbf w})}[\log[f(\mathbf X,\mathbf w^{(m+1)},\boldsymbol\theta^{(m)}_{-\mathbf w})]-\log[q(\boldsymbol\theta^{(m)}_{-\mathbf w})]] \label{eq:optstep} \\
    &\leq&  E_{q(\boldsymbol\theta_{-\mathbf w})}[\log[f(\mathbf X,\mathbf w^{(m+1)},\boldsymbol\theta^{(m+1)}_{-\mathbf w})]-\log[q(\boldsymbol\theta^{(m+1)}_{-\mathbf w})]] \label{eq:vbstep} \\
    &=& \log f(\mathbf X,\mathbf w^{(m+1)}; q^{(m+1)}). \nonumber
          \end{eqnarray}
          
The inequality in \eqref{eq:optstep} is guaranteed by step 2 of the adapted variational Bayes algorithm, and the inequality in \eqref{eq:vbstep} is the result of using the traditional variational Bayes algorithm with $\mathbf w$ considered known (step 3).   \\

The lower bound of the marginal distribution of $\mathbf X$ and $\mathbf w$ can be monitored until changes in this function are under some threshold. The specific form of this function can be found in Appendix B.2.  However, as the algorithm is guaranteed to converge, it is in practice more prudent to instead monitor changes in the parameter estimates from iteration to iteration and stop the algorithm when these changes are below a specified threshold. 

Convergence of the AVB algorithm is guaranteed.  However, convergence to a global maximum is not guaranteed.  In the maximization step of the AVB algorithm, a function proportional to the approximate posterior for the base functions is maximized in the base functions.  This function can be multimodal and occasionally the estimated base functions reflect a local maximum of the approximated posterior.  To circumvent this problem, in practice it is sometimes necessary to adjust the registration and warping penalties as the functions become registered.  An unregistered function that requires a substantial amount of warping can cause convergence to a local maximum due to the small penalty on warping.  The flexibility in warping allowed by this small penalty can cause the function to deform rather than register.  This can be remedied in two ways.  The first option is to perform a simple initial warping for this function that prevents the optimization from falling into a local mode.  The second option is to adjust the registration and warping parameters over time.  Initially a stronger warping penalty is employed to prevent function deformation.  Then, as the functions register, the warping penalty can be reduced to allow for a more complete registration.  When initializing an MCMC sampler, the final penalties on warping and registration from the adapted variational Bayes algorithm should be used.

\begin{table}
\centering
\caption{Computational Time (seconds)}
\begin{tabular}{ c c c c }
\hline\hline
Method & Sim & BGV & RBGV \\ [0.5ex]
\hline
ME & 176 & 1636 & NA \\
F-R & 1.3 & 2.9 & NA \\
AVB  & 1820 & 36540 & 28692 \\
MCMC  & 222 & 271296 & 46037 \\  
\hline
\end{tabular}
\label{table:time}
\end{table}

While AVB estimates are often close to MCMC estimates, often it will still be desirable to characterize the posterior distributions of all parameters through MCMC sampling. This is particularly prudent when noise is present in the observed unregistered functions which introduces a significant amount of variability in the posterior samples (see Appendix C.3).  We can express the computational savings due to using the AVB algorithm as the amount of time saved in burn-in iterations.  When applying the adapted variational Bayes algorithm to ``real" data such as the \textit{Berkeley Boys Growth Velocity Data}, we have found the adapted variational Bayes algorithm saves a significant amount of computational time in the burn-in period.  For simulated data, there may not be any savings in computational time due to the following: 1) MCMC samples move very quickly towards the optimal solution when the registration problem is perfectly defined and 2) the maximization step of the AVB algorithm is inefficient, especially in the first few iterations.  In Table \ref{table:time}, we compare the computational cost of obtaining estimates through MCMC sampling, AVB, the F-R algorithm, \citet{sriv:11}, and the ME algorithm, \citet{ram:98}, for simulated data, the original  \textit{Berkeley Boys Growth Velocity Data} (BGV), and the boys growth velocity data corrupted with noise (RBGV).  Full descriptions of these datasets and all proposed registration methods are described in Sections \ref{sec:comp} and Appendix C.3.  Since the ME, F-R and AVB algorithms do not include variability estimates, for this comparison the MCMC sampler is run long enough to obtain estimates but not necessarily long enough to characterize the posterior distributions.  Note: since the ME and F-R algorithms treat smoothing as a pre-processing step, for the noisy growth velocity data we only compare the computational time needed to burn-in an MCMC sampler to the computational time required to obtain AVB estimates (that can then be used to initialize an MCMC sampler).

\section{COMPARISON TO CURRENT METHODS}
\label{sec:comp}

\begin{figure}[!ht]
\begin{tabular}{cc}
\centering
\includegraphics[width=8cm]{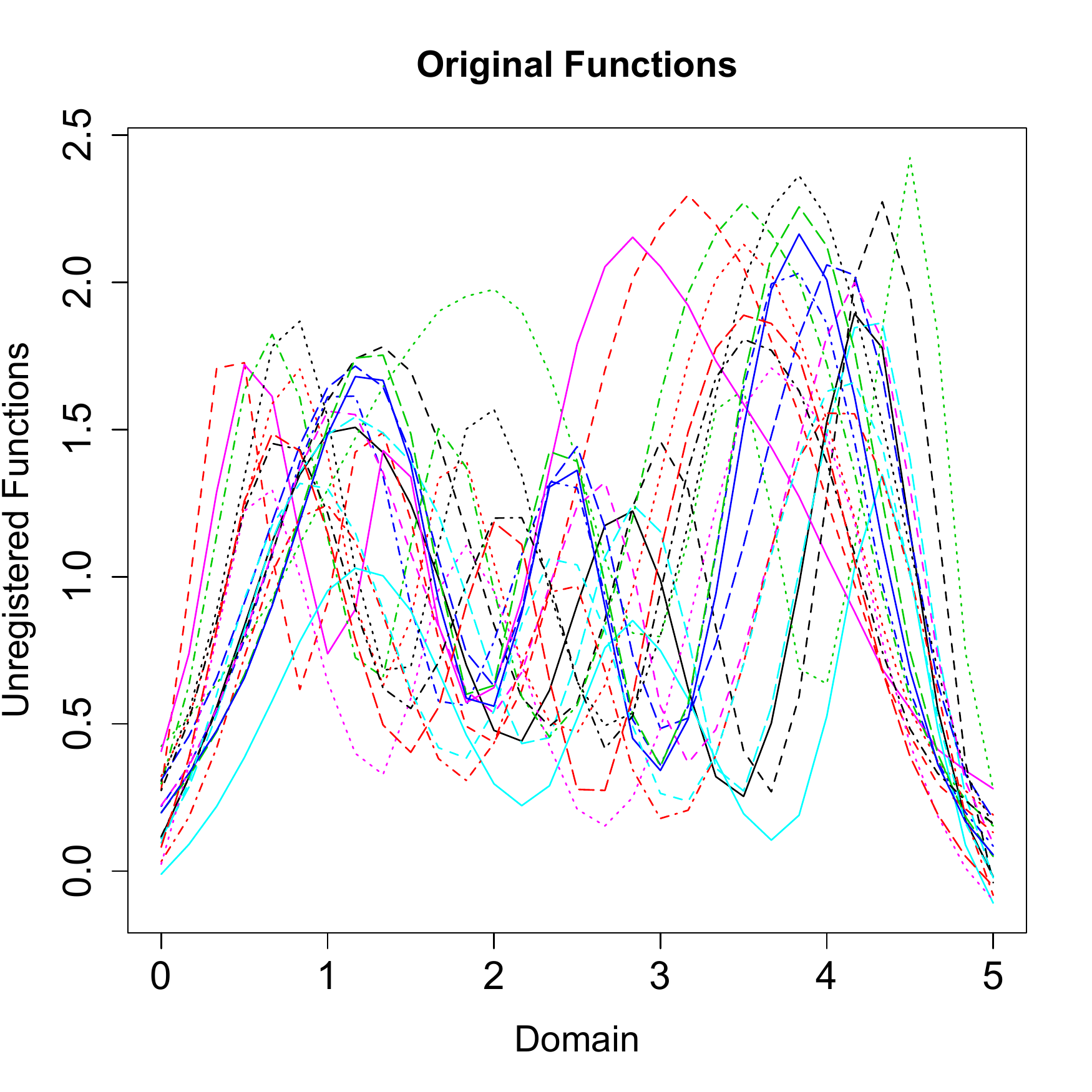} &
\includegraphics[width=8cm]{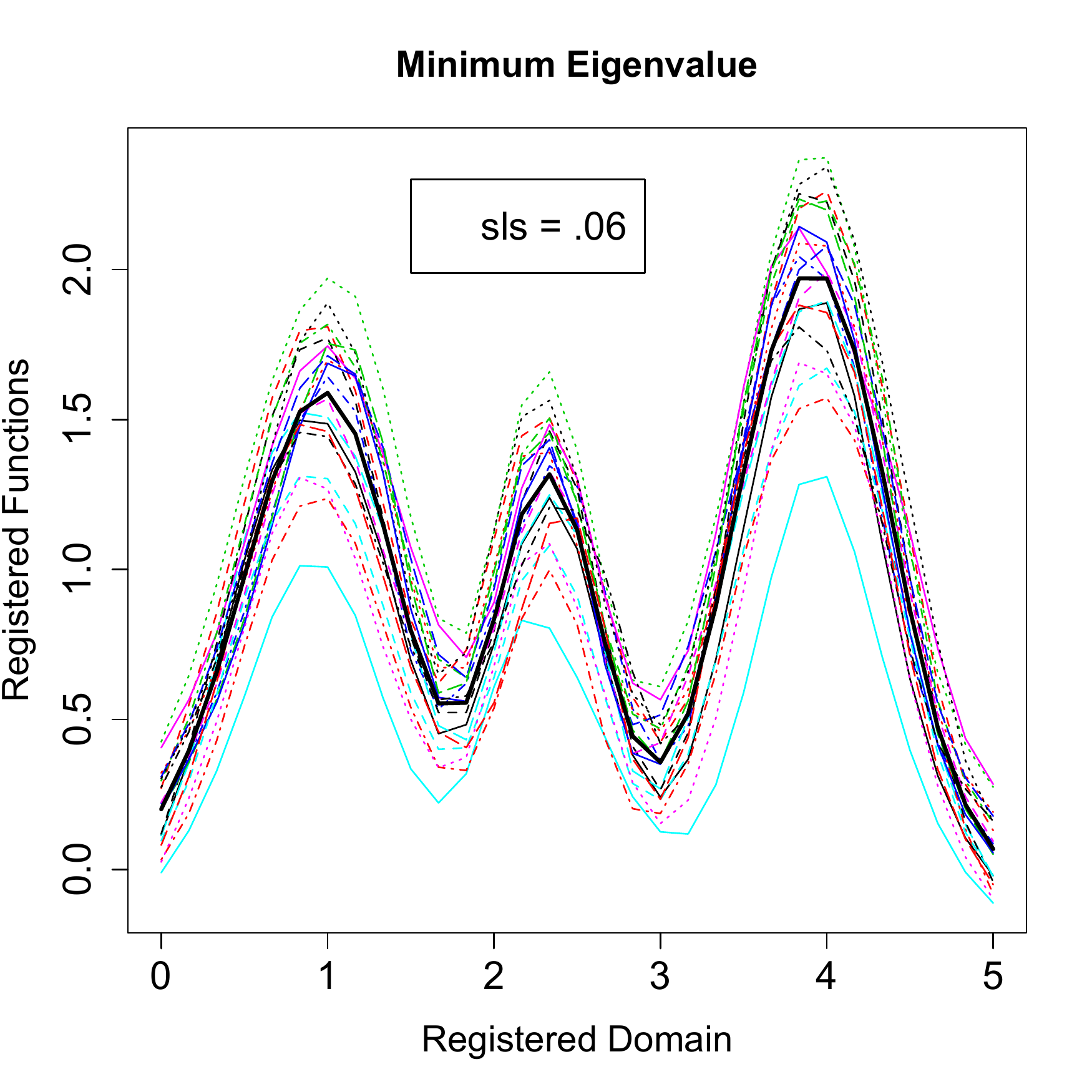}\\
\includegraphics[width=8cm]{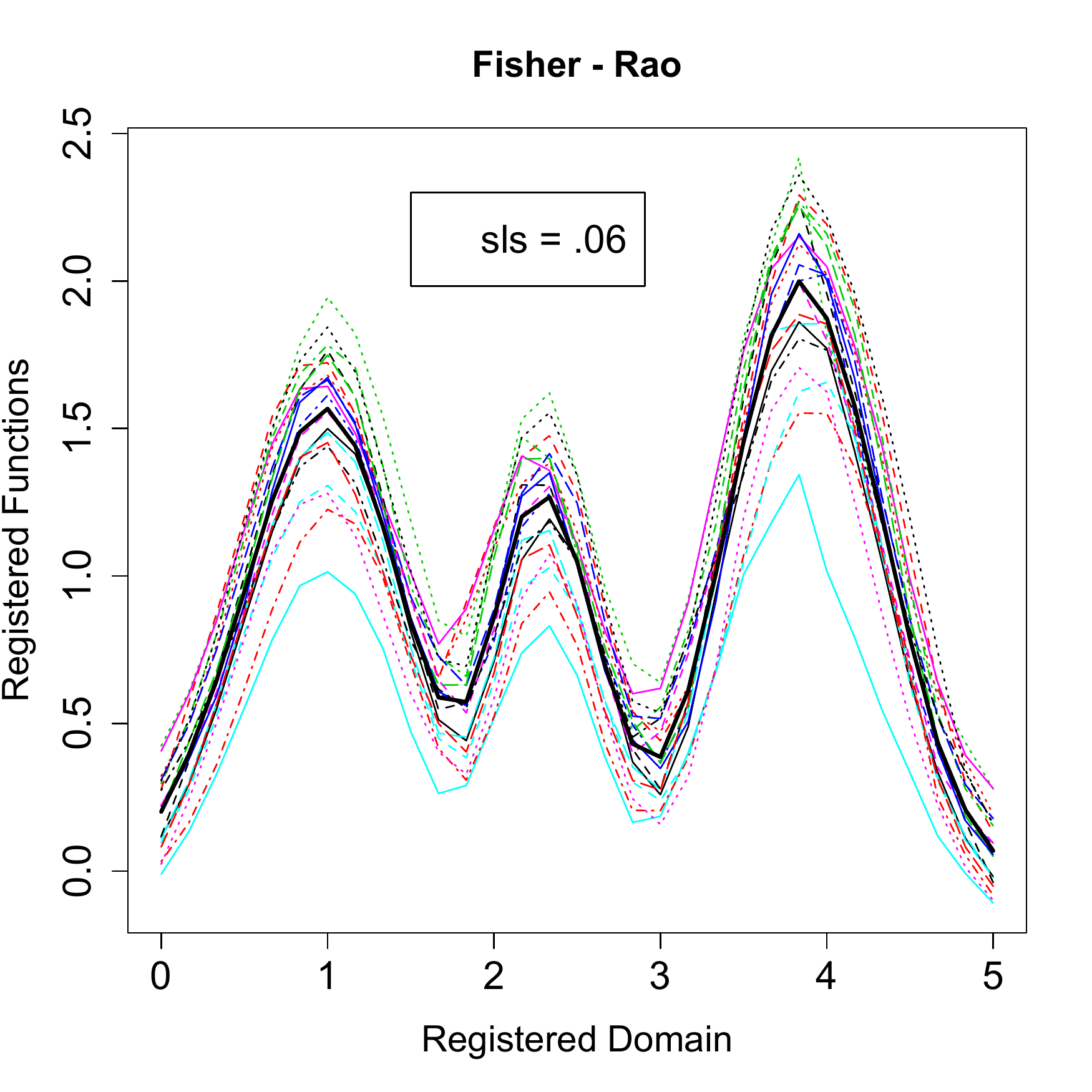}   &
\includegraphics[width=8cm]{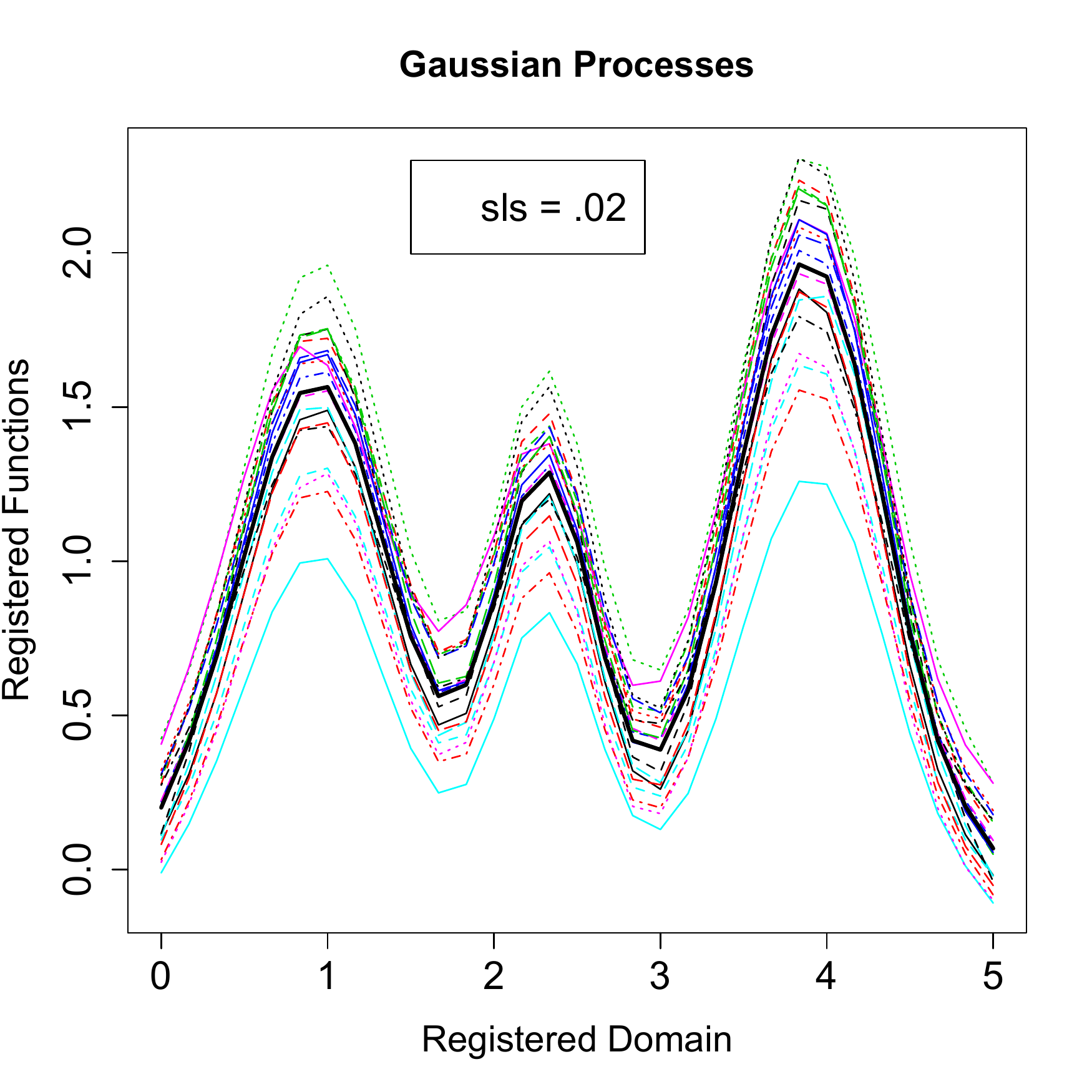}
\end{tabular}
\caption{Simulated Data Set 2.  \textbf{Top Left} Unregistered functions.  \textbf{Top Right} Registered functions using the minimum eigenvalue criteria (R package 'fda'). \textbf{Lower Left}  Functions registered by F-R (R package 'fdasrvf'). \textbf{Lower Right } Functions registered by the GP model.}
\label{fig:S3}
\end{figure}

\subsection{Comparison to Other Registration Procedures}
\label{sec:OM}
Our registration criterion minimizes all variation in the warped functions that is not in the direction of  the target function (allowing for vertical shifts).  In this respect, the underlying registration principle driving our model is similar to that proposed by \citet{ram:98}.   Here we will compare our registration results to those using Ramsey's method as well as the registration procedure proposed by \citet{sriv:11}.   Srivistava et. al.  propose a geometric framework for functional data registration using the Fisher-Rao Riemannian metric, \citet{rao:45}.  In this paper we will refer to Ramsey and Li's registration procedure as ME (minimum eigenvalue, \citet{ram:98}) and Srivistava's procedure as F-R (Fisher-Rao, \citet{sriv:11}), and the model proposed here as GP (Gaussian Processes).  In the paper by Srivistava, et. al., several comparisons of registration under the F-R framework to the registration methods proposed by \citet{gerv:04}, \citet{jam:07}, \citet{liu:04}, and \citet{tang:08} are provided.  In all cases, F-R appears to provide the most complete registration of the given set of functions.  In light of this illustration, we will consider their method as the current frontrunner in registration procedures and use it as the standard for our comparisons.

\begin{figure}[!ht]
\begin{tabular}{cc}
\centering
\includegraphics[width=8cm]{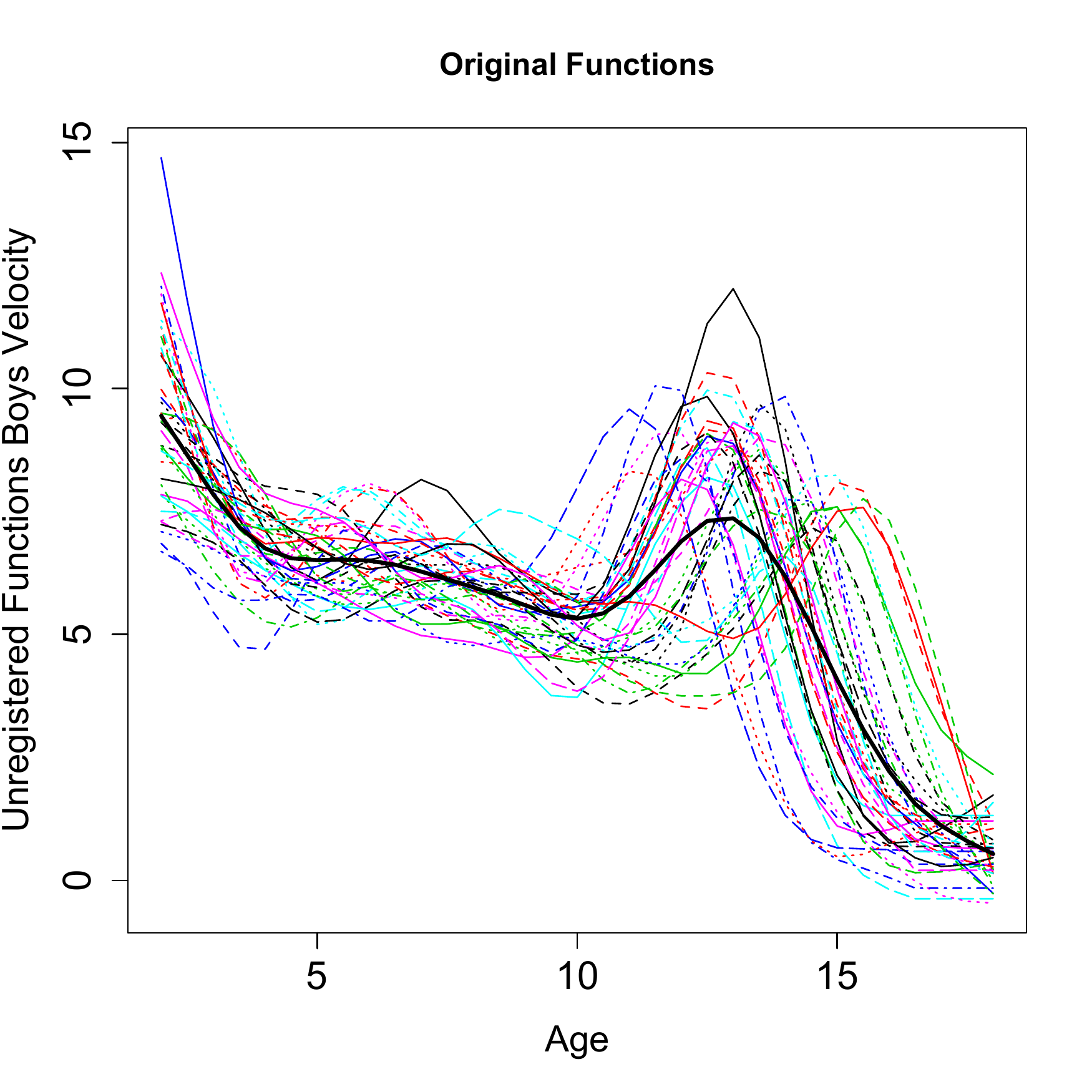} &
\includegraphics[width=8cm]{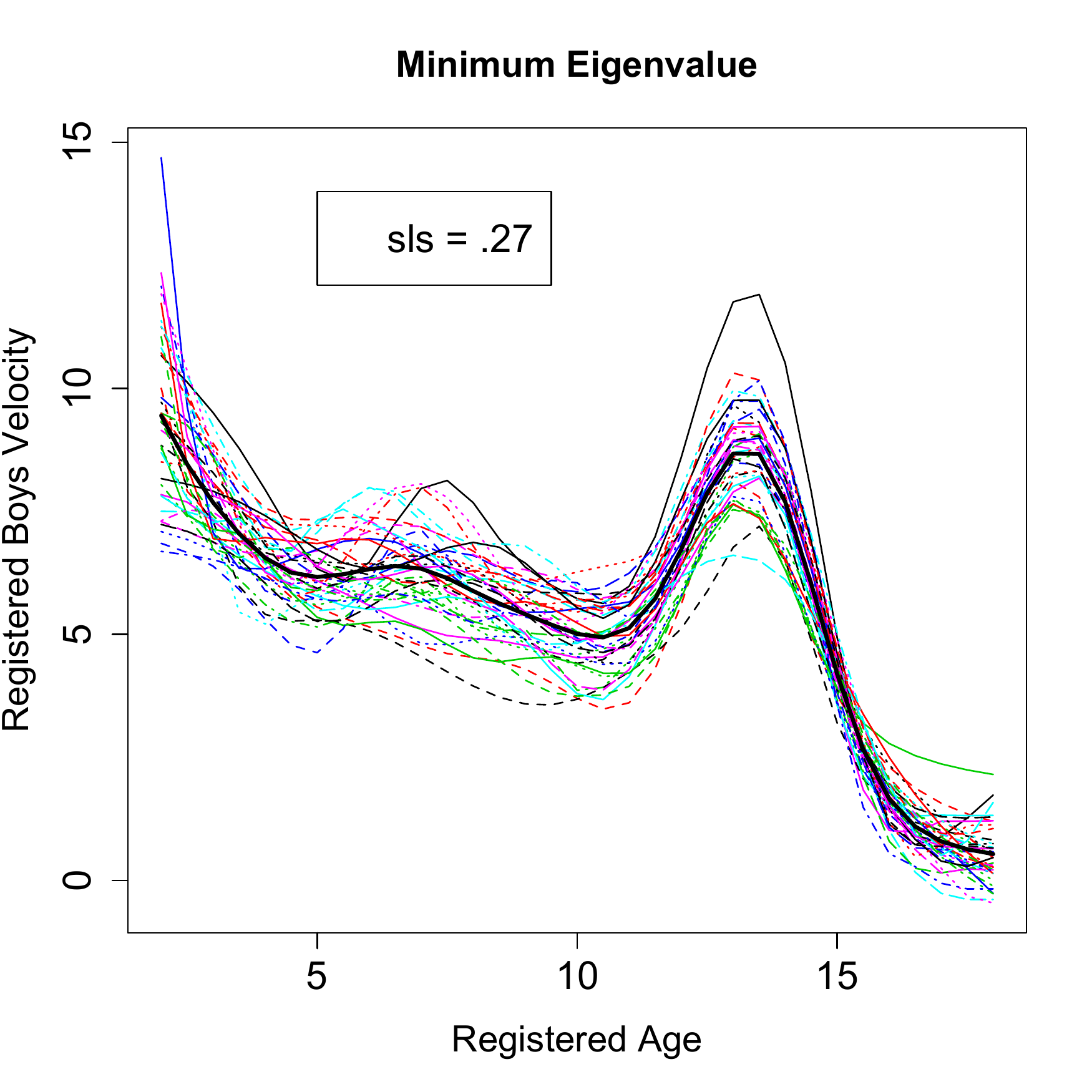}\\
\includegraphics[width=8cm]{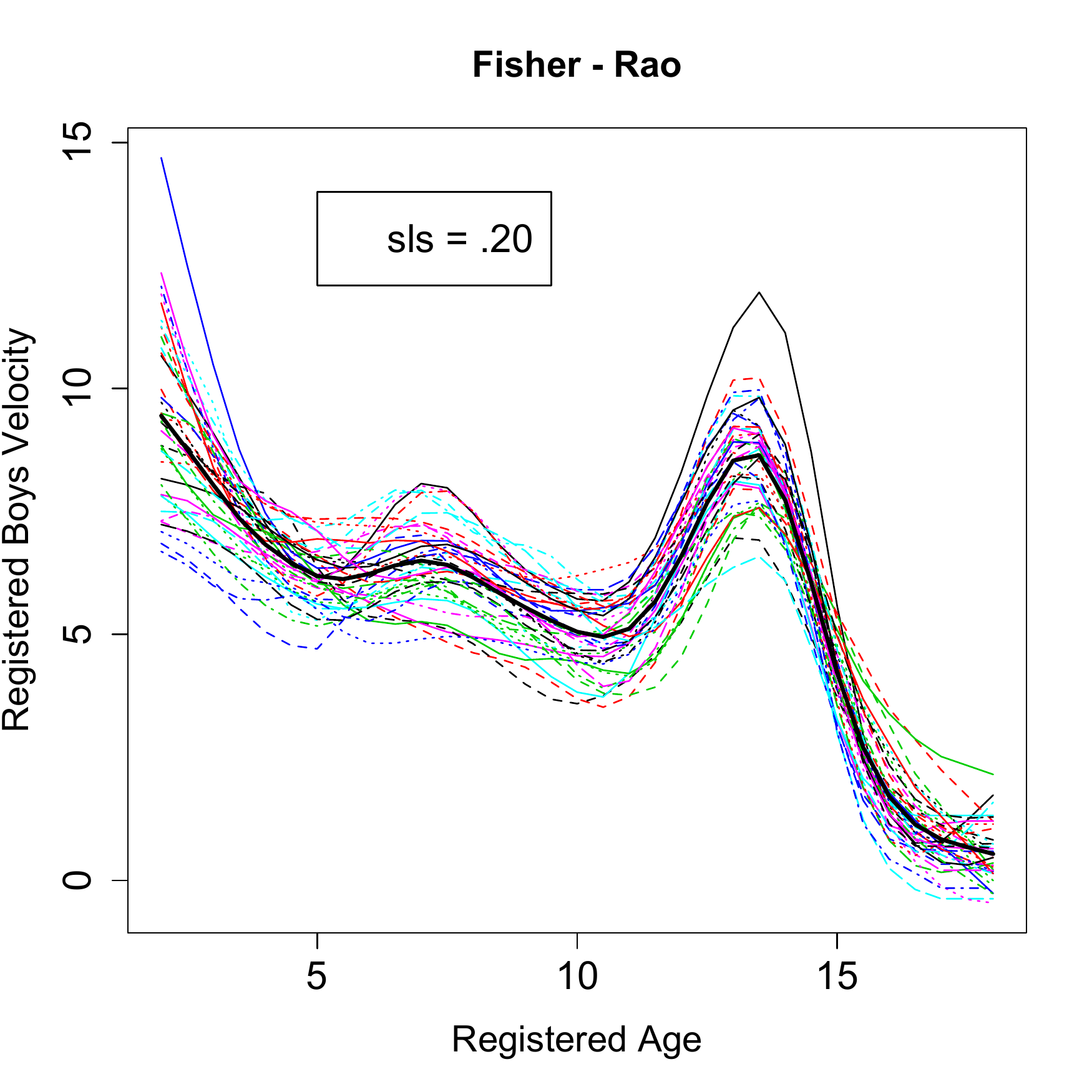}   &
\includegraphics[width=8cm]{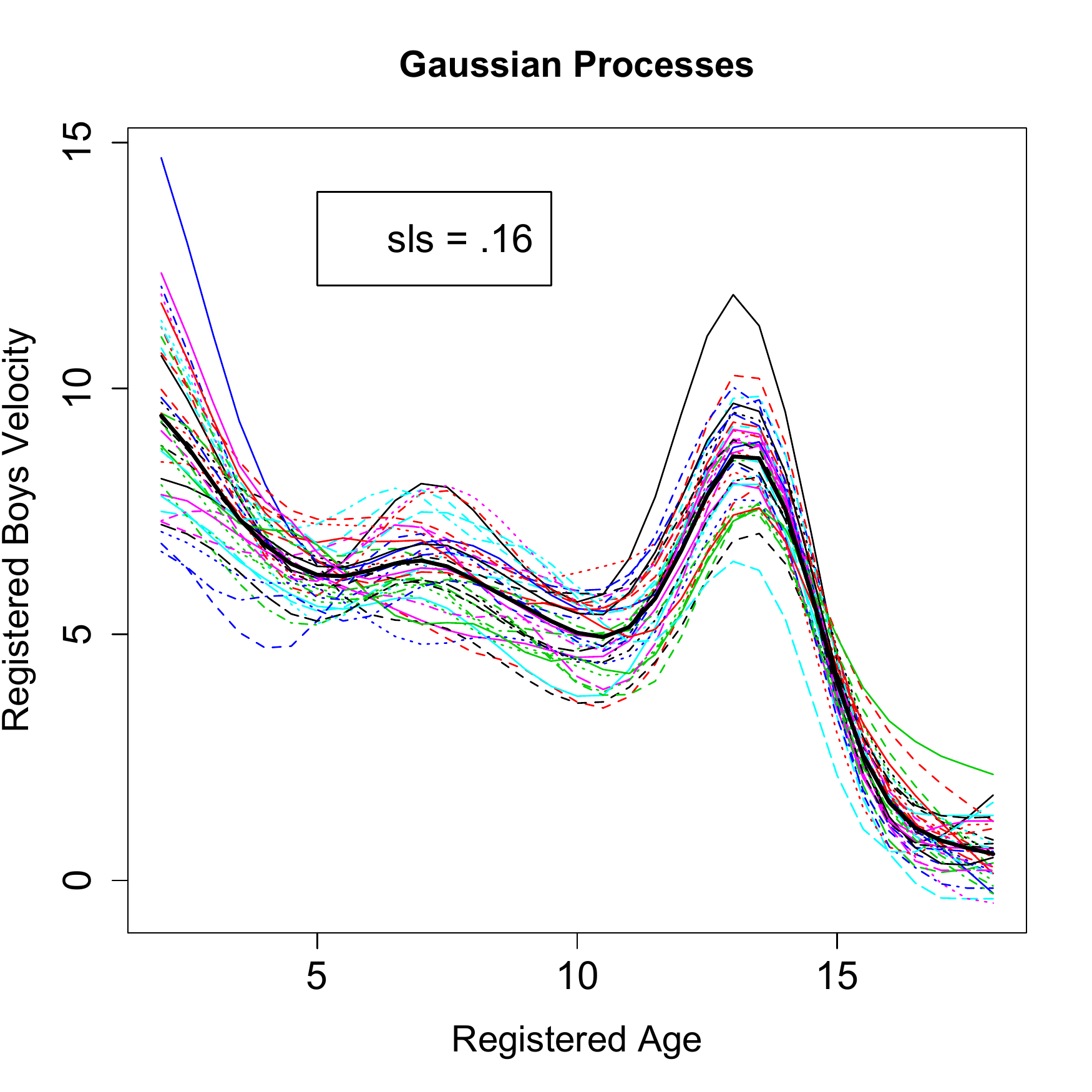}
\end{tabular}
\caption{Registered Boys Growth Velocity.  \textbf{Top Left} Original unregistered boys velocity data functions.  \textbf{Top Right} Boys velocity functions registered using the minimum eigenvalue criteria (R package 'fda'). \textbf{Lower Left}  Boys velocity functions registered by F-R (R package 'fdasrvf'). \textbf{Lower Right } Boys velocity functions registered by the GP model.}
\label{fig:SB}
\end{figure}

Figures \ref{fig:S3} and \ref{fig:SB} contain the datasets used for this analysis.  Each figure includes the original unregistered data along with plots of the functions registered using the three proposed methods.  For all three registration methods, a range of parameter values were explored for optimal registration.  

We have chosen to use the Sobolev Least Squares (\textit{sls}) criterion to compare the three registration methods for each dataset as advocated by \citet{sriv:11}.  The Sobolev Least Squares criterion compares the total cross-sectional variance of the first derivatives of the registered functions to that of the original functions.  Explicitly, 

 \begin{eqnarray}
sls &=& \frac{\sum_{i=1}^N \int (X_i'(h_i(t)) - \frac{1}{N} \sum_{j=1}^{N} X_j'(h_j(t)))^2dt}{\sum_{i=1}^N \int (X_i'(t) - \frac{1}{N} \sum_{j=1}^{N} X_j'(t))^2dt}.\label{eq:sls}
\end{eqnarray}

Lower values of \textit{sls} correspond to better function alignment.

\textbf{Simulated Data Set}
Figure \ref{fig:S3} contains the functions of a simulated data set.  These data consist of 20 unregistered scaled mixtures of three Gaussian probability density functions.  All three registration procedures result in similar alignments.   However, the GP method does a better job of recovering the original shape of the functions and results in the lowest \textit{sls}. Note: The ME registered functions are based on 5 complete runs of the ME algorithm where in each run the previous runs results were used as the `unregistered' functions.

\textbf{Berkeley Boys Growth Velocity Data}
Figure \ref{fig:SB} contains 39 velocity of growth functions for boys from the Berkeley Growth Study, \citet{tud:54}.  For this analysis, the original data are slightly changed to eliminate some erratic behavior at the beginning of each function.  Here, GP and F-R yield similar registration results.  However, the GP algorithm results in the lowest \textit{sls}.  ME registers the most significant peak in growth velocity but does not align lesser features as well as GP.  Note: The ME registered functions are based on 2 complete runs of the ME algorithm where in each run the previous runs results were used as the `unregistered' functions.  Running this algorithm more than twice resulted in function distortion due to over-warping and a larger \textit{sls}.

While the GP and F-R methods result in a similar alignment of functions, these results are achieved in very different environments that are specialized to satisfy specific inferential preferences.  The F-R registration method is convenient (using R package `fdasrvf') and provides fast high-quality estimates.  On the other hand, while providing comparable registration results, our method expands inferential capability by providing 1) variability estimates for all unknown parameters and 2) a probability framework in which future partially observed unregistered functions are considered.  In contrast to traditional functional prediction methods, our model not only provides an estimate of the complete unregistered function, but also estimates the complete warping function and the complete registered function.  Details of the prediction model are found in Section \ref{sec:pred}.  

 \subsection{Comparison to MCMC Results}
\label{sec:MCMC}
To establish the utility of the adapted variational Bayes algorithm, here we compare the estimates of registered functions using adapted variational Bayes versus those obtained through MCMC sampling.  For this exposition, the simulated data set and the Boys Growth Velocity data set described in Section \ref{sec:OM} are used to look at the discrepancies between the estimated registered functions  from MCMC sampling versus those determined by the AVB algorithm.

\begin{figure}
\begin{tabular}{cc}
\centering
\includegraphics[width=6cm]{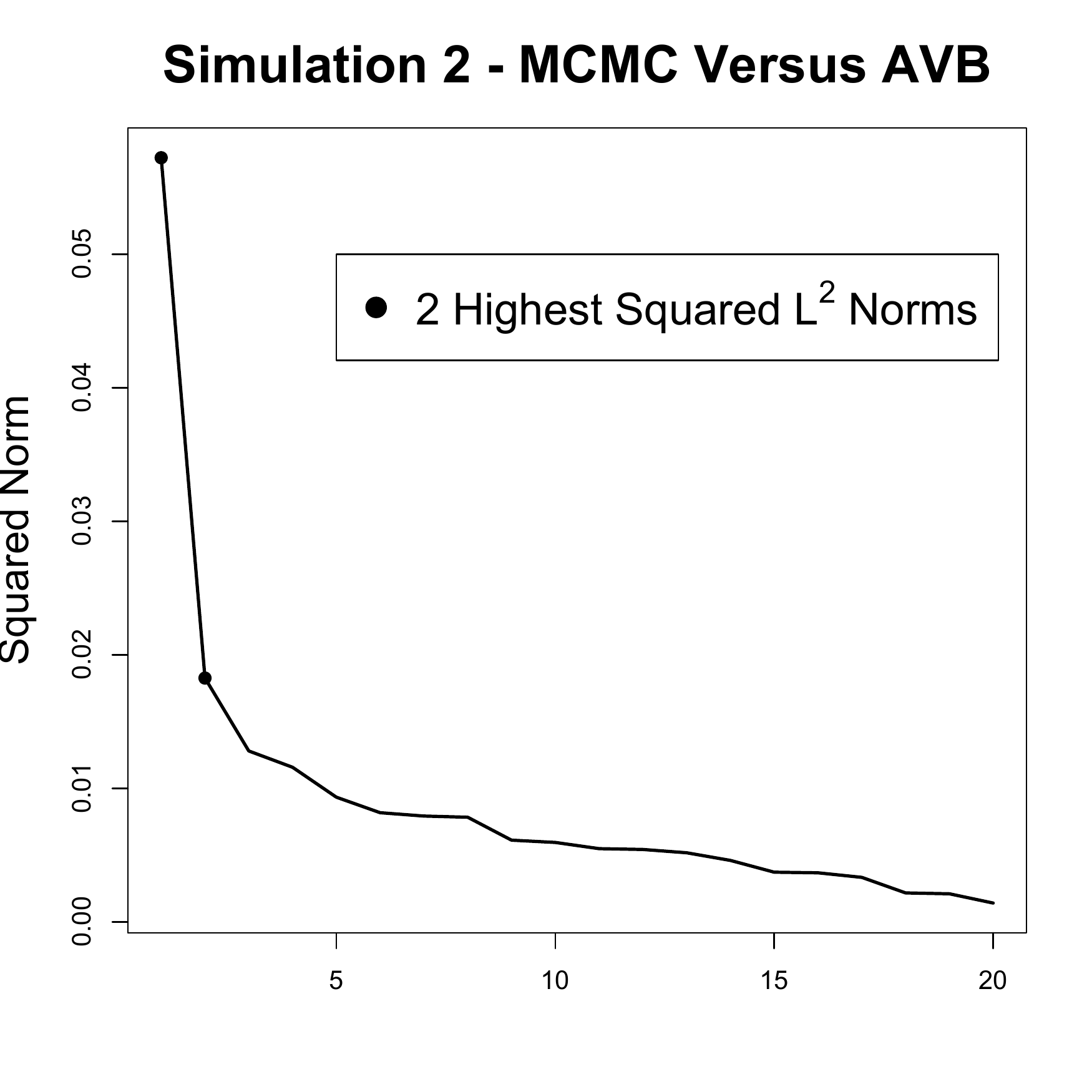} &
\includegraphics[width=6cm]{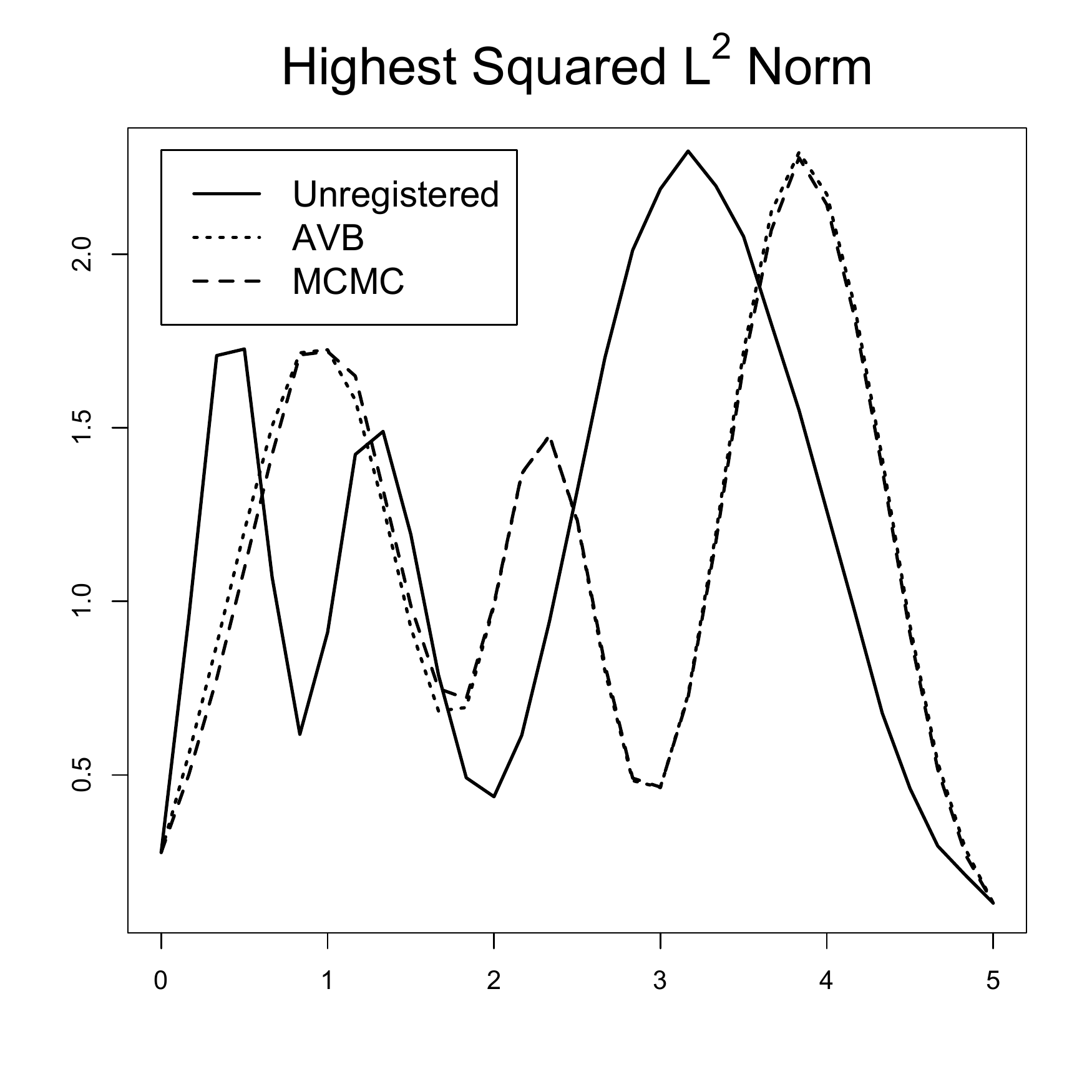}
\includegraphics[width=6cm]{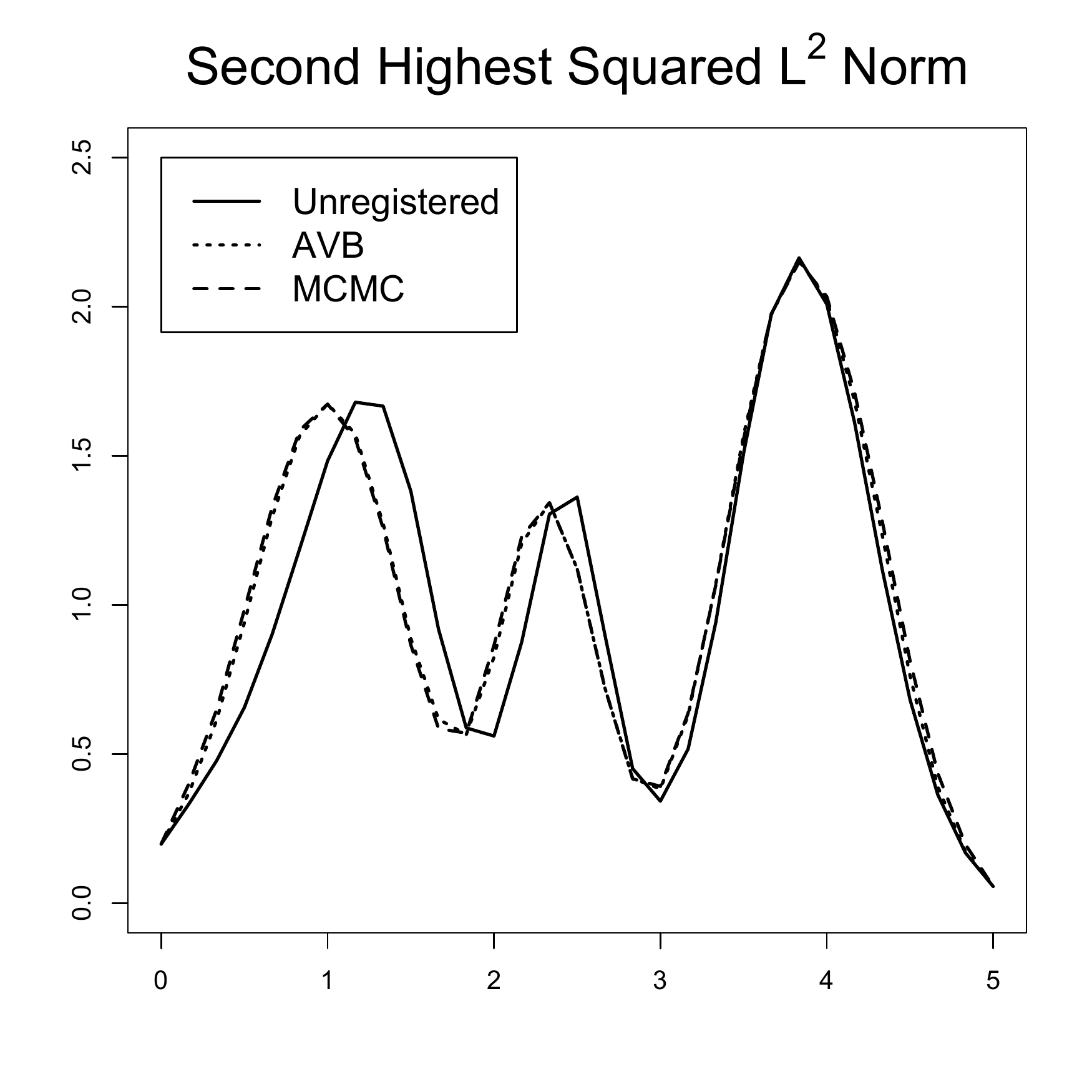}
\end{tabular}
\caption{Simulated Data - Difference Between MCMC and AVB Estimates.  \textbf{Left} Plot of the squared $L^2$ norm of the difference between the MCMC and AVB estimates for each observation in decreasing order of magnitude.  \textbf{Center and Right}  The original unregistered function plotted with the MCMC and AVB estimates of the registered functions for the observations with the two largest discrepancies between the MCMC and AVB estimates.}
\label{fig:NORMS23}
\end{figure}

\begin{figure}[!ht]
\begin{tabular}{cc}
\centering
\includegraphics[width=6cm]{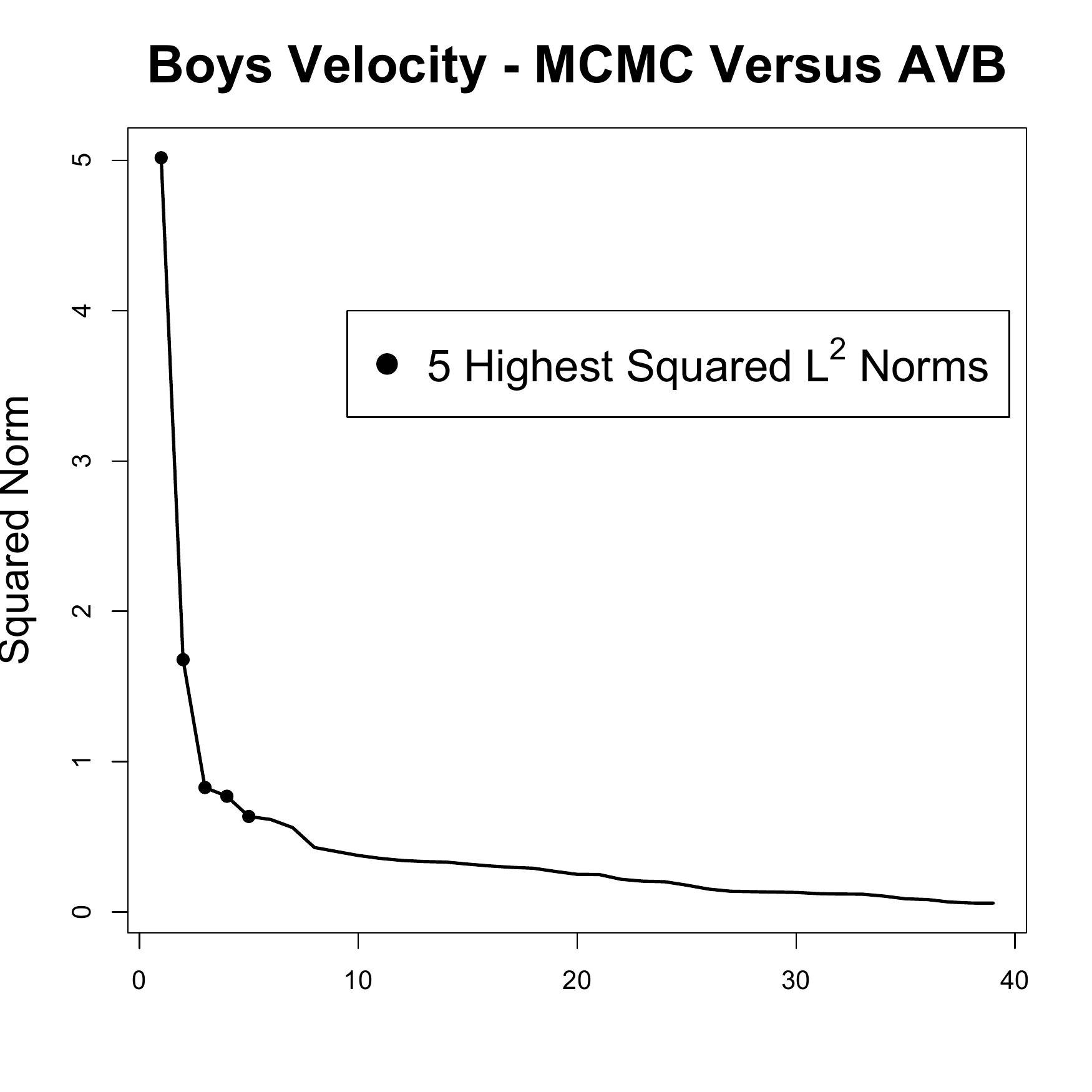} &
\includegraphics[width=6cm]{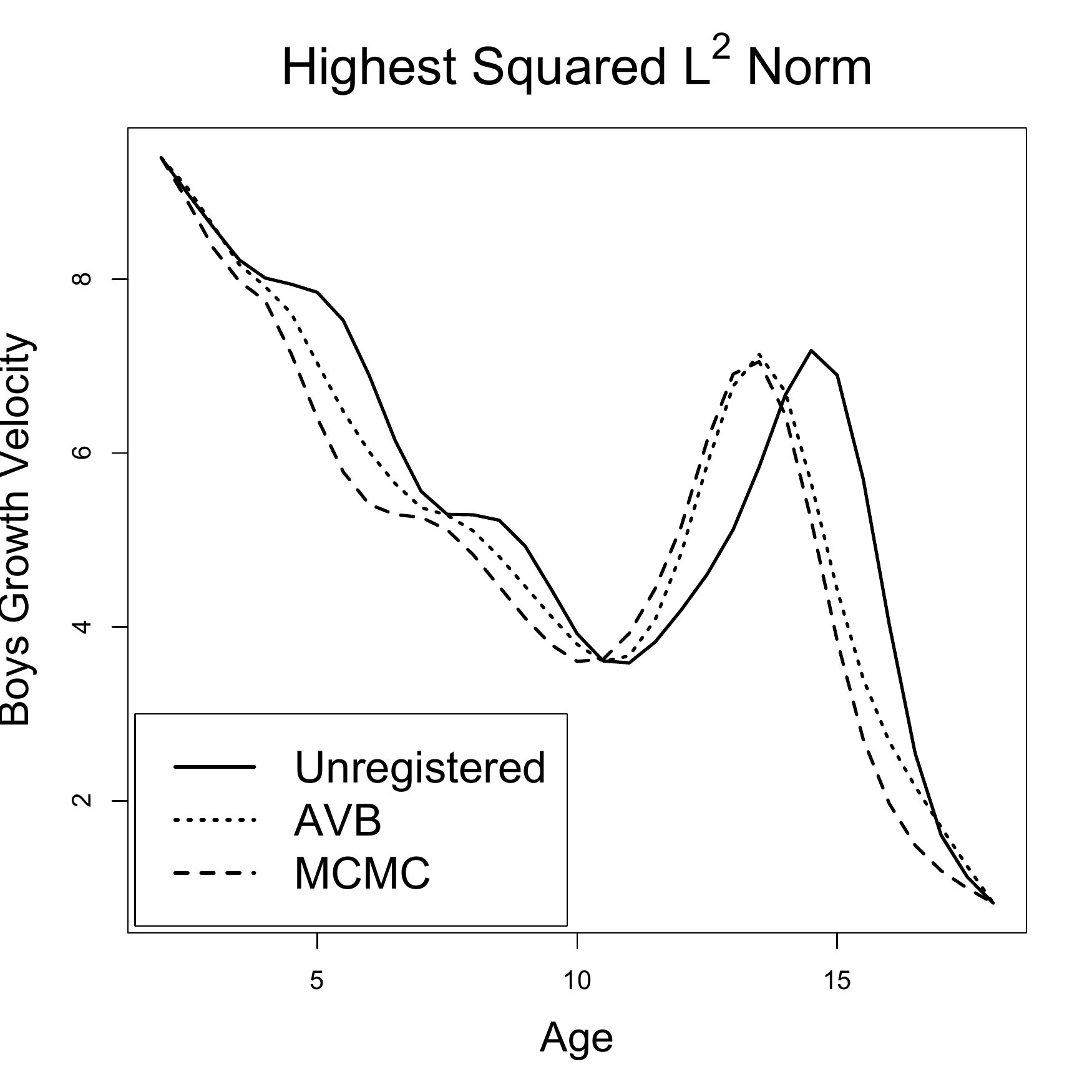} 
\includegraphics[width=6cm]{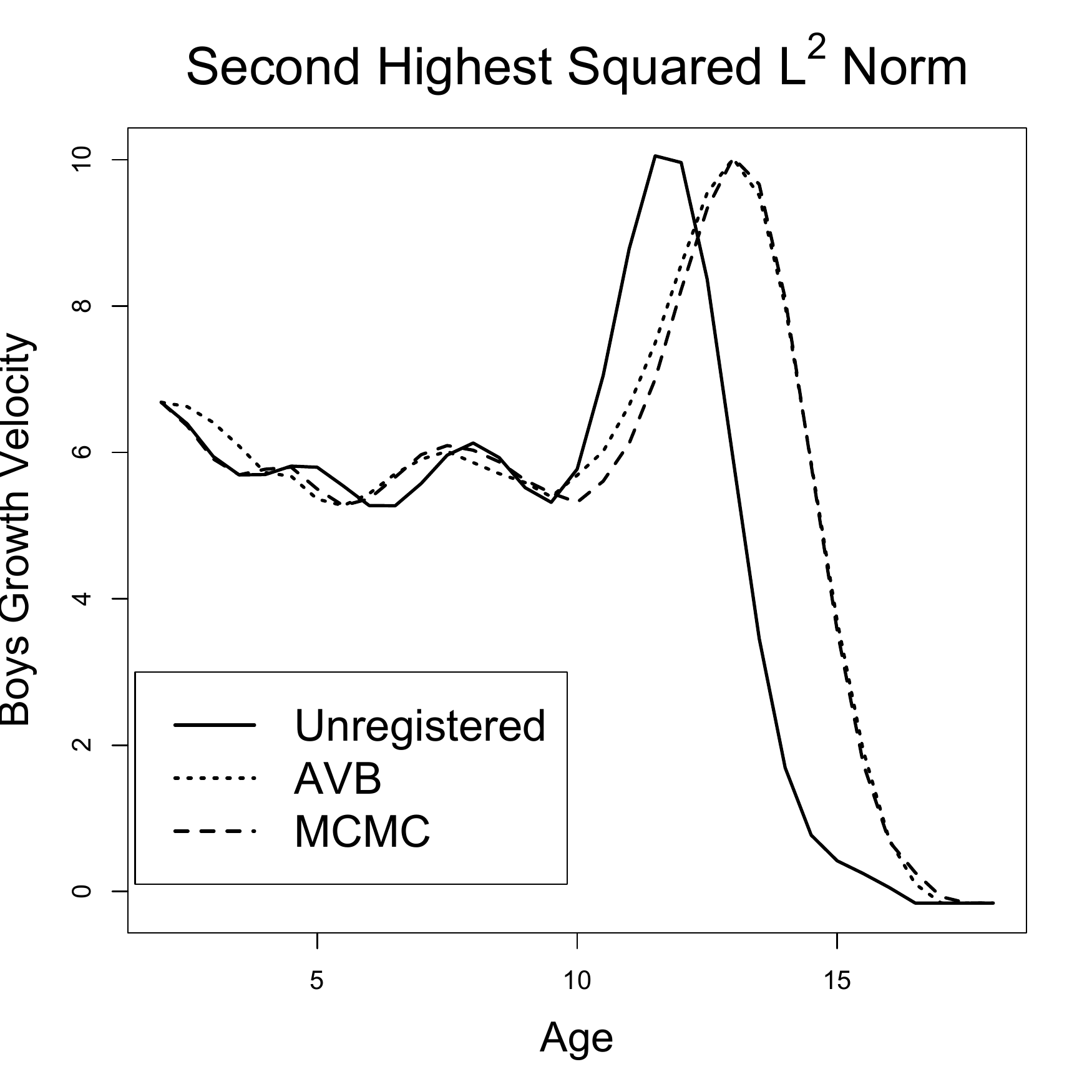}\\
\includegraphics[width=6cm]{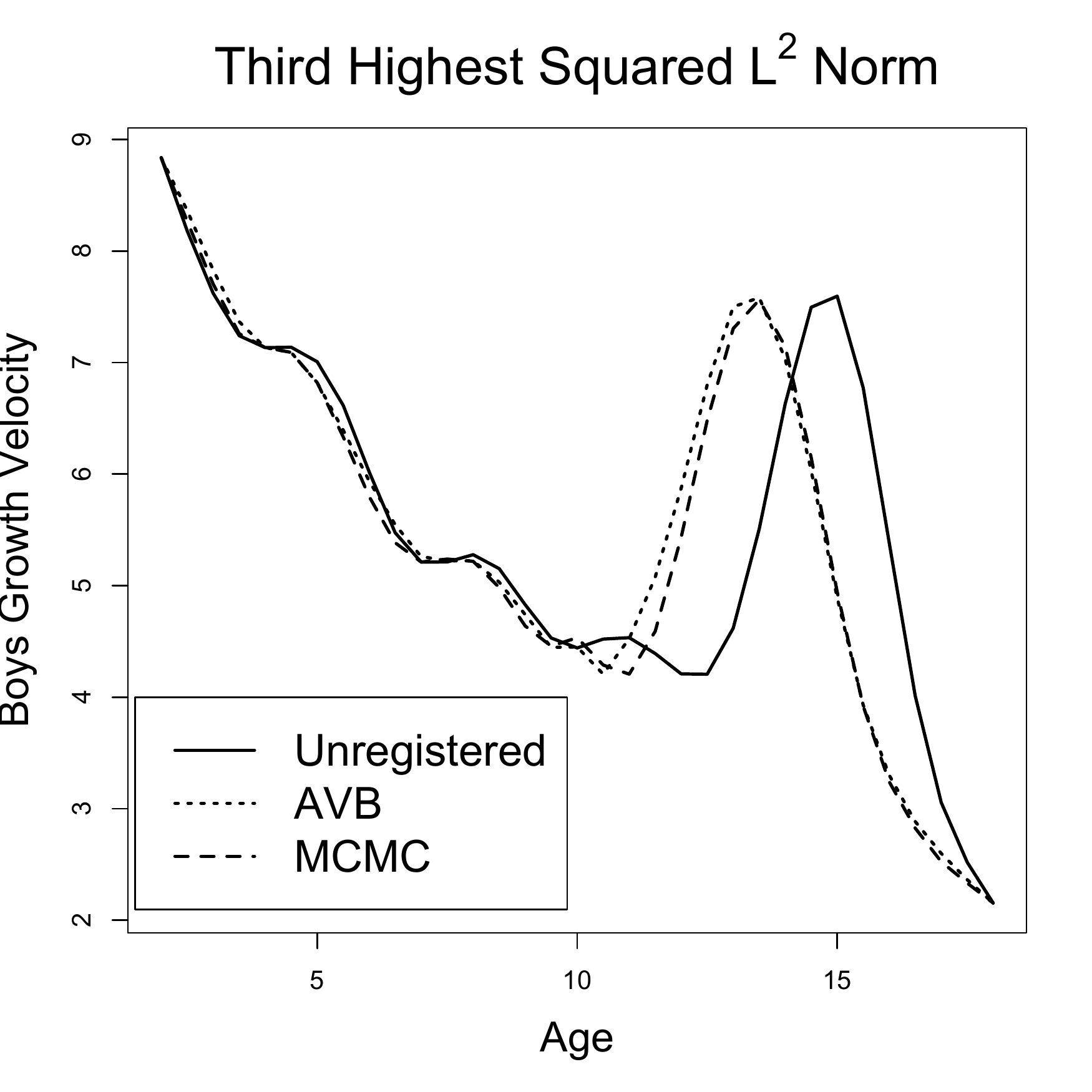} &
\includegraphics[width=6cm]{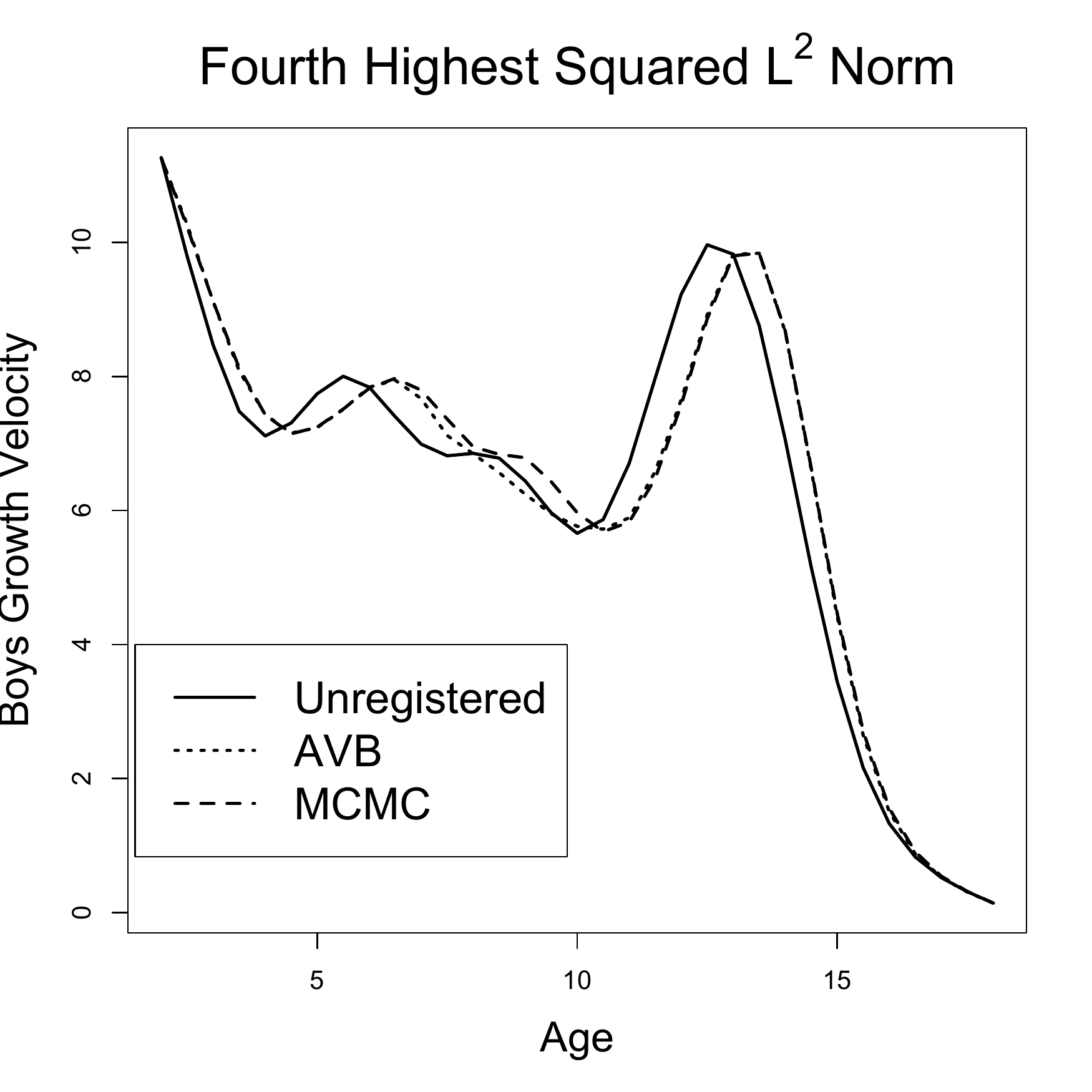}
\includegraphics[width=6cm]{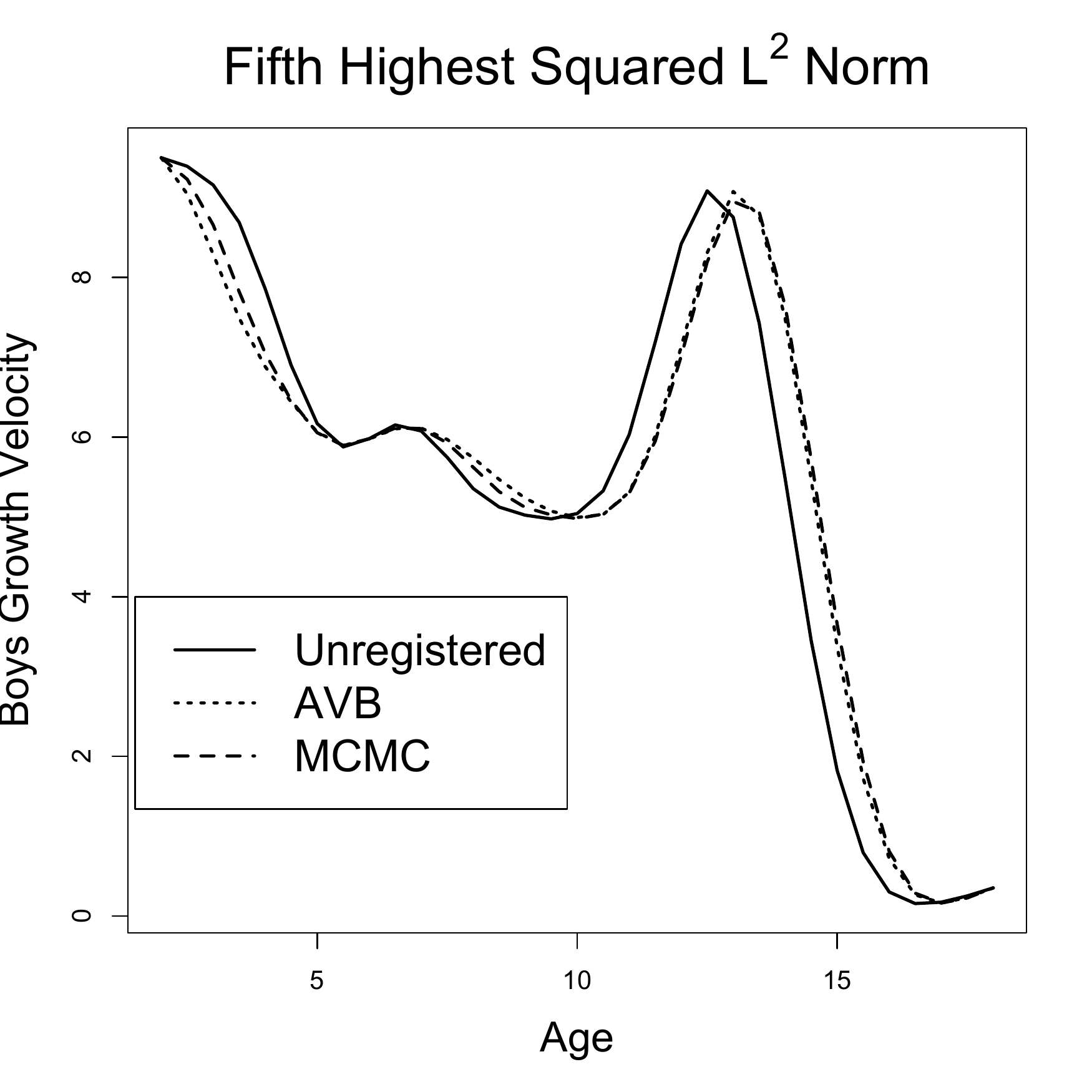}
\end{tabular}
\caption{Registered Boys Growth Velocity - Differences Between MCMC and AVB Estimates.  \textbf{Top Left} Plot of the squared $L^2$ norm of the difference between the MCMC and AVB estimates for each observation in decreasing order of magnitude .  \textbf{Top Center and Right}  The original unregistered function plotted with the MCMC and AVB estimates of the registered functions for the observations with the first two largest discrepancies between the MCMC and AVB estimates. \textbf{Lower}  Plots of the next three observations with the highest squared $L^2$ norms of the difference between the MCMC and AVB estimates.  The squared $L^2$ norm associated with the lower right plot is about .64.  As can be seen in this illustration, at this level there are only small differences between the MCMC and AVB estimates. }
\label{fig:NORMSB}
\end{figure}

The squared $L^2$ norm of the difference between the AVB and MCMC estimate of a registered function is used to quantify the differences between these estimates.  Figure \ref{fig:NORMS23} illustrates for the simulated data set how closely the AVB estimates follow the MCMC estimates.  Even the largest squared $L^2$ norms of the differences between these two estimates correspond to minor changes in the estimates.  These simulated data represent rather ideal conditions for registration where there is almost no variation in the registered functions beyond a scaling and vertical shift of the target function.  Consequently, as we might expect, the MCMC and AVB estimation procedures are primarily in agreement.  Figure \ref{fig:NORMSB} is a more realistic look at the differences between the MCMC and AVB registration results for data that has significant variation in the registered functions beyond a scaling and vertical shift of the target function.  However, even here we see the AVB algorithm performs well.  Of the 39 observations, in only 2 or 3 are there notable discrepancies between the AVB and MCMC estimated registered functions.

These examples show that while ideally AVB estimates are used to initialize a MCMC sampler, they often deviate only in minor ways from the posterior mean estimates obtained from MCMC sampling.  The next logical step is to compare the variance of the approximated posterior distributions to that present in the MCMC samples.  The AVB algorithm does not include approximate posterior distributions for the base functions or the registered functions.  However, we can compare the estimated credible intervals obtained through AVB and MCMC sampling respectively for the target function.  Here we will provide this comparison for the target functions associated with 1)  the original (noiseless) Berkeley Boys Growth Velocity data and 2) the Berkeley Boys Growth Velocity data corrupted with Gaussian noise (see  Appendix C.3 for more information on these data).  

\begin{figure}
\begin{tabular}{cc}
\centering
\includegraphics[width=8cm]{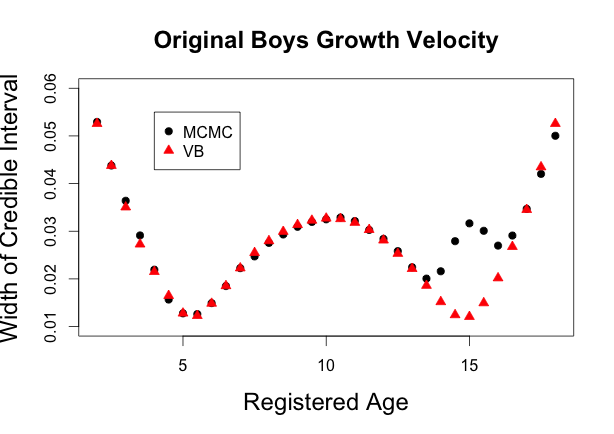} &
\includegraphics[width=8cm]{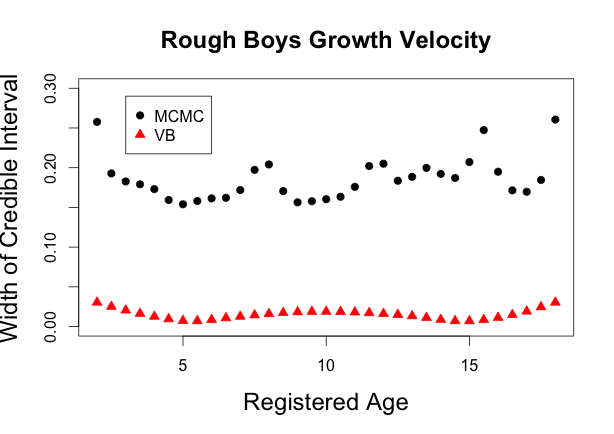} 
\end{tabular}
\caption{Pointwise Comparison of Credible Interval Width for the Target Function.  Both plots contain the width of the credible interval  based on the approximate posterior distribution for the target function obtained via AVB (triangles) and the width of the estimated credible intervals determined by the empirical quantiles of the posterior sample of the target function obtained through MCMC sampling (circles) at each time point . \textbf{Left} This is a comparison of the credible intervals for the target function when the data are recorded without noise.  For most time points in this plot, the width of the credible interval at that time point using the approximate posterior is almost identical to that obtained through MCMC sampling.  The biggest difference can be seen at ages 14-16 where the MCMC credible interval is slightly wider. \textbf{Right} This is a comparison of the credible intervals for the target function when the data are recorded with noise.  Here it can be seen that the approximated posterior distribution of the target function significantly underrepresents the variability in the true posterior distribution. }
\label{fig:CIWidth}
\end{figure}

Generally the concern with using variational Bayes to approximate the posterior distributions is that variability is underestimated in the approximated posteriors, \citet{wang:05} and \citet{bish:06}.  In the appendix, Figure C.1 contains credible bands determined through MCMC sampling for two of the registered functions from the noiseless growth velocity data.  As can be seen in this figure, for noiseless observations, once the registration parameters have been set, the result of specifying a highly informative prior on the registered functions is that most of the variability from the mean is eliminated resulting in very narrow credible bands. Of course, if the credible bands associated with the registered functions are narrow, it makes sense that credible bands for the target function are even narrower.  Thus, for the original boys growth velocity data, very little variability is exhibited in the MCMC samples of the target function and the differences between the estimated 95\% credible band determined from the q distribution of the target function and that determined from the quantiles of the posterior MCMC sample are so small that on a graph they are indistinguishable. In the plot on the left in Figure \ref{fig:CIWidth} is a comparison of the width of the respective credible intervals for the target function at each time point. Here it can be seen that not only are the credible intervals very narrow, but there are only minor differences in the width of the credible intervals determined from the approximate posterior distribution (AVB) and the MCMC sample.  

As can be seen in Figure C.2 of the appendix, if noisy data are recorded, the variability due to the noise process results in wider credible intervals for the registered functions (and hence the target function).  On the right hand side of  Figure \ref{fig:CIWidth} is a comparison of the width of the respective credible intervals for the target function at each time point for the model that both smooths and registers the noisy boys growth velocity data.  In this illustration, we can see the variability present in the posterior MCMC sample for the target function is not captured well by the q distribution associated with the target function.  This is not surprising and is an example of when performing MCMC sampling with an AVB initialization may be preferred to using AVB alone. Note: for these analyses both samplers were initialized with AVB estimates which made burn-in unnecessary. Post sampling analysis showed low autocorrelation in the final 999 samples of the target function (after thinning) for both the noiseless and noisy data.

\section{VARIATIONAL APPROXIMATION FOR FUNCTIONAL PREDICTION}
\label{sec:pred}

\subsection{Functional Prediction Algorithm}
\label{sec:predmeth}
The probabilistic framework of our registration model provides a natural structure in which we can consider new observations.  Functional prediction has previously been considered by \citet{fer:06}.  Here we extend current methods by taking into account the phase variability of a partially observed function.   

We will make the following assumptions.
\begin{enumerate}
\item{We have a sample of approximated unregistered functions, $\mathbf X_i = (X_i(t_1),\hdots, X_i(t_p))'$, $i = 1,\hdots,  N.$}
\item{$\mathbf X_i = (X_i(t_1),\hdots, X_i(t_p))'$, $i = 1,\hdots, N$ are registered using the registration method outlined in Section \ref{sec:meth} via a MCMC sampler or adapted variational Bayes.}
\item{From (2) we have obtained estimates for the target function, $f(t)$, the registered functions, $X_i(h_i(t))$, $i=1, \hdots, N$, the warping functions, $h_i(w_i(t))$, $i=1, \hdots, N$, $\sigma_{z0}^2$, and $\sigma_{z1}^2$.}
\item{A new function, $X_{N+1}(t)$ has been observed at the time points $(t_1, \hdots , t_r)'$, $r$ $<$ $p$.}
\item{$(X(h(t_1)), \hdots X(h(t_p)))'$ $\approx$ $N_p(\hat{\boldsymbol \mu}_{\mathbf X(\mathbf h)}, \widehat{\boldsymbol\Sigma}_{\mathbf X(\mathbf h)})$, the distribution of the registered functions can be approximated by a multivariate normal distribution using the sample mean, $\hat{\boldsymbol \mu}_{\mathbf X(\mathbf h)}$, and sample covariance matrix, $\widehat{\boldsymbol\Sigma}_{\mathbf X(\mathbf h)}$, of the estimated registered functions obtained in (2).   }
\item{$(w(t_1), \hdots w(t_{p-1}))'$ $\approx$ $N_{p-1}(\hat{\boldsymbol \mu}_{\mathbf w}, \widehat{\boldsymbol\Sigma}_{\mathbf w})$, the distribution of the base functions can be approximated by a multivariate normal distribution using the sample mean, $\hat{\boldsymbol \mu}_{\mathbf w}$, and sample covariance matrix, $\widehat{\boldsymbol\Sigma}_{\mathbf w}$, of the estimated base functions obtained in (2).  }
\end{enumerate} 

Under these assumptions, we will proceed as follows.

\begin{enumerate}
\item{Register the partially observed function, $\mathbf {X^P}_{N+1} = (X_{N+1}(t_1), \hdots , X_{N+1}(t_r))'$ to the estimated target function, $\hat f(t)$, truncated to an appropriate registration time, $t_f$, $f \in \{1,\hdots, p\}$, so that $h_{N+1}(t_f) = t_r$.}
\item{Using the distributions from assumptions (5) and (6) above, the estimate of the partial registered function, $\mathbf{X^P}_{N+1}(\mathbf {\hat h}_{N+1})$ $=$ $(X^P_{N+1}(\hat h_{N+1}(t_1)), \hdots , X^P_{N+1}(\hat h_{N+1}(t_f)))'$ and the estimate of the partial base function, $\mathbf {\widehat {w^P}}_{N+1}$ $=$ $(\widehat{w^P}_{N+1}(t_1), \hdots \widehat {w^P}_{N+1}(t_{f-1}))'$, estimate the registered and base functions to time $t_p$ and $t_{p-1}$ respectively using the conditional expectation of the multivariate normal distribution. Accordingly, denoting future registered observations and future warping function values , $\mathbf {X^F}_{N+1}(\mathbf {h}_{N+1})$ and $\mathbf {w^F}_{N+1}$, respectively, the estimates of these future values are\\ $\widehat{\mathbf{X^F}}_{N+1}(\mathbf {\hat h}_{N+1})$ $=$ $E(\mathbf{X^F}(\mathbf{h})|\mathbf{X^P}_{N+1}(\mathbf {\hat h}_{N+1})$, $\hat{\boldsymbol \mu}_{\mathbf X(\mathbf h)}$$, \widehat{\boldsymbol\Sigma}_{\mathbf X(\mathbf h)})$   and\\ $\widehat{\mathbf{w^F}}_{N+1}$ $=$ $E(\mathbf{w^F}|\widehat{\mathbf{w^P}}_{N+1}$, $\hat{\boldsymbol \mu}_{\mathbf w}$$, \widehat{\boldsymbol\Sigma}_{\mathbf w})$}.
\item{Estimate the complete unregistered function, $X_{N+1}(t)$, using the inverse of the estimated warping function and the estimated registered function. }
\end{enumerate} 

\subsection{Determining the Last Registered Time}

An additional random element in the prediction model is the last registered time of the truncated target function, $t_f$, used to register the partial observation.  To obtain the best possible registration of the partial observation, a range of final registration times are considered over a finer domain.  The efficiency of the adapted variational Bayes algorithm makes it possible to consider several possible partial registrations as follows.

\begin{enumerate}
\item{For each of the time points $t_j, j \in \{m, \hdots, (m+k-1)\}$, $t_{m+k-1} < t_p$, the partially observed function, $X^P_{N+1}(t)$, is registered to the estimated target function, $\hat f(t)$, truncated to time, $t_j$, so that $\hat h_{N+1(j)}(t_j) = t_r$, where $\hat h_{N+1(j)}(t)$ is the estimated warping function determined by registering $X^P_{N+1}(t)$ to the proposed final registration time $t_j$.  Note, the first and last times considered in this interval are chosen by plotting the partial unregistered function and the target function together and determining a generous interval that contains the appropriate final registration time.  This interval is subsequently made finer to allow this time to fall between two of the original time points.}
\item{Calculate $d_{t_j} = ||\mathbf{X^P}_{N+1} - (\hat z_{0(j)}\mathbf 1 +\hat z_{1(j)}\mathbf {f^U}_{(j)})||_2$ for each $t_j$, $j\in \{m, \hdots, m+k-1\}$ where \\ $\mathbf {f^U}_{(j)} = (\hat f(t_1), \hat f(\hat h^{-1}_{N+1(j)}(t_2)), \hdots, \hat f(\hat h^{-1}_{N+1(j)}(t_r) = t_j))'.$}
\item{Set $t_f =\underset{t_j, j \in\{m,\hdots, m+k-1\}}{\arg\min} d_{t_j}.$ }
\end{enumerate}   

This algorithm determines the final registered time, $t_f$, that results in the minimum $L^{2}$ norm between the partially recorded unregistered function and the target function evaluated at the inverse of the warping function estimated using that final time.  Note, for all j, $\mathbf{f^U}_{(j)}$ shares the same domain as the partially recorded unregistered function, $\mathbf{X^P}_{N+1}$.

\subsection{Confidence Intervals for Predicted Functions}

The efficiency of adapted variational Bayes for prediction also makes it possible to characterize variability in the estimates of the complete registered function, unregistered function, and base function via bootstrapping.  To capture variability in the predicted functions, for each of $m = 1, \hdots, M$ iterations, perform the following steps.

\begin{enumerate}
\item{Draw a new sample of $N$ registered functions from $N_p(\hat{\boldsymbol \mu}_{\mathbf X(\mathbf h)}, \widehat{\boldsymbol\Sigma}_{\mathbf X(\mathbf h)})$}.
\item{Draw a new sample of $N$ base functions from $N_{p-1}(\hat{\boldsymbol \mu}_{\mathbf w}, \widehat{\boldsymbol\Sigma}_{\mathbf w})$}.
\item{From the bootstrapped samples determined in (1) and (2), compute the sample mean and covariance matrix for the approximated registered functions, $\hat{\boldsymbol \mu}_{\mathbf X(\mathbf h)}^{(m)}$ and $\widehat{\boldsymbol\Sigma}_{\mathbf X(\mathbf h)}^{(m)}$ and also for the approximated base functions, $\hat{\boldsymbol \mu}_{\mathbf w}^{(m)}$ and $\widehat{\boldsymbol\Sigma}_{\mathbf w}^{(m)}$.} 
\item{Register the partially observed function as in prediction step (1) in Section \ref{sec:predmeth} above by setting the approximated target function, $\hat{\mathbf f}$, equal to $\hat{\boldsymbol \mu}_{\mathbf X(\mathbf h)}^{(m)}$ .} 
\item{Draw a sample of size $S$ from the distribution of $\mathbf{X^F}(\mathbf{h})|\mathbf{X^P}_{N+1}(\mathbf {\hat h}^{(m)}_{N+1})$, $\hat{\boldsymbol \mu}^{(m)}_{\mathbf X(\mathbf h)}$$, \widehat{\boldsymbol\Sigma}^{(m)}_{\mathbf X(\mathbf h)}$. Combine each of these samples of future values of the registered function with the estimated partially registered function determined in (4) to get a sample of $S$ estimated registered functions.}
\item{Draw a sample of size $S$ from the distribution of $\mathbf{w^F}|\widehat{\mathbf{w^P}}^{(m)}_{N+1}$, $\hat{\boldsymbol \mu}^{(m)}_{\mathbf w}$$, \widehat{\boldsymbol\Sigma}^{(m)}_{\mathbf w}$. Combine each of these samples of future values of the base function with the estimated partial base function determined in (4) to get a sample of S estimated base functions.}
\item{Determine the $S$ warping functions that result from step (6).}
\item{Determine the $S$ unregistered functions that result from combining each of the $S$ registered functions drawn in (5) with the corresponding inverse warping function from (7).}
\end{enumerate} 

This process results in $M\times S$ bootstrapped samples of the registered function, warping function and unregistered function.  From these samples, point wise bootstrapped confidence intervals can be determined for each function.
 
Here we have approximated the distributions of the base and registered functions by fitting a multivariate normal distribution.  However, given the small sample size, approximating these distributions by a multivariate t distribution as suggested in \citet{lange:89} may provide confidence intervals with better coverage properties.  \citet{beran:90} also provides a method for constructing robust confidence intervals in the context of univariate prediction problems.

 \subsection{Functional Prediction - El-Ni\~no Data}
 \label{sec:elnino}
 
The El-Ni\~no data consist of weekly readings of sea surface temperature with the first observation in June of 1950.  Complete data can be found at NOAA's Climate Prediction Center website (http://www.cpc.ncep.noaa.gov/data/indices/).  The  data that we are using for this analysis are found through Professor Frederic Ferraty's (Mathematics; University of Toulouse, France) website (http://www.math.univ-toulouse.fr/~ferraty/
SOFTWARES/NPFDA/npfda-datasets.html).  These data are a subset of the original data with monthly sea surface temperature records from June of 1950 to May of 2004.  For this analysis, the bi-monthly observations are added to the data to prevent significant changes to the shape of a given function due to interpolation error.  Also, light smoothing is applied to all functions.
 
The goal of our study is to predict how high temperatures will stay in the remaining part of the year in conjunction with how long temperatures will drop before they rise again based on the first seven months of temperature recordings from the lowest temperature recording in the previous year.  

For this purpose, the data are restructured to define a``year" as the period of time between the lowest temperatures in consecutive calendar years.  For example, the first year in our data set ranged from September 1950 to September 1951.  Note, these ``years" will not all be 12 months in length, and our final data had ``years"  that ranged from 11 to 14 months.  For our analysis, we will concentrate on a subset of this group of restructured functions where the previous year's lowest temperature for each selected temperature profile is between $19.5\,^{\circ}\mathrm{C}$ and  $21\,^{\circ}\mathrm{C}$.  Twenty-nine of the original temperature profiles fit this criterion.   We will use the first 28 functions to predict the remaining portion of the 29th function based on the first 7 months of sea surface temperature observed in that year.

For the purpose of registration, all functions need to be recorded over the same interval of time.  As mentioned above, in this particular case our data is recorded over a time periods that range from 11 to 14 months.  An easy remedy to this situation is to perform a simple initial warping to each function that rescales every observation to an 11 month time frame.  In our final analysis, this initial warping is accounted for when determining the final base functions used for the prediction algorithm.  

\begin{figure}[!ht]
\begin{tabular}{cc}
\centering
\includegraphics[width=8cm]{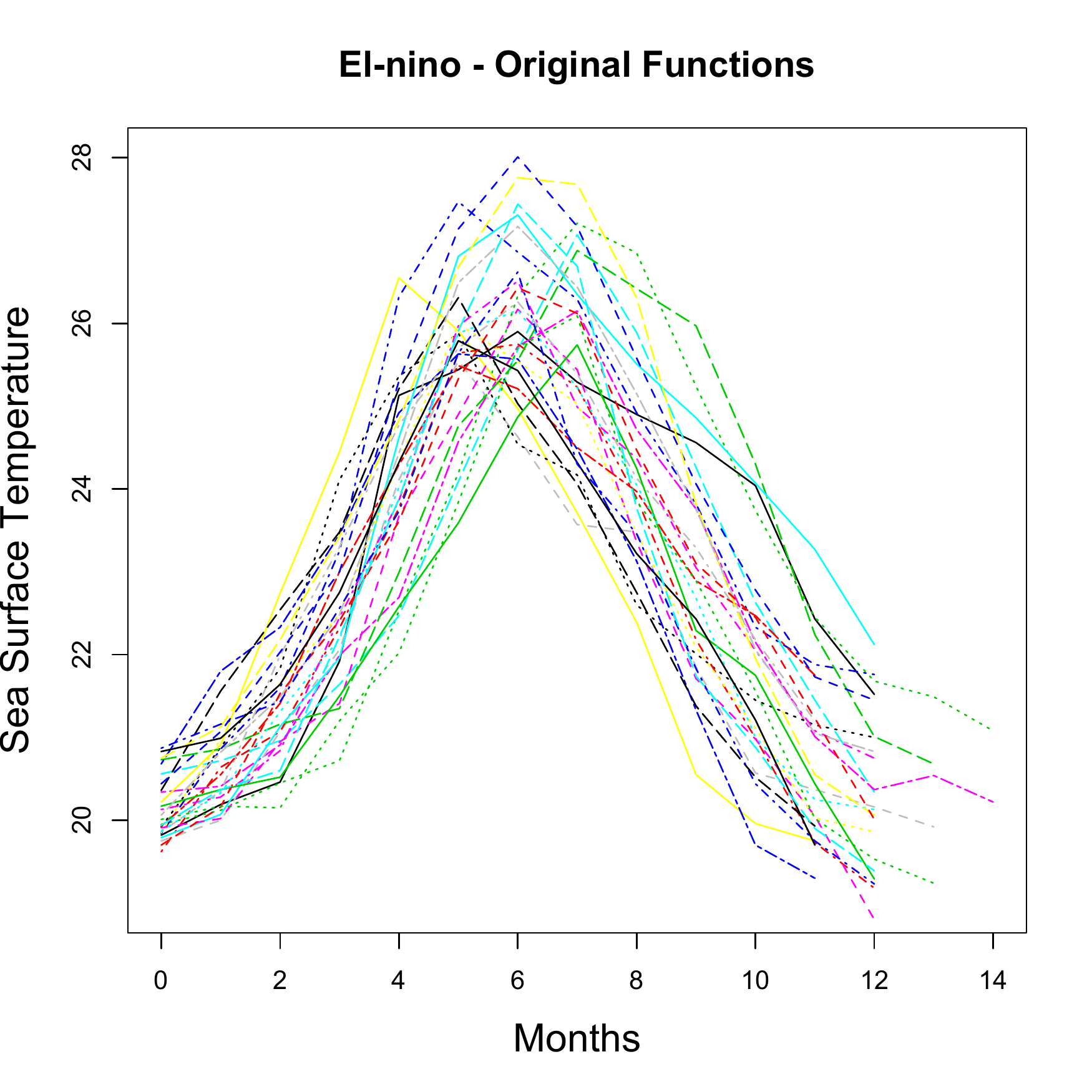} &
\includegraphics[width=8cm]{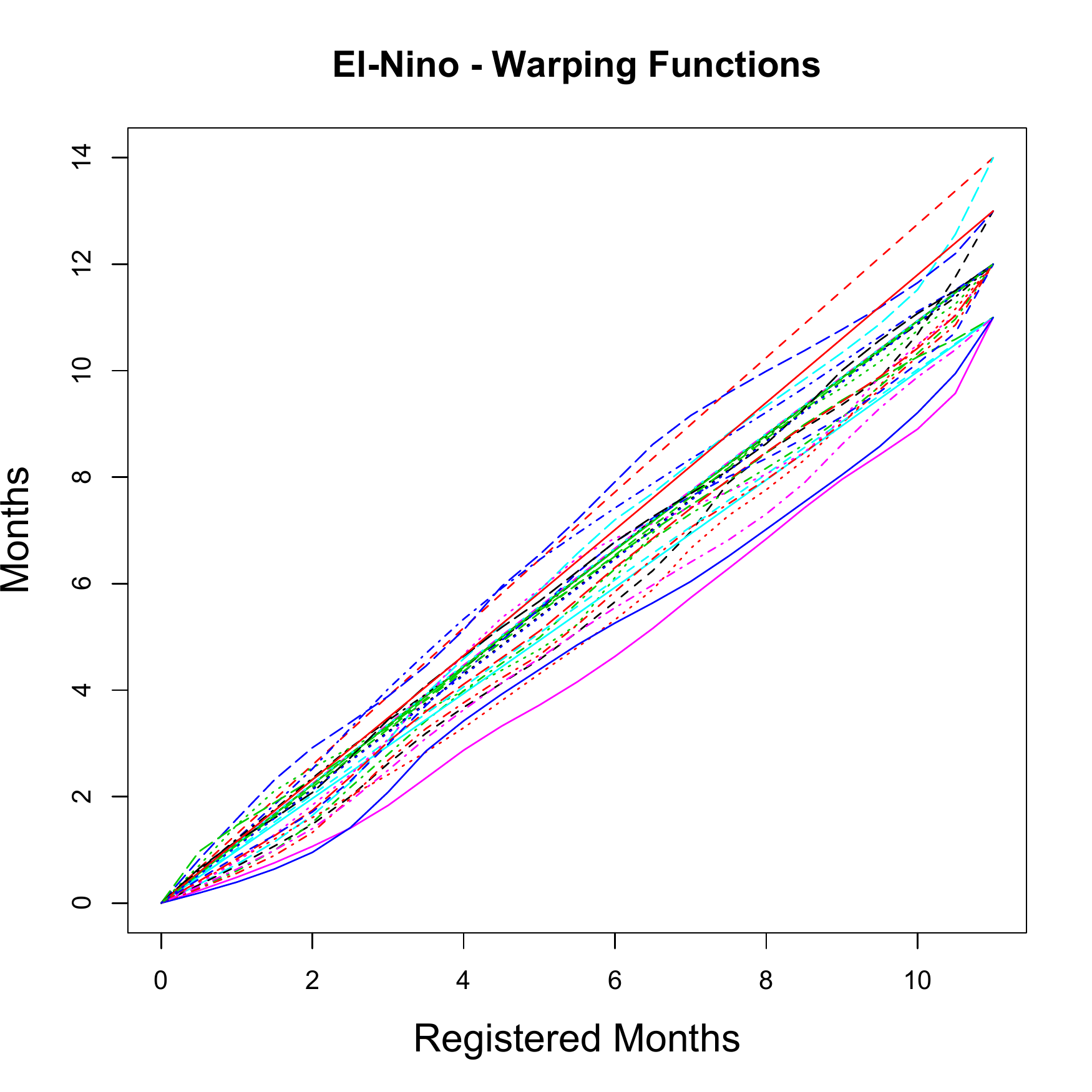}\\
\includegraphics[width=8cm]{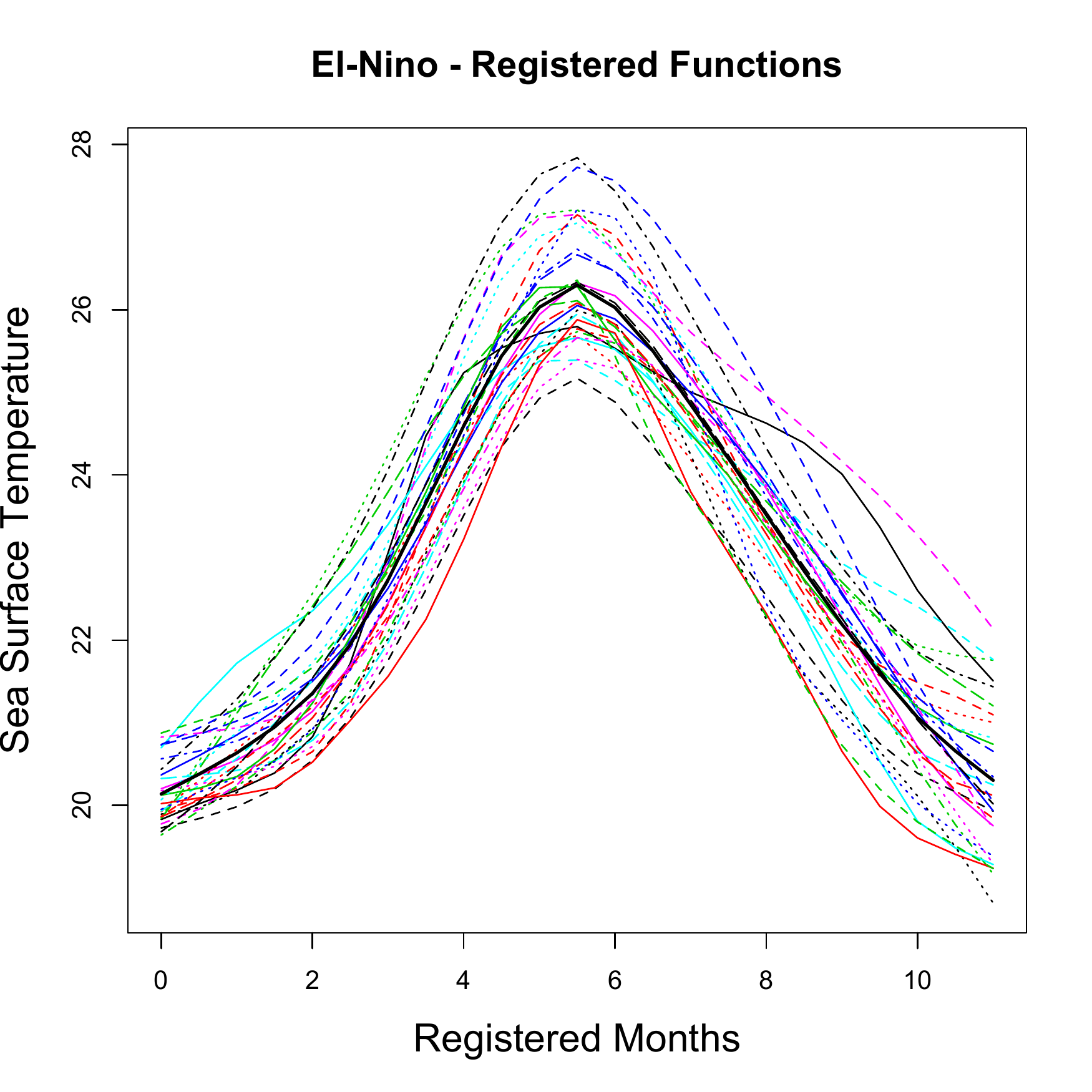}   &
\includegraphics[width=8cm]{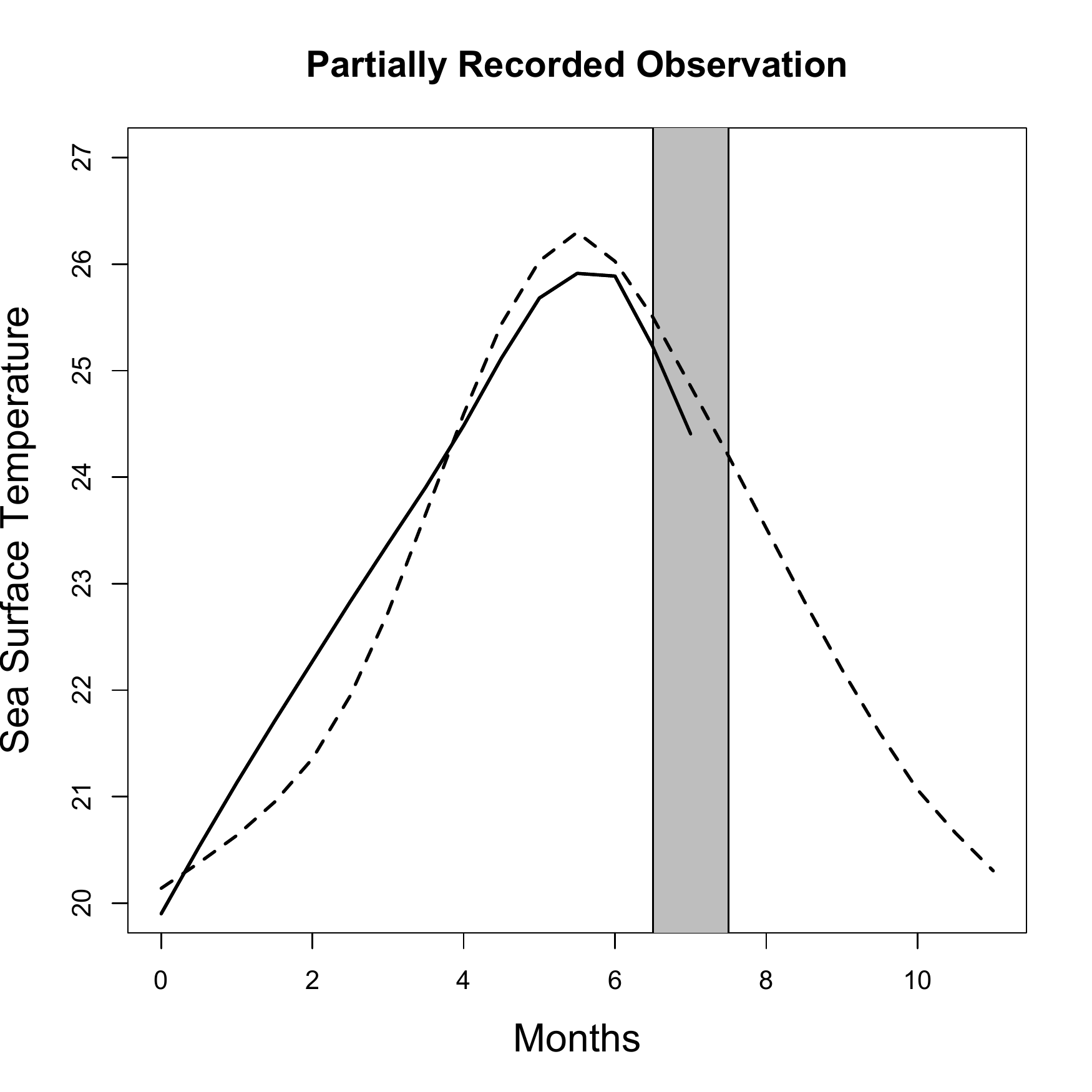}
\end{tabular}
\caption{El-ni\~no Data.  \textbf{Top Left} Original 28 profiles of sea surface temperature.  \textbf{Top Right} Estimated warping functions.  As can be seen here, the time period of the original data ranged from 11 to 14 months. \textbf{Lower Left}  Estimated registered temperature profiles. \textbf{Lower Right } The solid line is observation 29 recorded for 7 months.  The dashed line is the estimated target function.  The grey shaded area spans the 5 time points that are considered for the final time of the partial registration.}
\label{fig:REGEN}
\end{figure}

The original unregistered functions and the functions registered using the GP model described in Section \ref{sec:meth} are plotted in Figure \ref{fig:REGEN}.  For this data set, to register significant features in the sample while retaining function variation beyond a scaling and vertical shift of the target function, individual warping parameters, $\gamma_{w_i}$, $i=1,\hdots, 28$ were utilized instead of $\gamma_w$ in \eqref{eq:Prior}.  Significant differences in the amplitude variation in the original functions that is unassociated with temporal variation prevented the use of a global parameter.  However, only 3 unique warping parameters in total were necessary.

Using the empirical mean of the 28 original registered functions as the target function, the first 7 months of sea surface temperature records from observation 29 are registered to a piece of the target function where the final registered time is allowed to vary from 6.5 to 7.5 months.  Between these months, a finer time interval corresponding to weekly records is used to allow for additional flexibility in determining the final registered time.  The partially recorded function is plotted with the target function in the lower right panel of Figure \ref{fig:REGEN}. The grey shaded area includes the time points considered for the final time of the partial registration.  After the optimal registration of the partially recorded observation is determined, estimates of the entire registered function, warping function, and unregistered function are determined using the model outlined in Section \ref{sec:predmeth}.  

\begin{figure}
\begin{tabular}{cc}
\centering
\includegraphics[width=6cm]{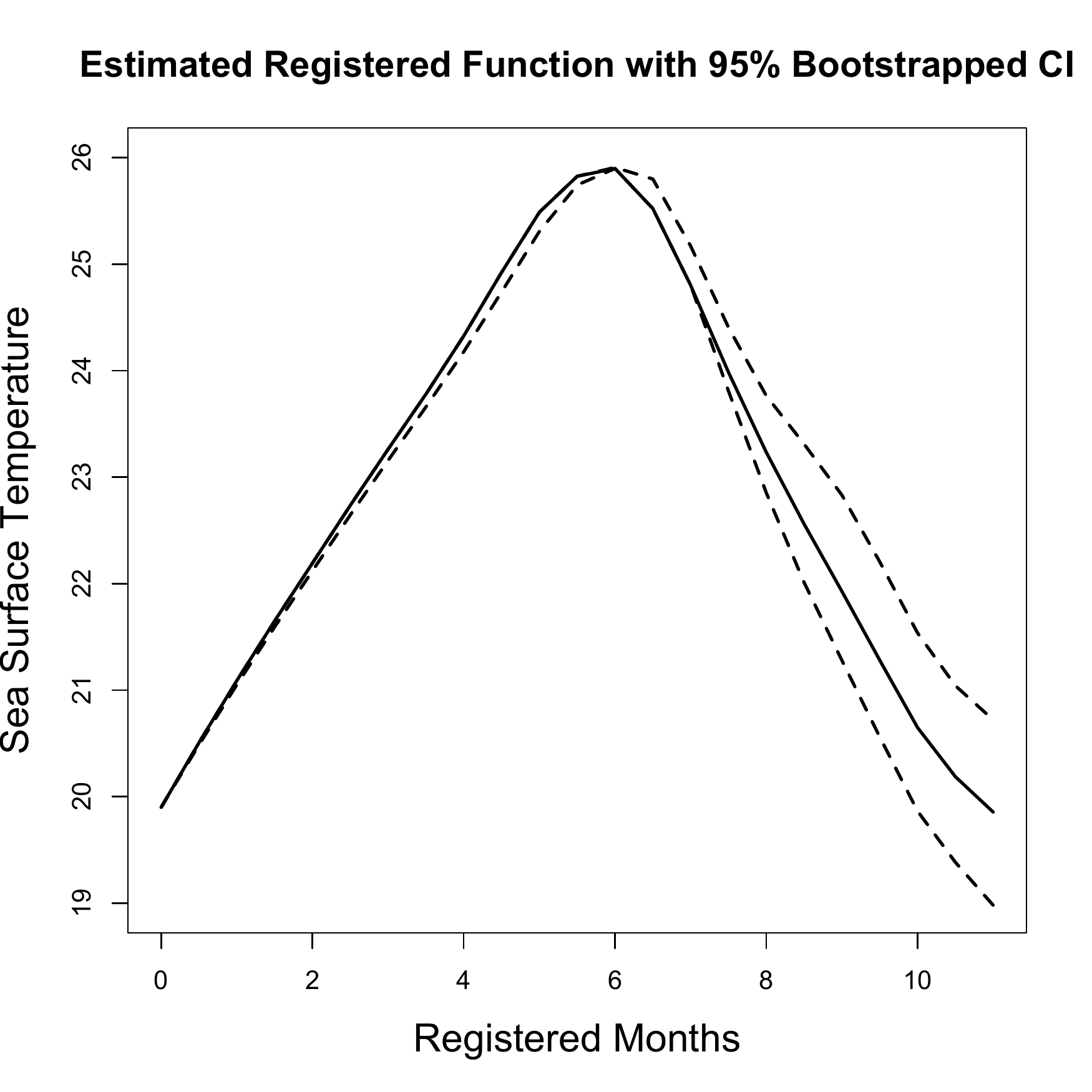} &
\includegraphics[width=6cm]{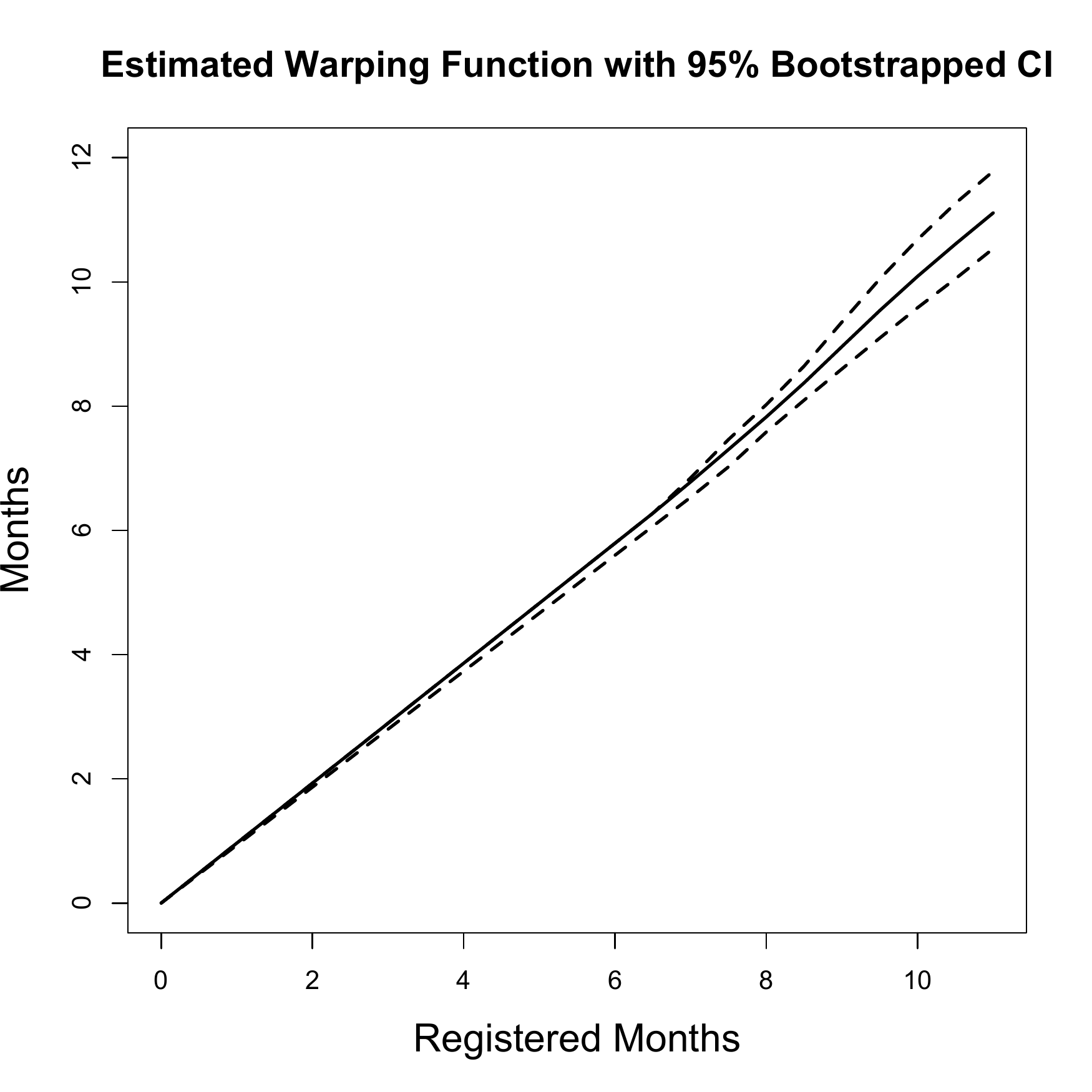} 
\includegraphics[width=6cm]{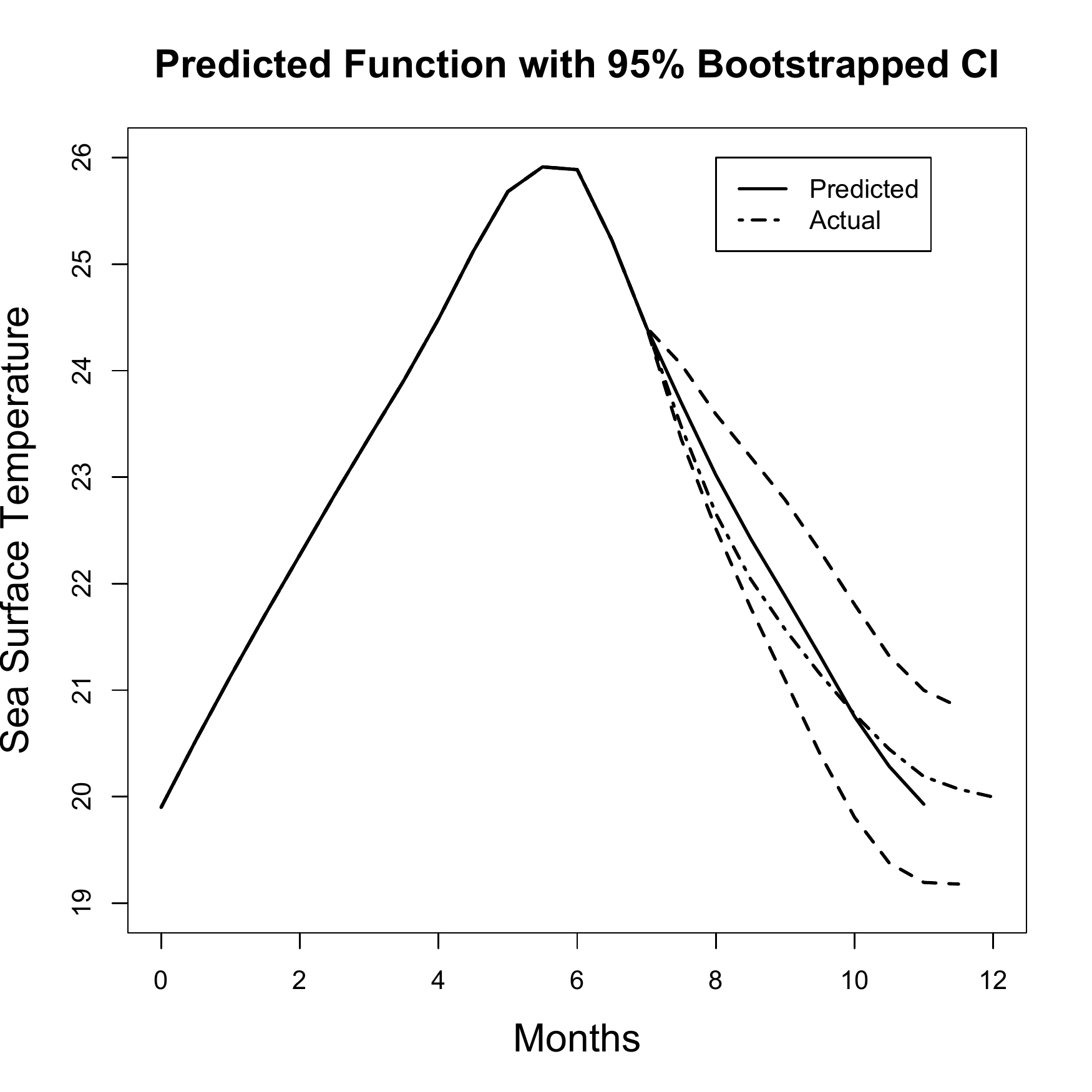}
\end{tabular}
\caption{Estimates and Bootstrapped Confidence Intervals.  \textbf{Left}  Estimated registered function with 95\% bootstrapped confidence interval. Note: In parts of the domain, the estimated confidence interval and estimated registered function overlap.  This is largely due to a bimodal distribution of the last registered time.  \textbf{Center}  Estimated warping function with 95\% bootstrapped confidence interval. \textbf{Right}  Estimated unregistered function with 95\% bootstrapped confidence interval.  The dashed and dotted line is the true unregistered function.}
\label{fig:CIS}
\end{figure}

One-hundred bootstrapped samples were used to estimate the variability in the predictions of all three estimated functions.  Figure \ref{fig:CIS} plots the initial estimates with the 95\% bootstrapped confidence intervals.  In addition, the plot of the estimated unregistered function also includes the true value of this function.

The primary advantage of registering the partially recorded observation before estimating future values is that we can capture variation in amplitude and timing separately. In Figure \ref{fig:CIS}, the first plot captures the variability in the future level of sea surface temperature (amplitude variation), and answers the question,``How high can we expect sea surface temperatures to stay?".  The second plot captures the variability in the timing of future observations (temporal variation) which addresses the question of, ``When can we expect sea level temperatures to begin rising again?". The confidence interval for the unregistered function seen in the last frame of Figure \ref{fig:CIS}, combines both amplitude and temporal variation to estimate the future trajectory of sea surface temperature for this year.  In this illustration it can be seen that the main difference in the estimated and actual temperature profiles lies in the timing of the lowest observation.  However, for this observation, the sea surface temperature at 12 months is not much different than the sea surface temperature at 11.5 months.  The predicted timing of the lowest temperature was 11.1 months.

One of the most notable features of this analysis is that there is little uncertainty in the registration of the first 7 months of sea surface temperatures.  The most prominent feature in the data is the peak temperature that occurs anywhere from 4 to 8 months in the original data.  In our partially recorded observation, as seen in Figure \ref{fig:REGEN}, the peak of the target function and the partially recorded observation are already closely aligned.  Additionally, this observation happens to be similar in shape to the target function.  The combination of these features resulted in only a minimal amount of variation in the estimated registered and warping functions in the first 7 months.  However, we note here, this phenomenon is an artifact of these particular data, and in other analyses more variation in the registered timing of the partially recorded observation would be expected.  

The El-ni\~no data set provides a challenging registration problem.  The registered functions vary significantly in directions beyond the target function.  Choosing curve specific registration parameters enabled features common to all functions to be registered while retaining prominent features in each individual curve.  This is just one example of the difficulties that can arise in registering functional data and in turn how these challenges can be addressed to analyze data that does not fit the ``ideal" registration problem.

\section{DISCUSSION}
\label{sec:conc}
In this paper we have developed a methodology for Bayesian registration that accounts for uncertainty in the registered functions, the warping functions, and the parameters associated with them.  The hierarchical structure of this model allows multiple inferential procedures to be included in one analysis.  We give an example where functional regularization and registration are performed in one model.  However, these models will accommodate any combination of inferential procedures for which an appropriate prior can be defined. For instance, the models proposed here can easily be extended to a functional linear regression model where the registered functions or registered latent functions are considered as covariates.

While our registration algorithm provides high quality estimates in a highly flexible model, the associated computational costs due to running a MCMC sampling scheme for a high-dimensional model are considerable.  To address these costs, we have proposed the Adapted Variational Bayes algorithm.  This algorithm has been shown to converge in a similar way to traditional variational Bayes, but may not converge to a global maximum.  However, an initial warping can be performed to move a function closer to its optimal registered value if the algorithm converges to a local maximum. 

The bijective relationship between the base and warping functions in this model makes it possible to perform functional prediction in the context of registration.  Using the estimated values of the base and registered functions from an initial registration of $N$ functions, we use approximate distributions for these functions to predict future values of the registered, warping, and unregistered functions of a new function that is only partially observed.  Furthermore, the AVB algorithm makes it possible to re-sample from these distributions to bootstrap confidence intervals for these predicted values.

While the AVB algorithm provides estimates that are similar to their MCMC counterparts, for some models MCMC sampling remains the optimal inferential procedure.  For example, in the model that combines smoothing and registration AVB estimates are approximate and preferably are only used to initialize an MCMC sampler.  While we have used Metropolis within Gibbs samplers, these are likely not optimal.  Determining the most efficient method of sampling from these joint posterior distributions is left for future work.  

For simplicity, we have used inverse-Gamma or Gamma priors for the variance components in these models. The best choice of priors for these components is a uniform prior on the square root of the variance components as suggested by \citet{gelman:06}.  This is particularly a problem when the variance component is small or there is very little data to estimate these components.  For the models presented here, Gamma priors for the smoothing parameters are sufficient as small changes in these parameters do not significantly affect the model.  However, in the analysis of the noisy boys' growth velocity data in Appendix C, we saw some evidence that the ``uninformative"  inverse-Gamma prior resulted in a slightly upward biased estimate of the noise variance.

Modeling functions as Gaussian processes in a Bayesian hierarchical model offers a unified approach to performing multiple inference procedures for functional data within one model.  In \citet{earls:14}, we established the properties of functional data estimates determined by approximating functional distributions over a finite subset of observed time points and estimating functions at unobserved time points with linear interpolation.  Furthermore, we demonstrated that providing smoothing information in  scale or covariance matrices results in regularized functional estimates.  Here we extend this work by using informative covariance matrices to register functions where the registered functions are modeled as Gaussian processes.  Future work includes adapting these models for other areas of inference for functional data and continuing to improve on the models we have proposed here.

\noindent%
\bigskip
\appendix
\counterwithin{figure}{section}

\section{}

Below, in detail, are the specifications for the hierarchical Bayesian registration model discussed in this paper.  The first section includes the basic model for functional data registration also found in Section 2.   Section A.2 describes the MCMC sampling scheme for this model. \\

\subsection{Functional Registration}

As discussed in Section 2, the initial assumption of this model is that we are interested in registering functional data, $X_i(t), i=1, \hdots , N$, where these data are either observed directly over a set of time points, $\mathbf t = (t_1 \hdots t_p)'$, or are estimated from noisy observations, $\mathbf Y_i$$=$ $(Y_i(t_1), \hdots, Y_i(t_p))'$.  We assume a Gaussian noise process such that for each observation $\mathbf Y_i$,\\

$f(Y_{i}(t_{j})\mid X_{i}(t_{j}), \sigma^{2})$ $\mathbf{=}$ $ N(X_{i}(t_{j}),\sigma^{2})$ for $ i=1, \hdots, N, \quad j=1, \hdots,  p,$\\

which results in the joint distribution of all observations,\\

$f(\mathbf Y \mid \mathbf X,\sigma^{2})\mathbf{=}\prod_{i=1}^{N} N_p(\mathbf  X_{i},\sigma^{2}I),$\\

where $\mathbf Y$ is the matrix such that the observation for function $X_i(t)$ at time point $t_j$ is in the $i$th row and the $j$th column, $\mathbf X$ is the matrix of the corresponding means for each entry in $\mathbf Y$, and $\mathbf X_i = (X_i(t_1),\hdots, X_i(t_p))'$, the vector of evaluations of the functions $X_i(t)$ at time points $\mathbf t = (t_1,\hdots , t_p)'$.

When the observations are observed noisily, the registered functions and noise variance are characterized by the following prior distributions:
\begin{eqnarray}
\mathbf{X}_i(\mathbf{h}_i)\mid z_{0i},z_{1i}, \mathbf f,\gamma_R,\lambda_X  &\sim& N_p(z_{0i}\mathbf{1}+z_{1i}\mathbf {f},\gamma_R^{-1}\boldsymbol{\Sigma}+\boldsymbol\Sigma_X), \quad i=1\hdots N, \label{datamod}\\
\boldsymbol\Sigma_X &=& \eta_X^{-1}\mathbf P_1+ \lambda_X^{-1}\mathbf{P}_2, \text{ and} \nonumber\\
\sigma^2_y &\sim& IG(a,b).\nonumber
\end{eqnarray}

However, If each function $X_i(t)$ is observed directly over $\mathbf t$, \eqref{datamod} assumes the roll of the distribution of the observed data and the covariance matrix, $\boldsymbol\Sigma_X$, designed to penalize roughness in the unregistered functions is excluded. This results in the following data distribution,

\begin{eqnarray}
\mathbf{X}_i(\mathbf{h}_i)\mid z_{0i},z_{1i}, \mathbf f,\gamma_R,\lambda_X  &\sim& N_p(z_{0i}\mathbf{1}+z_{1i}\mathbf {f},\gamma_R^{-1}\boldsymbol{\Sigma}), \quad i=1\hdots N.\label{eq:datadist}
\end{eqnarray}

In both scenarios, we assume the following additional priors:
\begin{eqnarray}
\eta_X &\sim& G(c,d),  \nonumber\\
\lambda_X &\sim& G(c,d),  \nonumber\\
\mathbf{h}_i(t_j) &=& t_1 + \sum_{k=2}^{j} (t_k-t_{k-1})\exp({w_i(t_k)}),  \quad  i=1\hdots N, \quad j = 1 \hdots p, \nonumber \\
\mathbf w_i \mid \gamma_w &\propto& N_{p-1}(\mathbf 0,\gamma_w^{-1}\boldsymbol\Sigma+\lambda_w^{-1}\mathbf P_2)\mathbf{1}\{  t_1 + \sum_{k=2}^{p} (t_k-t_{k-1})\exp({w_i(t_k)})=t_p\},   \quad i=1\hdots N, \label{eq:WPrior} \quad s,t\in \mathcal {T}, \nonumber \\
z_{0i}\mid \sigma_{z0}^2 &\sim& N(0,\sigma_{z0}^2) \quad i=1\hdots (N-1), \quad z_{0N}=-\sum_{i=1}^{N-1} z_{0i},  \nonumber\\
\sigma_{z0}^2 &\sim& IG(a,b), \nonumber \\
z_{1i}\mid \sigma_{z1}^2 &\sim& N(1,\sigma_{z1}^2), \quad i=1\hdots N, \nonumber\\
\sigma_{z1}^2 &\sim& IG(a,b), \nonumber\\
\mathbf f \mid \eta_f, \lambda_f &\sim& N_p(0,\boldsymbol\Sigma_f), \nonumber\\
\boldsymbol\Sigma_f &=&\eta_f^{-1}\mathbf{P}_1+\lambda_f^{-1}\mathbf{P}_2, \nonumber\\
\eta_f &\sim& G(c,d), \text{ and}  \nonumber\\
\lambda_f &\sim& G(c,d),  \nonumber
\end{eqnarray}

where a, b, c, and d are fixed hyperparameters.  \\

$\boldsymbol\Sigma$ is a fixed matrix designed to penalize variation in any direction from the corresponding mean of the distribution in which it is utilized.  It is composed of two matrices, $\mathbf P_1$ and $\mathbf P_2$, such that  $\boldsymbol\Sigma$ $=$ $\mathbf P_1$ + $\mathbf P_2$.  $\mathbf P_1$ penalizes variation from the mean in constant and linear directions, and $\mathbf P_2$ penalizes variation from the mean in directions of curvature.  For the distribution on the registered functions, $\boldsymbol\Sigma$ penalizes variation from a vertical shift and scaling of the target function.  In the distribution of the base functions, $\boldsymbol\Sigma$ penalizes variation from the identity warping.  The fixed parameters $\gamma_R$ and $\gamma_w$ determine the degree of these penalties for the registered  functions and the base functions, respectively. 

 $\mathbf P_2$ is also used to penalize curvature in the registered functions, base functions, and the target function with associated smoothing parameters  $\lambda_X$, $\lambda_w$, and $\lambda_f$.   Further details of the construction of $\mathbf P_1$ and $\mathbf P_2$ are found in \citet{earls:14}.  Also note, in the prior on the base functions \eqref{eq:WPrior}, we are allowing $\mathbf P_2$ to be more generally interpreted as a penalty on severe deformations of the unregistered functions.  Here, $\mathbf P_2$ may be either a penalty on the squared second derivative of the base functions (as above) or a penalty on the squared first derivative of the base functions.
 
\subsection{MCMC Sampling}
 
Using these assumptions, the following full conditional distributions are derived to run a MCMC  sampler:

\begin{eqnarray}
\mathbf X_i \mid rest  &\sim& N_p((\sigma^{-2}_y\mathbf I_p + \boldsymbol\Sigma_X^{-1})^{-1}(\sigma^{-2}_y\mathbf I_p\mathbf Y_i + \boldsymbol\Sigma^{-1}_X(z_{0i}\mathbf 1_p + z_{1i}\mathbf {f(h_i^{-1})}), (\sigma^{-2}_y\mathbf I_p + \boldsymbol\Sigma_X^{-1})^{-1}),\nonumber\\
\mathbf f \mid rest &\sim& N_p(\boldsymbol\mu_{\mathbf f\mid rest},\boldsymbol\Sigma_{\mathbf f\mid rest}),\nonumber\\
\boldsymbol\Sigma_{\mathbf f\mid rest}&=& (\sum_{i=1}^{N}z_{1i}^2(\gamma_R^{-1}\boldsymbol{\Sigma}+\boldsymbol\Sigma_{X})^{-1}+ \boldsymbol\Sigma_f^{-1})^{-1}, \nonumber \\
\boldsymbol\mu_{\mathbf f\mid rest} &=& \boldsymbol\Sigma_{\mathbf f\mid rest}((\gamma_R^{-1}\boldsymbol{\Sigma}+\boldsymbol\Sigma_{X})^{-1}\sum_{i=1}^{N} z_{1i}(\mathbf{X_i(h_i)}-z_{0i}\mathbf 1_p)),\nonumber \\
\sigma_Y^{2}\mid rest &\sim& IG(a + Np/2, b + 1/2\sum_{i=1}^{N} (\mathbf Y_{i} - \mathbf X_{i})'(\mathbf Y_{i} - \mathbf X_{i})), \nonumber\\
\eta_{X}\mid rest &\sim& G(c + N, d +1/2\sum_{i=1}^{N} \textrm{tr}((\mathbf {X_i(h_i)}-(z_{0i}\mathbf 1_p + z_{1i}\mathbf f))(\mathbf {X_i(h_i)}-(z_{0i}\mathbf 1_p + z_{1i}\mathbf f))'\mathbf P_1^{-}),\nonumber\\
\lambda_{X}\mid rest &\sim& G(c + N, d +1/2\sum_{i=1}^{N} \textrm{tr}((\mathbf {X_i(h_i)}-(z_{0i}\mathbf 1_p + z_{1i}\mathbf f))(\mathbf {X_i(h_i)}-(z_{0i}\mathbf 1_p + z_{1i}\mathbf f))'\mathbf P_2^{-}),\nonumber\\
z_{0i}\mid rest &\sim& N(\mu_{z_{0i}\mid rest},\sigma^2_{z_{0i}\mid rest} ), \nonumber\\
\sigma^2_{z_{0i}\mid rest} &=& (\sigma^{-2}_{z_{0}} + 2*\mathbf 1_p'(\lambda_R^{-1}\boldsymbol\Sigma + \boldsymbol\Sigma_X)^{-1}\mathbf 1_p)^{-1},\nonumber\\
\mu_{z_{0i}\mid rest} &=& \sigma^2_{z_{0i}\mid rest}(\mathbf{X_i(h_i)}-\mathbf{X_N(h_N)}+(z_{1N}-z_{1i})\mathbf f - \sum_{j=1}^{N-1}z_{0j}\mathbf{1}\{ j\neq i\}\mathbf 1_p)'(\gamma_R^{-1}\boldsymbol\Sigma + \boldsymbol\Sigma_X)^{-1}\mathbf 1_p),\nonumber\\
\sigma^2_{z_{0}}\mid rest &\sim& IG(a + (N-1)/2, b +1/2 \sum_{i=1}^{N-1}z_{0i}^2 ),\nonumber\\
z_{1i}\mid rest &\sim& N(\mu_{z_{1i}\mid rest},\sigma^2_{z_{1i}\mid rest} ), \nonumber\\
\sigma^2_{z_{1i}\mid rest} &=& (\sigma^{-2}_{z_{1}} + \mathbf f'(\gamma_R^{-1}\boldsymbol\Sigma + \boldsymbol\Sigma_X)^{-1}\mathbf f)^{-1},\nonumber\\
\mu_{z_{1i}\mid rest} &=& \sigma^2_{z_{1i}\mid rest}((\mathbf{X_i(h_i)}-z_{0i}\mathbf 1_p)'(\lambda_R^{-1}\boldsymbol\Sigma + \boldsymbol\Sigma_X)^{-1}\mathbf f),\nonumber\\
\sigma^2_{z_{1}}\mid rest &\sim& IG(a + N/2, b +1/2 \sum_{i=1}^{N}z_{1i}^2 ),\nonumber\\
\eta_{f}\mid rest &\sim& G(c + 1, d + (\mathbf f'\mathbf P_{1}^{-}\mathbf f )/2 ), \text{ and}\nonumber\\
\lambda_{f}\mid rest &\sim& G(c + (p-2)/2, d + (\mathbf f'\mathbf P_{2}^{-}\mathbf f)/2 ).\nonumber
\end{eqnarray}

Note, this list does not include a full conditional for the base functions or registered functions as their priors are not conjugate.  Instead, the base and registered functions are sampled via a Metropolis step. \\

\subsection{Forcing Registered Time to Equal Standard Time}

\begin{enumerate}
\item{If the functions are registered so that registered features occur at the average time that they appear in the unregistered sample, for all $t$, $t \in \{t_1,\hdots,  t_p\}$, the average warping at that time point, $\overline{h_\mathbf{{\cdot}}(t)} = \frac{1}{N}\sum_{i=1}^{N} h_i(t),$ is the identity.  Over all observed time points this implies $\overline{\mathbf h_\mathbf{{\cdot}}}$ = $(\overline{h_\mathbf{{\cdot}}(t_1)}=t_1, \hdots ,\overline{h_\mathbf{{\cdot}}(t_p)}=t_p)'$.}
\item{Generally, after the registration process, this property will not hold, and instead  $\overline{\mathbf h_\mathbf{{\cdot}}}$ = $(\overline{h_\mathbf{{\cdot}}(t_1)}=\tilde{t}_{1}, \hdots ,\overline{h_\mathbf{{\cdot}}(t_p)}=\tilde{t}_{p})'$ where $t_j\neq \tilde t_j$ for at least one $j \in \{1, \hdots, p\}$.}
\item{The goal is to shift the functions so that these average warpings correspond to the correct registered times, i.e. $\overline{\tilde{\mathbf h}_\mathbf{{\cdot}}}$ = $(\overline{h_\mathbf{{\cdot}}(\tilde t_1)}=\tilde{t}_{1}, \hdots ,\overline{h_\mathbf{{\cdot}}(\tilde t_p)}=\tilde{t}_{p})'$}.
\item{(3) implies that after the initial registration, we have the correct registered function values over the new set of times, $\mathbf{\tilde t} = (\tilde t_1, \hdots, \tilde t_p)'$, i.e. we have estimates of \\$\mathbf X_i(\tilde{\mathbf h}_i) = (X_i(h_i(\tilde t_1)), \hdots, X_i(h_i(\tilde t_p)))'$, for all $i = 1, \hdots, N$}.
\item{If it is desired, the estimated values of the registered functions at the original time points, $\mathbf t$, can be obtained by interpolating values between the new set of time points, $\mathbf{\tilde t}$.}
\end{enumerate}

For example, after the initial registration process, suppose $\overline{h_\mathbf{{\cdot}}(2)}=2.25$ and $\overline{h_\mathbf{{\cdot}}(3)}=3.1$, where 2 and 3 are in the set of original time points.  We can alter registered time so that  $\overline{h_\mathbf{{\cdot}}(2.25)}=2.25$ and $\overline{h_\mathbf{{\cdot}}(3.1)}=3.1$, as desired.  Using the notation above, this implies $\{2.25,3.1\}\subset \tilde{\mathbf t}$ and for all $i, i=1\, \hdots , N$, from the initial registration, we estimated the following values,  $X_i(h_i(2.25))$ and $X_i(h_i(3.1))$. From these we can estimate $X_i(h_i(3))$ by interpolating the values $X_i(h_i(2.25))$ and $X_i(h_i(3.1))$.

\section{}

\subsection{Adapted Variational Bayes}

Below are the approximate posterior distributions for the adapted variational Bayes estimation procedure outlined in Section 3.1.  

Based on the full conditional distributions found in Appendix A.2, the following approximate posterior distributions are updated in each iteration for the registration model that accounts for noisy observations:

\begin{eqnarray}
q(\mathbf X_i) &\sim& N_p(\boldsymbol\mu_{q(\mathbf X_i)},\boldsymbol\Sigma_{q(\mathbf X_i)}),\nonumber\\
q(\mathbf f) &\sim& N_p(\boldsymbol\mu_{q(\mathbf f)}, \boldsymbol\Sigma_{q(\mathbf f)}),\nonumber\\
q(\sigma_Y^2) &\sim& IG(a_{q(\sigma_Y^2)}, b_{q(\sigma_Y^2)}),\nonumber\\
q(\eta_X) &\sim& G(c_{q(\eta_X)}, d_{q(\eta_X)}), \nonumber\\
q(\lambda_X) &\sim& G(c_{q(\lambda_X)}, d_{q(\lambda_X)}), \nonumber\\
q(z_{0i}) &\sim&  N(\mu_{q(z_{0i})},\sigma^2_{q(z_{0i})})\nonumber\\
q(\sigma_{z_0}^2) &\sim& IG(a_{q(\sigma_{z_0}^2)}, b_{q(\sigma_{z_0}^2)}), \nonumber\\
q(z_{1i}) &\sim&  N(\mu_{q(z_{1i})},\sigma^2_{q(z_{1i})}),\nonumber\\
q(\sigma_{z_1}^2) &\sim& IG(a_{q(\sigma_{z_1}^2)}, b_{q(\sigma_{z_1}^2)}), \nonumber\\
q(\eta_f) &\sim& G(c_{q(\eta_f)}, d_{q(\eta_f)}), \text{ and} \nonumber\\
q(\lambda_f) &\sim& G(c_{q(\lambda_f)}, d_{q(\lambda_f)}). \nonumber
\end{eqnarray}

If the observations are recorded without noise, i.e. we have observations $\mathbf X_i$, $i=1\hdots N$ as described in \eqref{eq:datadist}, the following $q$ distributions are omitted,
\begin{eqnarray}
q(\mathbf X_i) \text{ for } i = 1\hdots N, q(\sigma^2_y), q(\eta_X), \text{ and } q(\lambda_X). \nonumber
\end{eqnarray}

As the $q$ densities are all of known distributional forms, updating these densities is equivalent to updating their parameters. First, assuming the data are recorded without noise, for each iteration, the following parameters are updated for the $q$ densities listed above:  
\begin{flalign}
&\boldsymbol\Sigma_{q(\mathbf f)} = \Big[\sum_{i=1}^{N}(\sigma_{q(z_{1i})}^2 + \mu_{q(z_{1i})}^2)\gamma_R\boldsymbol{\Sigma}^{-1}+\mu_{q(\eta_{\mathbf f})}\mathbf P_{1}^{-} + \mu_{q(\lambda_{\mathbf f})}\mathbf P_2^{-}\Big]^{-1},&\nonumber\\
&\boldsymbol\mu_{q(\mathbf f)} =\boldsymbol\Sigma_{q(\mathbf f)}\gamma_R\boldsymbol{\Sigma}^{-1}\Big[\sum_{i=1}^{N}\mu_{q(z_{1i})}(\mathbf X_i(\mathbf h_i)-\mu_{q(z_{0i})}\mathbf 1_p)\Big],&\nonumber\\
&\sigma_{q(z_{0i})}^2 = (\mu_{q(\sigma^{-2}_{z_0})} + 2\mathbf  1_p'\gamma_R\boldsymbol{\Sigma}^{-1}\mathbf 1_p)^{-1},&\nonumber\\
&\mu_{q(z_{0i})} = \Big[\sigma_{q(z_{0i})}^2\big(\mathbf X_i(\mathbf h_i)'-\mathbf X_N(\mathbf h_N)'+(\mu_{q(z_{1N})}-\mu_{q(z_{1i})})\mu_{q(\mathbf f)}'-\sum_{j=1}^{N-1} \mu_{q(z_{0j})}\mathbf{1}\{i\neq j\}\mathbf 1_p'\big)\Big]\gamma_R\boldsymbol\Sigma^{-1}\mathbf 1_p,&\nonumber\\
&\sigma_{q(z_{1i})}^2 = (\mu_{q(\sigma^{-2}_{z_1})} + \textrm{tr}((\boldsymbol\Sigma_{q(\mathbf f)} + \mu_{q(\mathbf f)}\mu_{q(\mathbf f)}')\gamma_R\boldsymbol{\Sigma}^{-1}))^{-1},&\nonumber\\
&\mu_{q(z_{1i})} = \sigma_{q(z_{1i})}^2(\mu_{q(\sigma^{-2}_{z_{1}})}+\mu_{q(\mathbf f)}'\gamma_R\boldsymbol\Sigma^{-1}(\mathbf X_i(\mathbf h_i) - \mu_{q(z_{0i})}\mathbf 1_p)),&\nonumber\\
&d_{q(\eta_{\mathbf f})} = d+1/2*\textrm{tr}(\mathbf P_1^{-}(\Sigma_{q(\mathbf f)}+\mu_{q(\mathbf f)}\mu_{q(\mathbf f)}')),&\nonumber\\
&d_{q(\lambda_{\mathbf f})} = d+1/2*\textrm{tr}(\mathbf P_2^{-}(\Sigma_{q(\mathbf f)}+\mu_{q(\mathbf f)}\mu_{q(\mathbf f)}')),&\nonumber\\&b_{q(\sigma_{z_{0}}^{2})} = b+1/2\sum_{i=1}^{N-1}( \sigma^2_{q(z_{0i})} + \mu_{q(z_{0i})}^2), \text{ and}&\nonumber\\
&b_{q(\sigma_{z_{1}}^{2})} = b+1/2\sum_{i=1}^{N}( \sigma^2_{q(z_{1i})} + \mu_{q(z_{1i})}^2).&\nonumber
\end{flalign}

Note, these updates are listed in an order that allows the convergence criterion to be calculated.  Further details on the convergence criterion can be found in the next section.

For the model where observations are recorded with noise, in all of the updates above $\mathbf X_i(\mathbf h_i)$ and  $\mathbf X_N(\mathbf h_N)$ are replaced by $\boldsymbol\mu_{q(\mathbf X_i(\mathbf h_i))}$ and  $\boldsymbol\mu_{q(\mathbf X_N(\mathbf h_N))}$, respectively.  Additionally, $\gamma_R\boldsymbol\Sigma^{-1}$ is replaced by $(\gamma_R^{-1}\boldsymbol\Sigma+\boldsymbol\Sigma_X)^{-1}$.  For each $i$,  $\boldsymbol\mu_{q(\mathbf X_i(\mathbf h_i))}$ is determined by using the update for the mean of the $q$ distribution of the unregistered function, $\boldsymbol\mu_{q(\mathbf X_i)}$ below, and registering it using the current value of $\mathbf h_i$.  In addition to these modified updates, the following additional updates necessary:

\begin{flalign}
&\boldsymbol\Sigma_{q(\mathbf X_i)} = (\mu_{q(\frac{1}{\sigma^2_Y})}\mathbf I_p+ \mu_{q(\eta_X)}\mathbf P_1^{-} + \mu_{q(\lambda_X)}\mathbf P_2^{-})^{-1}, &\nonumber\\
&\boldsymbol\mu_{q(\mathbf X_i)} = \boldsymbol\Sigma_{q(\mathbf X_i)} [\mu_{q(\frac{1}{\sigma^2_Y})}\mathbf Y_i + ( \mu_{q(\eta_X)}\mathbf P_1^{-} + \mu_{q(\lambda_X)}\mathbf P_2^{-})
 (\mu_{q(z_{0i})}\mathbf 1_p+\mu_{q(z_{1i})}E_{(\mathbf\theta_{-\mathbf X_i})}[\mathbf f(\mathbf  h_i^{-1})])],& \nonumber\\
&b_{q(\sigma_Y^{2})} =b+\frac{1}{2}\sum_{i=1}^{N}\Big(\mathbf Y_i'\mathbf Y_i-2\boldsymbol\mu_{q(\mathbf X_i)}'\mathbf Y_i + \sum_{j=1}^{p} \boldsymbol\Sigma_{q(\mathbf X_i)}[j,j]+\boldsymbol\mu_{q(\mathbf X_i)}[j]^2\Big),&\nonumber\\
&d_{q(\eta_{\mathbf X})} = d+\frac{1}{2}\textrm{tr}\Big[\sum_{i=1}^{N}\Big(\boldsymbol\Sigma_{q(\mathbf X_i)} + \boldsymbol\mu_{q(\mathbf X_i)}\boldsymbol\mu_{q(\mathbf X_i)}'-2\boldsymbol\mu_{q(\mathbf X_i)}(\mu_{q(z_{0i})}\mathbf 1_p + \mu_{q(z_{1i})}E_{(\mathbf\theta_{-\eta_X})}[\mathbf f(\mathbf  h_i^{-1})])'&\nonumber \\
& \quad\quad\quad+( \sigma^2_{q(z_{0i})}+\mu_{q(z_{0i})}^2)\mathbf 1_p\mathbf 1_p' + 2\mu_{q(z_{0i})}\mu_{q(z_{1i})}\mathbf 1_pE_{(\mathbf\theta_{-\eta_X})}[\mathbf f(\mathbf  h_i^{-1})]'&\nonumber\\
 & \quad\quad\quad+ \textit{ }( \sigma^2_{q(z_{1i})}+\mu_{q(z_{1i})}^2)E_{(\mathbf\theta_{-\eta_X})}[\mathbf f(\mathbf  h_i^{-1})\mathbf f(\mathbf  h_i^{-1})']\Big)\mathbf P_1^{-} \Big], \text{ and}\nonumber&\nonumber\\
&d_{q(\lambda_{\mathbf X})} =  d+\frac{1}{2}\textrm{tr}\Big[\sum_{i=1}^{N}\Big(\boldsymbol\Sigma_{q(\mathbf X_i)} + \boldsymbol\mu_{q(\mathbf X_i)}\boldsymbol\mu_{q(\mathbf X_i)}'-2\boldsymbol\mu_{q(\mathbf X_i)}(\mu_{q(z_{0i})}\mathbf 1_p + \mu_{q(z_{1i})}E_{(\mathbf\theta_{-\lambda_X})}[\mathbf f(\mathbf  h_i^{-1})])'&\nonumber \\
 &\quad\quad\quad+( \sigma^2_{q(z_{0i})}+\mu_{q(z_{0i})}^2)\mathbf 1_p\mathbf 1_p' + 2\mu_{q(z_{0i})}\mu_{q(z_{1i})}\mathbf 1_pE_{(\mathbf\theta_{-\lambda_X})}[\mathbf f(\mathbf  h_i^{-1})]'&\nonumber\\
 &\quad\quad\quad + ( \sigma^2_{q(z_{1i})}+\mu_{q(z_{1i})}^2)E_{(\mathbf\theta_{-\lambda_X})}[\mathbf f(\mathbf  h_i^{-1})\mathbf f(\mathbf  h_i^{-1})']\Big)\mathbf P_2^{-} \Big].&\nonumber
\end{flalign}

Note, these updates contain terms that cannot be evaluated.  For instance, $E_{(\mathbf\theta_{-\eta_X })}[\mathbf f(\mathbf  h_i^{-1})\mathbf f(\mathbf  h_i^{-1})']$ cannot  be determined because the approximate distribution of $\mathbf f(\mathbf  h_i^{-1})$ is unknown.  These terms can however be approximated.  Appendix C.2 provides details of the approximated values used for this analysis.

\subsection{Convergence Criterion}

When the functional observations, $\mathbf X$ = $\mathbf X_i$, $i\hdots N$, are recorded without noise, \\$E_{q(\boldsymbol\theta_{-\mathbf w})}\big[\log f(\mathbf X, \mathbf w,\boldsymbol\theta_{-\mathbf w})- \log q(\boldsymbol\theta_{-\mathbf w})\big]$ is monitored until the desired threshhold is met, where

\begin{eqnarray}
E_{q(\boldsymbol\theta_{-\mathbf w})}\big[\log f(\mathbf X, \mathbf w,\boldsymbol\theta_{-\mathbf w})- \log q(\boldsymbol\theta_{-\mathbf w})\big]&=&E_{q(\boldsymbol\theta_{-\mathbf w})}\big[\log (f(\mathbf X, \mathbf w\mid \boldsymbol\theta_{-\mathbf w})f( \boldsymbol\theta_{-\mathbf w})) - \log q(\boldsymbol\theta_{-\mathbf w})\big]\nonumber\\
&=&E_{q(\boldsymbol\theta_{-\mathbf w})}\big[\log f(\mathbf X, \mathbf w\mid \boldsymbol\theta_{-\mathbf w}) + \log f( \boldsymbol\theta_{-\mathbf w}) - \log q(\boldsymbol\theta_{-\mathbf w})\big]\nonumber\\
&=&E_{q(\boldsymbol\theta_{-\mathbf w})}\big[\log f(\mathbf X, \mathbf w\mid \boldsymbol\theta_{-\mathbf w})\big] \nonumber \\
&&  +\text{ } E_{q(\boldsymbol\theta_{-\mathbf w})}\big[\log f(\mathbf f)-\log q(\mathbf f)\big]\nonumber \\
&& + \sum_{i=1}^{(N-1)}E_{q(\boldsymbol\theta_{-\mathbf w})}\big[\log f(z_{0i})-\log q(z_{0i})\big]\nonumber\\
&& + \sum_{i=1}^{(N)}E_{q(\boldsymbol\theta_{-\mathbf w})}\big[\log f(z_{1i})-\log q(z_{1i})\big]\nonumber\\
&& + \textit{ } E_{q(\boldsymbol\theta_{-\mathbf w})}\big[\log f(\sigma^2_{z_0})-\log q(\sigma^2_{z_0})\big]\nonumber\\
&& + \textit{ } E_{q(\boldsymbol\theta_{-\mathbf w})}\big[\log f(\sigma^2_{z_1})-\log q(\sigma^2_{z_1})\big]\nonumber\\
&& + \textit{ } E_{q(\boldsymbol\theta_{-\mathbf w})}\big[\log f(\eta_f)-\log q(\eta_f)\big]\nonumber\\
&& + \textit{ }E_{q(\boldsymbol\theta_{-\mathbf w})}\big[\log f(\lambda_f)-\log q(\lambda_f)\big].\nonumber
\end{eqnarray}

Now looking at each piece individually,

$E_{q(\boldsymbol\theta_{-\mathbf w})}\big[\log f(\mathbf X, \mathbf w\mid \boldsymbol\theta_{-\mathbf w})\big] $
\begin{eqnarray}
&=& E_{q(\boldsymbol\theta_{-\mathbf w})}\Big[\sum_{i=1}^{N}\big(\log [(2\pi)^{-p/2}\mid\gamma_R^{-1}\boldsymbol\Sigma\mid^{-1/2}]\big)\Big]\nonumber \\
&& + \textit{ }E_{q(\boldsymbol\theta_{-\mathbf w})}\Big[\sum_{i=1}^{N}-\frac{1}{2}[(\mathbf X_i(\mathbf h_i)-(z_{0i}\mathbf 1_p+z_{1i}\mathbf f))'\gamma_R\boldsymbol\Sigma^{-1}(\mathbf X_i(\mathbf h_i) -(z_{0i}\mathbf 1_p+z_{1i}\mathbf f))]\Big]\nonumber\\
&=&\sum_{i=1}^{N}\big(\log[(2\pi)^{-p/2}\mid\gamma_R^{-1}\boldsymbol\Sigma\mid^{-1/2}]\big)\nonumber\\
&& + \sum_{i=1}^{N} -\frac{1}{2}\Big[(\mathbf X_i(\mathbf h_i)'\gamma_R\boldsymbol\Sigma^{-1}\mathbf X_i(\mathbf h_i))-\nonumber\\
&& \textit{         }2\mathbf X_i(\mathbf h_i)'\gamma_R\boldsymbol\Sigma^{-1}\mu_{q(z_{0i})}\mathbf 1_p-2\mathbf X_i(\mathbf h_i)'\gamma_R\boldsymbol\Sigma^{-1}\mu_{q(z_{1i})}\boldsymbol\mu_{q(\mathbf f)} +\nonumber\\
&&\textit{ } (\sigma^2_{q(z_{1i})} + \mu_{q(z_{1i})}^2)\textrm{tr}((\boldsymbol\Sigma_{q(\mathbf f)} + \boldsymbol\mu_{q(\mathbf f)}\boldsymbol\mu_{q(\mathbf f)}')\gamma_R\boldsymbol\Sigma^{-1})+\nonumber\\
&&\textit{ }2\mu_{q(z_{0i})}\mu_{q(z_{1i})}\mathbf 1_p'\gamma_R\boldsymbol\Sigma^{-1}\boldsymbol\mu_{q(\mathbf f)}\Big] -\nonumber\\
&&\textit{ } \Big[\sum_{i=1}^{N-1} (\sigma^2_{q(z_{0i})} + \mu_{q(z_{0i})}^2) + \frac{1}{2} \sum_{i=1}^{N-1}\sum_{j=1}^{N-1}\mu_{q(z_{0i})}\mu_{q(z_{0j})}\mathbf{1}\{j \neq i\}\Big]\mathbf 1_p'\gamma_R\boldsymbol\Sigma^{-1}\mathbf 1_p, \nonumber
\end{eqnarray}

\begin{eqnarray}
E_{q(\boldsymbol\theta_{-\mathbf w})}\big[\log f(\mathbf f)-\log q(\mathbf f)\big] &=& E_{q(\boldsymbol\theta_{-\mathbf w})}\Big[-\frac{p}{2}\log 2\pi + \frac{1}{2}\log \mid\eta_f\mathbf P_1^{-} + \lambda_f\mathbf P_2^{-}\mid\Big] - \nonumber \\
&& \textit{ } E_{q(\boldsymbol\theta_{-\mathbf w})}\Big[\frac{1}{2}(\textrm{tr}[\mathbf f\mathbf f'(\eta_f\mathbf P_1^{-}+\lambda_f\mathbf P_2^{-})]\Big] +\nonumber\\
&& \textit{ }  E_{q(\boldsymbol\theta_{-\mathbf w})}\Big[\frac{p}{2}\log 2\pi + \frac{1}{2} \log \mid\boldsymbol\Sigma_{q(\mathbf f)}\mid\Big] +\nonumber\\
&& \textit{ } E_{q(\boldsymbol\theta_{-\mathbf w})}\Big[\frac{1}{2}\textrm{tr}(\mathbf f\mathbf f '\boldsymbol\Sigma_{q(\mathbf f)}^{-1}) - \mathbf f'\boldsymbol\Sigma_{q(\mathbf f)}^{-1}\boldsymbol\mu_{q(\mathbf f)}\Big] +\nonumber\\
&& \textit{ } E_{q(\boldsymbol\theta_{-\mathbf w})}\Big[\frac{1}{2}\boldsymbol\mu_{q(\mathbf f)}'\boldsymbol\Sigma_{q(\mathbf f)}^{-1}\boldsymbol\mu_{q(\mathbf f)}\Big]\nonumber\\
&=& C + \frac{1}{2} E_{q(\boldsymbol\theta_{-\mathbf w})}\big[2\log \eta_f\big]+ \frac{1}{2} E_{q(\boldsymbol\theta_{-\mathbf w})}\big[(p-2)\log \lambda_f\big] - \nonumber\\
&& \textit{ } \frac{1}{2}\textrm{tr}\Big((\boldsymbol\Sigma_{q(\mathbf f)} + \boldsymbol\mu_{q(\mathbf f)}\boldsymbol\mu_{q(\mathbf f)}')(\mu_{q(\eta_f)}\mathbf P_1^{-}+\mu_{q(\lambda_f)}\mathbf P_2^{-})\Big)-\nonumber\\
&& \textit{ }\frac{1}{2}\log \mid\boldsymbol\Sigma_{q(\mathbf f)}^{-1}\mid + \frac{p}{2}, \nonumber
\end{eqnarray}
where $C$ is a constant that does not change from one iteration to the next. For $\mathbf z_0$ $=$ $(z_{01},\hdots ,z_{0(N-1)})', $

\begin{eqnarray}
E_{q(\boldsymbol\theta_{-\mathbf w})}\big[\log f(\mathbf z_0)-\log q(\mathbf z_0)\big] &=& E_{q(\boldsymbol\theta_{-\mathbf w})}\Big[-\frac{N-1}{2}\log 2\pi -\frac{N-1}{2}\log \sigma^2_{z_0} -\sum_{i=1}^{N-1}-\frac{1}{2\sigma^2_{z_0}}z_{0i}^2  + \nonumber\\
&& \textit{ }\frac{N-1}{2}\log 2\pi+\frac{N-1}{2}\log \sigma^2_{q({z_{0i}})}+\nonumber \\
&& \textit{ }\sum_{i=1}^{N-1}\frac{1}{2\sigma^2_{q(z_{0i})}}(z_{0i}-\mu_{q(z_{0i})})^2\Big] \label{eq:z0}\\
&=&\frac{N-1}{2}\log \sigma^2_{q({z_{0i}})} - E_{q(\boldsymbol\theta_{-\mathbf w})}\Big[\frac{N-1}{2} \log \sigma^2_{z_0}\Big]-\label{eq:z02} \\ 
&&\textit{ } \frac{1}{2}\mu_{q(\frac{1}{\sigma^2_{z_{0}}})}\Big(\sum_{i=1}^{N-1}(\sigma_{q(z_{0i})}^2+\mu_{q(z_{0i})}^2)\Big) +\frac{N-1}{2}.\nonumber
\end{eqnarray}

For $\mathbf z_1$ $=$ $(z_{11},\hdots ,z_{1N})',$

\begin{eqnarray}
E_{q(\boldsymbol\theta_{-\mathbf w})}\big[\log f(\mathbf z_1)-\log q(\mathbf z_1)\big] &=& E_{q(\boldsymbol\theta_{-\mathbf w})}\Big[-\frac{N}{2} \log 2\pi -\frac{N}{2}\log \sigma^2_{z_1} -\sum_{i=1}^{N}-\frac{1}{2\sigma^2_{z_1}}z_{1i}^2  + \nonumber\\
&& \textit{ }\frac{N}{2}\log 2\pi+\frac{N}{2}\log \sigma^2_{q({z_1})}+\sum_{i=1}^{N}\frac{1}{2\sigma^2_{q(z_{1i})}}(z_{1i}-\mu_{q(z_{1i})})^2\Big] \nonumber\\
&=&\frac{N}{2}\log \sigma^2_{q({z_{1i}})} - E_{q(\boldsymbol\theta_{-\mathbf w})}\Big[\frac{N}{2}\log \sigma^2_{z_1}\Big]-\nonumber \\
&&\textit{ } \frac{1}{2}\mu_{q(\frac{1}{\sigma^2_{z_1}})}\Big(\sum_{i=1}^{N}(\sigma_{q(z_{1i})}^2+\mu_{q(z_{1i})}^2)\Big) +\frac{N}{2}.\nonumber
\end{eqnarray}

For $\sigma_{z_0}^2$,

\begin{eqnarray}
E_{q(\boldsymbol\theta_{-\mathbf w})}\big[\log f(\sigma_{z_0}^2)-\log q(\sigma_{z_0}^2)\big] &=& E_{q(\boldsymbol\theta_{-\mathbf w})}\Big[\log \frac{b^a}{\Gamma(a)}-(a+1)\log \sigma_{z_0}^2-b\frac{1}{\sigma_{z_0}^2}-\nonumber\\
&& \log \frac{b_{q(\sigma_{z_0}^2)}^{a_{q(\sigma_{z_0}^2)}}}{\Gamma(a_{q(\sigma_{z_0}^2)})}+(a_{q(\sigma_{z_0}^2)}+1)\log \sigma_{z_0}^2+\nonumber\\
&& \textit{ } b_{q(\sigma_{z_0}^2)}\frac{1}{\sigma_{z_0}^2}\Big]\nonumber\\
&=&E_{q(\boldsymbol\theta_{-\mathbf w})}\big[-(a+1)\log \sigma_{z_0}^2\big]  -b\mu_{q(\frac{1}{\sigma_{z_0}^2})} -\log  \frac{b_{q(\sigma_{z_0}^2)}^{a_{q(\sigma_{z_0}^2)}}}{\Gamma(a_{q(\sigma_{z_0}^2)})}+\nonumber \\
&& \textit{ } \log \frac{b^a}{\Gamma(a)}+E_{q(\boldsymbol\theta_{-\mathbf w})}\big[(a_{q(\sigma_{z_0}^2)}+1)\log \sigma_{z_0}^2\big]+ b_{q(\sigma_{z_0}^2)}\mu_{q(\frac{1}{\sigma_{z_0}^2})}.\nonumber
\end{eqnarray}

For $\sigma_{z_1}^2$,

\begin{eqnarray}
E_{q(\boldsymbol\theta_{-\mathbf w})}\big[\log f(\sigma_{z_1}^2)-\log q(\sigma_{z_1}^2)\big] &=& E_{q(\boldsymbol\theta_{-\mathbf w})}\Big[\log \frac{b^a}{\Gamma(a)}-(a+1)\log \sigma_{z_1}^2-b\frac{1}{\sigma_{z_1}^2}-\nonumber\\
&& \log \frac{b_{q(\sigma_{z_1}^2)}^{a_{q(\sigma_{z_1}^2)}}}{\Gamma(a_{q(\sigma_{z_1}^2)})}+(a_{q(\sigma_{z_1}^2)}+1)\log \sigma_{z_1}^2+\nonumber\\
&& \textit{ } b_{q(\sigma_{z_1}^2)}\frac{1}{\sigma_{z_1}^2}\Big]\nonumber\\
&=&E_{q(\boldsymbol\theta_{-\mathbf w})}\big[-(a+1)\log \sigma_{z_1}^2\big]  -b\mu_{q(\frac{1}{\sigma_{z_1}^2})} -\log  \frac{b_{q(\sigma_{z_1}^2)}^{a_{q(\sigma_{z_1}^2)}}}{\Gamma(a_{q(\sigma_{z_1}^2)})}+\nonumber \\
&& \textit{ } \log \frac{b^a}{\Gamma(a)}+E_{q(\boldsymbol\theta_{-\mathbf w})}\big[(a_{q(\sigma_{z_1}^2)}+1)\log \sigma_{z_1}^2\big]+ b_{q(\sigma_{z_1}^2)}\mu_{q(\frac{1}{\sigma_{z_1}^2})}.\nonumber
\end{eqnarray}

For $\eta_f$,

\begin{eqnarray}
E_{q(\boldsymbol\theta_{-\mathbf w})}\big[\log f(\eta_f)-\log q(\eta_f)\big] &=& E_{q(\boldsymbol\theta_{-\mathbf w})}\Big[\log \frac{d^c}{\Gamma(c)}+(c-1)\log \eta_f-d\eta_f-\nonumber\\
&& \log \frac{d_{q(\eta_f)}^{c_{q(\eta_f)}}}{\Gamma(c_{q(\eta_f)})}-c \log \eta_f +d_{q(\eta_f)}{\eta_f}\Big]\nonumber\\
&=& \log \frac{d^c}{\Gamma(c)} - \log \frac{d_{q(\eta_f)}^{c_{q(\eta_X)}}}{\Gamma(c_{q(\eta_f)})}-E_{q(\boldsymbol\theta_{-\mathbf w})}\big[\log \eta_f\big]  -d\mu_{q(\eta_f)} +\nonumber\\
&& d_{q(\eta_f)}\mu_{q(\eta_f)}.\nonumber 
\end{eqnarray}

For $\lambda_f$,

\begin{eqnarray}
E_{q(\boldsymbol\theta_{-\mathbf w})}\big[\log f(\lambda_f)-\log q(\lambda_f)\big] &=& E_{q(\boldsymbol\theta_{-\mathbf w})}\Big[\log \frac{d^c}{\Gamma(c)}+(c-1)\log \lambda_f-d\lambda_f-\nonumber\\
&& \log \frac{d_{q(\lambda_f)}^{c_{q(\lambda_f)}}}{\Gamma(c_{q(\lambda_f)})}-\Big(\frac{p-2}{2}+c-1\Big)\log \lambda_f +d_{q(\lambda_f)}{\lambda_f}\Big]\nonumber\\
&=& \log \frac{d^c}{\Gamma(c)} - \log \frac{d_{q(\lambda_f)}^{c_{q(\lambda_f)}}}{\Gamma(c_{q(\lambda_f)})}-\frac{p-2}{2}E_{q(\boldsymbol\theta_{-\mathbf w})}\big[\log \lambda_f\big]  -d\mu_{q(\lambda_f)} +\nonumber\\
&& d_{q(\lambda_f)}\mu_{q(\lambda_f)}.\nonumber 
\end{eqnarray}

The expression for $E_{q(\boldsymbol\theta_{-\mathbf w})}\big[\log f(\mathbf X, \mathbf w,\boldsymbol\theta_{-\mathbf w})- \log q(\boldsymbol\theta_{-\mathbf w})\big]$ can be simplified much further by combining terms that cancel out.  However, in some cases the ability to cancel terms depends on the order of the updates.  For instance, in the expression, $E_{q(\boldsymbol\theta_{-\mathbf w})}\big[\log f(\sigma_{z_0}^2)-\log q(\sigma_{z_0}^2)\big]$, the terms $-b\mu_{q(\frac{1}{\sigma^2_{z_0}})}$ and $b_{q(\sigma^2_{z_0})}\mu_{q(\frac{1}{\sigma^2_{z_0}})}$ cancel with  $- \frac{1}{2}\mu_{q(\frac{1}{\sigma^2_{z_0}})}\Big(\sum_{i=1}^{N-1}(\sigma_{q(z_{0i})}^2+\mu_{q(z_{0i})}^2)\Big)$ from $E_{q(\boldsymbol\theta_{-\mathbf w})}\big[\log f(\mathbf z_0)-\log q(\mathbf z_0)\big]$ as long as the parameters of $q(\mathbf z_0)$ are updated before $b_{q(\sigma^2_{z_0})}$.  For convenience,  we have taken account the ordering necessary to compute the convergence criterion in the updates given above. Additionally, note all components in this expression that do not change from one iteration to the next can be ignored.  

\section{}

\subsection{Functional Data Regularization and Registration}
 \label{sec:smooth}
 
If instead of the function itself, noisy observations of each unregistered function, $X_i(t)$ are observed over a finite number of time points, $\mathbf t = (t_1,\hdots, t_p)'$, we will additionally assume that the observations, $Y_i(t_j),  j=1,\hdots, p$ are iid, $N(X_i(t_j),\sigma_Y^2).$  Incorporating registration and smoothing into a single model has also been considered recently by \citet{rak:14}.  In their paper, each  registered noisy function, $Y_i(h_i(t))$ at time point $t_j$, $j = 1, \hdots, p$ is composed as follows:
 \begin{eqnarray}
 Y_i(h_i(t_j)) &=& f(t_j) + r_i(t_j) + \epsilon_i(t_j),\nonumber
\end{eqnarray}
where $f(t)$ is similar to our target function, $r_i(t)$ is a function-specific random effect that accounts for variation in individual noiseless functions beyond the target function, and $\epsilon_i(t_j)$ are iid Gaussian noise.  

The advantage of our model is that incorporating individual random effects is unnecessary.  Noting that the \textit{observations} are noisy, not the registered functions; smoothing in our model is applied to the observations, not to the functions after registration.  Under these conditions, variability in the estimated unregistered, smoothed functions, $X_i(t)$, can be looked at separately from variability in the estimated registered functions, $X_i(h_i(t))$.  Appendix C.3 provides an example of how treating smoothing as a pre-processing step underestimates variability in the posterior distributions of the registered functions.

In the presence of noisy observations, the following distributions are either altered or added to the registration model presented in Section 2,

 \begin{eqnarray}
 Y_i(t_j)\mid X_i(t_j) &\sim& N(X_i(t_j),\sigma_Y^2), \quad i=1\hdots N, \quad j=1,\hdots, p, \label{eq:datadist} \\
 \mathbf{X}_i(\mathbf{h}_i)\mid z_{0i},z_{1i}, \mathbf f,\eta_X, \lambda_X  &\sim& N_p(z_{0i}\mathbf{1}+z_{1i}\mathbf {f},\gamma_R^{-1}\boldsymbol{\Sigma}+\boldsymbol\Sigma_X), \quad i=1,\hdots, N, \label{eq:RegPrior}\\
\boldsymbol\Sigma_X &=& \eta_X^{-1}\mathbf P_1 + \lambda_X^{-1}\mathbf P_2,\nonumber\\
\eta_X &\sim& G(c,d),  \nonumber\\
\lambda_X &\sim& G(c,d), \text{ and}  \nonumber\\
\sigma_Y^2 &\sim& IG(a,b).\nonumber
\end{eqnarray}

The most significant change to the model is that we now include a smoothing penalty in the covariance specification for the registered functions.  Here specifying $\mathbf P_2$ (a roughness penalty - see Section 2 for more details) in the prior distribution for the registered functions establishes regularization in these functions. The associated smoothing parameter is $\lambda_X$.  As mentioned previously, $\mathbf P_1$ is required to define $\boldsymbol\Sigma_X$ as a proper covariance matrix.  More details on these matrices can be found in Appendix A.  As can be seen above, $\eta_X$ and $\lambda_X$ are considered as additional unknown parameters in this hierarchical model.
   
In the prior specifications of this model, equation \eqref{eq:RegPrior} incorporates penalties for both smoothing and registration within the prior for the registered functions. The full conditional distribution for each approximated registered function, $\mathbf X_i(\mathbf h_i)$, when data are noisily observed is the joint full-conditional of the unregistered function and the warping function,
\begin{eqnarray}
f(\mathbf{X}_i(\mathbf{h}_i)\mid rest) &=& f(\mathbf w_i, \mathbf X_i \mid rest).  \nonumber \\
&=& f(\mathbf w_i \mid \mathbf X_i, rest)f(\mathbf{X}_i\mid rest)\nonumber
\end{eqnarray}

 Instead of drawing from this joint full-conditional directly, we will proceed by first drawing from $f(\mathbf{X}_i\mid rest)$ and then given $\mathbf X_i$, draw from $ f(\mathbf w_i \mid \mathbf X_i, rest)$.
 
These full conditional distributions are determined in the standard way recognizing that the prior distribution for a registered function can be factored into two components. One component penalizes lack of registration given the approximated unregistered function, $\mathbf X_i$;  the other component penalizes roughness in the registered function which implicitly penalizes roughness in the unregistered function.  The roughness penalty is independent of the warping function and therefore also of $\mathbf w_i$.  Specifically, the prior distribution \eqref{eq:RegPrior} for each $\mathbf X_i(\mathbf h_i)$, $i= 1, \hdots, N$, is such that  \\

$f(\mathbf X_i(\mathbf h_i) \mid \mathbf X_i, \mathbf w_i, z_{0i},z_{1i},\mathbf f,\eta_X,\lambda_X) \propto$
 \begin{eqnarray}
  && \exp\left[-\frac{1}{2}\left((\mathbf X_i(\mathbf h_i)-(z_{0i}\mathbf {1} + z_{1i}\mathbf {f}))'\gamma_R\boldsymbol{\Sigma}^{-1} (\mathbf X_i(\mathbf h_i)-(z_{0i}\mathbf {1} + z_{1i}\mathbf {f}))\right) \right ] * \label{eq:RegPrior1}\\
   && \exp \left [-\frac{1}{2}\left((\mathbf X_i(\mathbf h_i)-(z_{0i}\mathbf {1} + z_{1i}\mathbf {f}))'\boldsymbol\Sigma_X^{-1} (\mathbf X_i(\mathbf h_i)-(z_{0i}\mathbf {1} + z_{1i}\mathbf {f}))\right) \right ].    \label{eq:RegPrior2}
  \end{eqnarray}

Accordingly, the components of the joint distribution of the data and all unknown parameters that are dependent on $\mathbf w_i$ are \eqref{eq:RegPrior1}, and the prior on the approximated base functions, 

\begin{eqnarray}
\mathbf w_i &\propto& N_{p-1}(\mathbf 0,\gamma_w^{-1}\boldsymbol\Sigma+\lambda_w^{-1}\mathbf P_w)\mathbf{1}\{  t_1 + \sum_{k=2}^{p} (t_k-t_{k-1})\exp({w_i(t_{k-1})})=t_p\}, \nonumber \\
&& i=1\hdots N. \label{eq:Prior}
\end{eqnarray}

The resulting full conditional distribution for the approximated functions $\mathbf w_i$ is such that

$f(\mathbf w_i \mid rest) \propto$
 \begin{eqnarray}
  && \exp\left[-\frac{1}{2}\left((\mathbf X_i(\mathbf h_i)-(z_{0i}\mathbf {1} + z_{1i}\mathbf {f}))'\gamma_R\boldsymbol{\Sigma}^{-1} (\mathbf X_i(\mathbf h_i)-(z_{0i}\mathbf {1} + z_{1i}\mathbf {f}))\right) \right ] * \nonumber\\
   && \exp \left [-\frac{1}{2}\left(\mathbf w_i'(\gamma_w^{-1}\boldsymbol{\Sigma} +  \lambda_w^{-1}\mathbf P_w)^{-1} \mathbf w_i\right) \right ] \mathbf{1}\{  t_1 + \sum_{k=2}^{p} (t_k-t_{k-1})\exp({w_i(t_{k-1})})=t_p\}. \nonumber  
\end{eqnarray}

This full conditional does not have a known distributional form and can be sampled from via a Metropolis step in a MCMC sampler.

The components of the joint distribution of the data and all unknown parameters that are dependent on $\mathbf X_i$ are the sampling distribution \eqref{eq:datadist} and equation \eqref{eq:RegPrior2}.  The resulting full conditional distribution is such that

$f(\mathbf X_i \mid rest) \propto$
 \begin{eqnarray}
  && \exp\left[-\frac{1}{2\sigma_Y^2}(\mathbf Y_i-\mathbf X_i)'(\mathbf Y_i-\mathbf X_i) \right ] * \nonumber\\
   && \exp \left [-\frac{1}{2}\left((\mathbf X_i(\mathbf h_i)-(z_{0i}\mathbf{1} + z_{1i}\mathbf {f}))'\boldsymbol\Sigma_X^{-1} (\mathbf X_i(\mathbf h_i)-(z_{0i}\mathbf {1} + z_{1i}\mathbf {f}))\right) \right ]. \nonumber
\end{eqnarray}

This full conditional distribution also is not of a known distributional form and can be sampled from using a Metropolis step.  However, as significant features of the unregistered function, $\mathbf X_i$,  should be unchanged by the registration.  Smoothness in the registered function, $\mathbf X_i(\mathbf h_i)$, implies the same level of smoothness in the unregistered function $\mathbf X_i$.  For ease of sampling, we will re-write \eqref{eq:RegPrior2} in terms of the unregistered function, $\mathbf X_i$ so that

$f(\mathbf X_i \mid rest) \propto$
 \begin{eqnarray}
  && \exp\left[-\frac{1}{2\sigma_Y^2}(\mathbf Y_i-\mathbf X_i)'(\mathbf Y_i-\mathbf X_i) \right ] * \nonumber\\
   && \exp \left [-\frac{1}{2}\left((\mathbf X_i-(z_{0i}\mathbf {1} + z_{1i}\mathbf {f}(\mathbf h_i^{-1})))'\boldsymbol\Sigma_X^{-1} (\mathbf X_i-(z_{0i}\mathbf{1} + z_{1i}\mathbf {f}(\mathbf h_i^{-1})))\right) \right ], \label{eq:transXR}
\end{eqnarray}
which results in a multivariate normal full conditional distribution for $\mathbf X_i$.
   
   When noisy observations, $\mathbf Y_i$, $i= 1, \hdots, N$ are recorded,  the approximation we make in \eqref{eq:transXR}, while preserving conjugacy, prevents exact variational Bayes updates to be performed on the approximate posterior distributions for the following parameters: $\mathbf X_i$, $i=1\hdots N $, $\sigma_Y^2$, $\eta_X$, and $\lambda_X$.  Hence, the adapted variational Bayes procedure proposed here requires special handling under this data assumption. 
  
\subsection{Adapted Variational Bayes For Noisy Functional Data}
 \label{sec:NAVB}
 
In the traditional variational Bayes and AVB algorithms, it is assumed that \\
$E_{(\boldsymbol\theta_{-k})}(\log  f(\boldsymbol\theta_k\mid rest)$ can be evaluated for all parameters with conditionally conjugate prior distributions.  For the registration model that accounts for noisy observations, some of these expectations cannot be determined exactly.  However, we can adjust the original AVB algorithm further to perform an approximate inference procedure under these conditions.  With these adjustments, the convergence properties of the adapted variational Bayes algorithm no longer hold.  Nevertheless, we have found in practice that the adjusted algorithm still results in useful estimates for initializing a MCMC sampler.  

Here we look at why the approximate posterior distributions for $\mathbf X_i$, $i= 1,\hdots ,N$, $\eta_X$, and $\lambda_X$ cannot be updated properly using the adapted variational Bayes algorithm.  In the $m^{th}$ iteration, the following update should be made to log $q(\mathbf X_i)$, for $i = 1\hdots N$:

\begin{eqnarray}
\log  [q^{(m)}(\mathbf X_i)] \propto E_{(\mathbf\theta_{-\mathbf X_i})}[\log f(\mathbf X_i \mid rest)],\nonumber
\end{eqnarray}
where
\begin{eqnarray}
 E_{(\mathbf\theta_{-\mathbf X_i})}[\log f(\mathbf X_i\mid rest)] &\propto& E_{(\mathbf\theta_{-\mathbf X_i})}\Big[-\frac{1}{2}[(\mathbf X_i-\boldsymbol\mu_{\mathbf X_i\mid rest})'\boldsymbol\Sigma^{-1}_{\mathbf X_i \mid rest}(\mathbf X_i-\boldsymbol\mu_{\mathbf X_i\mid rest})] \Big],\nonumber \\
 \boldsymbol\Sigma_{\mathbf X_i \mid rest} &=& (\frac{1}{\sigma_Y^2}\mathbf I_p + \eta_X\mathbf P_1^{-} + \lambda_X\mathbf P_2^{-})^{-1}, \text{ and} \nonumber \\
 \boldsymbol\mu_{\mathbf X_i\mid rest} &=& \boldsymbol\Sigma_{\mathbf X_i \mid rest} [\frac{1}{\sigma_Y^2}\mathbf Y_i + (\eta_X\mathbf P_1^{-} + \lambda_X\mathbf P_2^{-})(z_{0i}\mathbf 1_p + z_{1i}\mathbf f(\mathbf h_i^{-1}))].
 \nonumber
\end{eqnarray}

Taking the expectation over the $q$ distributions for all other parameters except for the base functions results to the following updated parameters of  $q(\mathbf X_i) = N_p(\boldsymbol\mu_{q(\mathbf X_i)}, \boldsymbol\Sigma_{q(\mathbf X_i)})$,
\begin{eqnarray}
\boldsymbol\Sigma^{(m)}_{q(\mathbf X_i)} &=& (\mu_{q(\frac{1}{\sigma^2_Y})}\mathbf I_p+ \mu_{q(\eta_X)}\mathbf P_1^{-} + \mu_{q(\lambda_X)}\mathbf P_2^{-})^{-1}, \text{ and} \nonumber\\
\boldsymbol\mu^{(m)}_{q(\mathbf X_i)} &=& \boldsymbol\Sigma^{(m)}_{q(\mathbf X_i)} [\mu_{q(\frac{1}{\sigma^2_Y})}\mathbf Y_i + ( \mu_{q(\eta_X)}\mathbf P_1^{-} + \mu_{q(\lambda_X)}\mathbf P_2^{-})
 (\mu_{q(z_{0i})}\mathbf 1_p+\mu_{q(z_{1i})}E_{(\mathbf\theta_{-\mathbf X_i})}[\mathbf f(\mathbf  h_i^{-1})])]. \nonumber\\
 && \quad \label{eq:expX}
 \end{eqnarray}

In \eqref{eq:expX}, the expectation of $\mathbf f(\mathbf  h_i^{-1})$ is unknown.  So, the first approximation we will make is that 
$E_{(\mathbf\theta_{-\mathbf X_i})}[\mathbf f(\mathbf  h_i^{-1})]$ $\approx$ $\boldsymbol\mu_{q(\mathbf f)}(\mathbf h_i^{-1})$. 

Similarly, to update log $q(\eta_X)$:
\begin{eqnarray}
\log [q^{(m)}(\eta_X)] \propto E_{(\mathbf\theta_{-\eta_X})}[\log f(\eta_X \mid rest)],\nonumber
\end{eqnarray}
where
\begin{eqnarray}
 E_{(\mathbf\theta_{-\eta_X)}}[\log  f(\eta_X \mid rest)] &\propto& E_{(\mathbf\theta_{-\eta_X})} [c_{\eta_X\mid rest} \log  \eta_X - d_{\eta_X\mid rest}\eta_X], \nonumber \\
 c_{\eta_X\mid rest} &=& N+c, \text{ and}\nonumber \\
 d_{\eta_X\mid rest} &=& d+\frac{1}{2}\sum_{i=1}^{N}\textrm{tr}[(\mathbf X_i - (z_{0i}\mathbf 1_p + z_{1i}\mathbf f (\mathbf h^{-1}_i)))(\mathbf X_i - (z_{0i}\mathbf 1_p + z_{1i}\mathbf f (\mathbf h^{-1}_i)))'\mathbf P_1^{-}].\nonumber 
 \end{eqnarray}

Taking the expectation over the $q$ distributions for all other parameters except for the base functions results to the following updated parameters of  $q(\eta_X) = G(c_{q(\eta_X)},d_{q(\eta_X)})$,

\begin{eqnarray}
 c^{(m)}_{q(\eta_X)} &=& N+c,\nonumber \\
 d^{(m)}_{q(\eta_X)} &=& d+\frac{1}{2}\textrm{tr}\Big[\Big(\sum_{i=1}^{N}\Big(\boldsymbol\Sigma_{q(\mathbf X_i)} + \boldsymbol\mu_{q(\mathbf X_i)}\boldsymbol\mu_{q(\mathbf X_i)}'-2\boldsymbol\mu_{q(\mathbf X_i)}(\mu_{q(z_{0i})}\mathbf 1_p + \mu_{q(z_{1i})}E_{(\mathbf\theta_{-\eta_X})}[\mathbf f(\mathbf  h_i^{-1})])',  \nonumber \\
 && +\quad 2\mu_{q(z_{0i})}\mu_{q(z_{1i})}\mathbf 1_pE_{(\mathbf\theta_{-\eta_X})}[\mathbf f(\mathbf  h_i^{-1})]' + \textit{ }( \sigma^2_{q(z_{1i})}+\mu_{q(z_{1i})}^2)E_{(\mathbf\theta_{-\eta_X})}[\mathbf f(\mathbf  h_i^{-1})\mathbf f(\mathbf  h_i^{-1})']\Big), \text{ and}\nonumber\\
 && + \quad \Big(2\sum_{i=1}^{N-1} (\sigma^2_{q(z_{0i})} + \mu_{q(z_{0i})}^2) + \sum_{i=1}^{N-1}\sum_{j=1}^{N-1}\mu_{q(z_{0i})}\mu_{q(z_{0j})}\mathbf{1}\{j \neq i\}\Big)\mathbf 1_p\mathbf 1_p' \Big)\mathbf P_1^{-} \Big].\nonumber
   \end{eqnarray} 

In the expression for  $d^{(m)}_{q(\eta_X)}$, $E_{(\mathbf\theta_{-\eta_X})}[\mathbf f(\mathbf  h_i^{-1})]$ and $E_{(\mathbf\theta_{-\eta_X})}[\mathbf f(\mathbf  h_i^{-1})\mathbf f(\mathbf  h_i^{-1})']$ are unknown.  Thus, we will make the following approximations,

$E_{(\mathbf\theta_{-\eta_X})}[\mathbf f(\mathbf  h_i^{-1})]$ $\approx$ $\boldsymbol\mu_{q(\mathbf f)}(\mathbf h_i^{-1})$ and  $E_{(\mathbf\theta_{-\eta_X})}[\mathbf f(\mathbf  h_i^{-1})\mathbf f(\mathbf  h_i^{-1})']$ $\approx$ $\boldsymbol\Sigma_{q(\mathbf X_i)}/N + \boldsymbol\mu_{q(\mathbf f)}(\mathbf h_i^{-1}) \boldsymbol\mu_{q(\mathbf f)}(\mathbf h_i^{-1})'.$\\

Note, $\boldsymbol\Sigma_{q(\mathbf X_i)}$ does not depend on $i$.

The variational Bayes algorithm update for $\lambda_X$ is similar to that of $\eta_X$ and requires the same approximations,  
\begin{eqnarray}
\log [q^{(m)}(\lambda_X)] \propto E_{(\mathbf\theta_{-\lambda_X})}[\log f(\lambda_X \mid rest)],\nonumber
\end{eqnarray}
where
\begin{eqnarray}
 E_{(\mathbf\theta_{-\lambda_X)}}[\log f(\lambda_X \mid rest)] &\propto& E_{(\mathbf\theta_{-\lambda_X})} [c_{\lambda_X\mid rest} \log  \lambda_X - d_{\lambda_X\mid rest}\lambda_X], \nonumber \\
 c_{\lambda_X\mid rest} &=& N\Big(\frac{p-2}{2}\Big) +c , \text{ and}\nonumber \\
 d_{\eta_X\mid rest} &=& d+\frac{1}{2}\sum_{i=1}^{N}\textrm{tr}[(\mathbf X_i - (z_{0i}\mathbf 1_p + z_{1i}\mathbf f (\mathbf h^{-1}_i)))(\mathbf X_i - (z_{0i}\mathbf 1_p + z_{1i}\mathbf f (\mathbf h^{-1}_i)))'\mathbf P_2^{-}]. \nonumber 
 \end{eqnarray}

Taking the expectation over the $q$ distributions for all other parameters except for the base functions results to the following updated parameters of  $q(\lambda_X) = G(c_{q(\lambda_X)},d_{q(\lambda_X)})$,

\begin{eqnarray}
 c^{(m)}_{q(\lambda_X)} &=& N\Big(\frac{p-2}{2}\Big) +c, \text{ and} \nonumber \\
 d^{(m)}_{q(\lambda_X)} &=& d+\frac{1}{2}\textrm{tr}\Big[\Big(\sum_{i=1}^{N}\Big(\boldsymbol\Sigma_{q(\mathbf X_i)} + \boldsymbol\mu_{q(\mathbf X_i)}\boldsymbol\mu_{q(\mathbf X_i)}'-2\boldsymbol\mu_{q(\mathbf X_i)}(\mu_{q(z_{0i})}\mathbf 1_p + \mu_{q(z_{1i})}E_{(\mathbf\theta_{-\lambda_X})}[\mathbf f(\mathbf  h_i^{-1})])' \nonumber \\
 && +\quad 2\mu_{q(z_{0i})}\mu_{q(z_{1i})}\mathbf 1_pE_{(\mathbf\theta_{-\lambda_X})}[\mathbf f(\mathbf  h_i^{-1})]' + \textit{ }( \sigma^2_{q(z_{1i})}+\mu_{q(z_{1i})}^2)E_{(\mathbf\theta_{-\lambda_X})}[\mathbf f(\mathbf  h_i^{-1})\mathbf f(\mathbf  h_i^{-1})']\Big) \nonumber\\
 && + \quad \Big(2\sum_{i=1}^{N-1} (\sigma^2_{q(z_{0i})} + \mu_{q(z_{0i})}^2) + \sum_{i=1}^{N-1}\sum_{j=1}^{N-1}\mu_{q(z_{0i})}\mu_{q(z_{0j})}\mathbf{1}\{j \neq i\}\Big)\mathbf 1_p\mathbf 1_p' \Big)\mathbf P_2^{-} \Big].\nonumber
   \end{eqnarray} 

Again, in the expression for  $d^{(m)}_{q(\lambda_X)}$, $E_{(\mathbf\theta_{-\lambda_X})}[\mathbf f(\mathbf  h_i^{-1})]$ and $E_{(\mathbf\theta_{-\lambda_X})}[\mathbf f(\mathbf  h_i^{-1})\mathbf f(\mathbf  h_i^{-1})']$ are unknown.  Thus, we will make the following approximations,

$E_{(\mathbf\theta_{-\lambda_X})}[\mathbf f(\mathbf  h_i^{-1})]$ $\approx$ $\boldsymbol\mu_{q(\mathbf f)}(\mathbf h_i^{-1})$ and  $E_{(\mathbf\theta_{-\lambda_X})}[\mathbf f(\mathbf  h_i^{-1})\mathbf f(\mathbf  h_i^{-1})']$ $\approx$ $\boldsymbol\Sigma_{q(\mathbf X_i)}/N + \boldsymbol\mu_{q(\mathbf f)}(\mathbf h_i^{-1}) \boldsymbol\mu_{q(\mathbf f)}(\mathbf h_i^{-1})'$.

Due to these modifications, if noisy observations are observed the convergence properties of the adapted variational Bayes algorithm are not guaranteed to hold, and $\log [f(\mathbf Y,\mathbf w; q)]$ cannot be monitored.  However, we can monitor convergence for this model as follows.  Taking advantage of the fact that functional smoothing converges more quickly than functional registration, fix the approximated unregistered functions, $\mathbf X_i$, $i =1, \hdots, N$, after a small number of iterations and proceed as if they are known.  Then, as in the model where the observations are recorded without noise, $\log  [f(\mathbf X,\mathbf w; q)]$ can be monitored.

\subsection{The Berkeley Growth Data}

\label{sec:RBVD}
 We refer back to the Berkeley Boys Growth Velocity dataset from Section 4.1.  In Section 4.1, these data were smoothed prior to registration.  Here, we again consider these functions with the added assumption that they are corrupted by simulated mean zero iid Gaussian noise, where the true noise variance, $\sigma_Y^2$, is .25.  
 
 \begin{figure}
\begin{tabular}{cc}
\centering
\includegraphics[width=8cm]{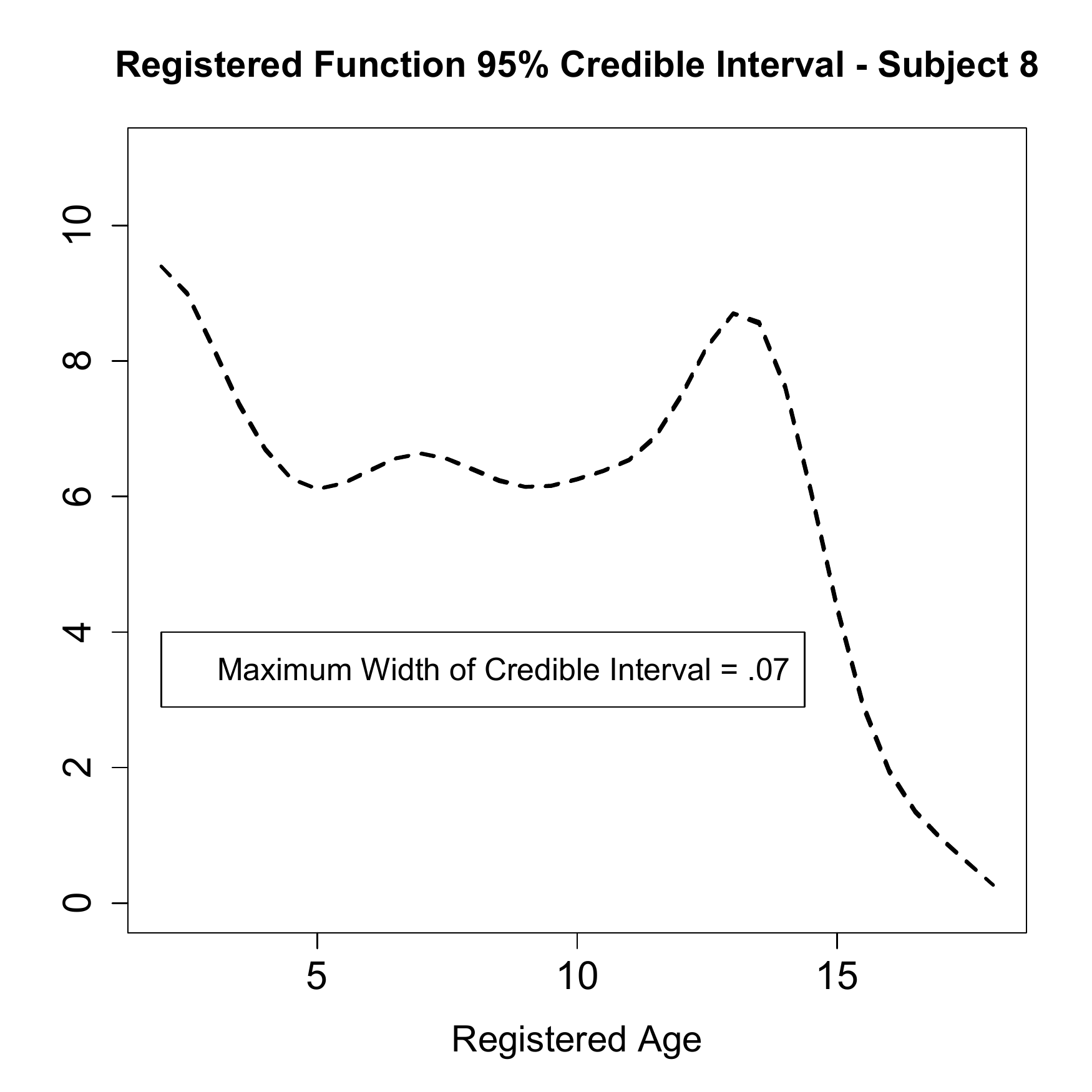} &
\includegraphics[width=8cm]{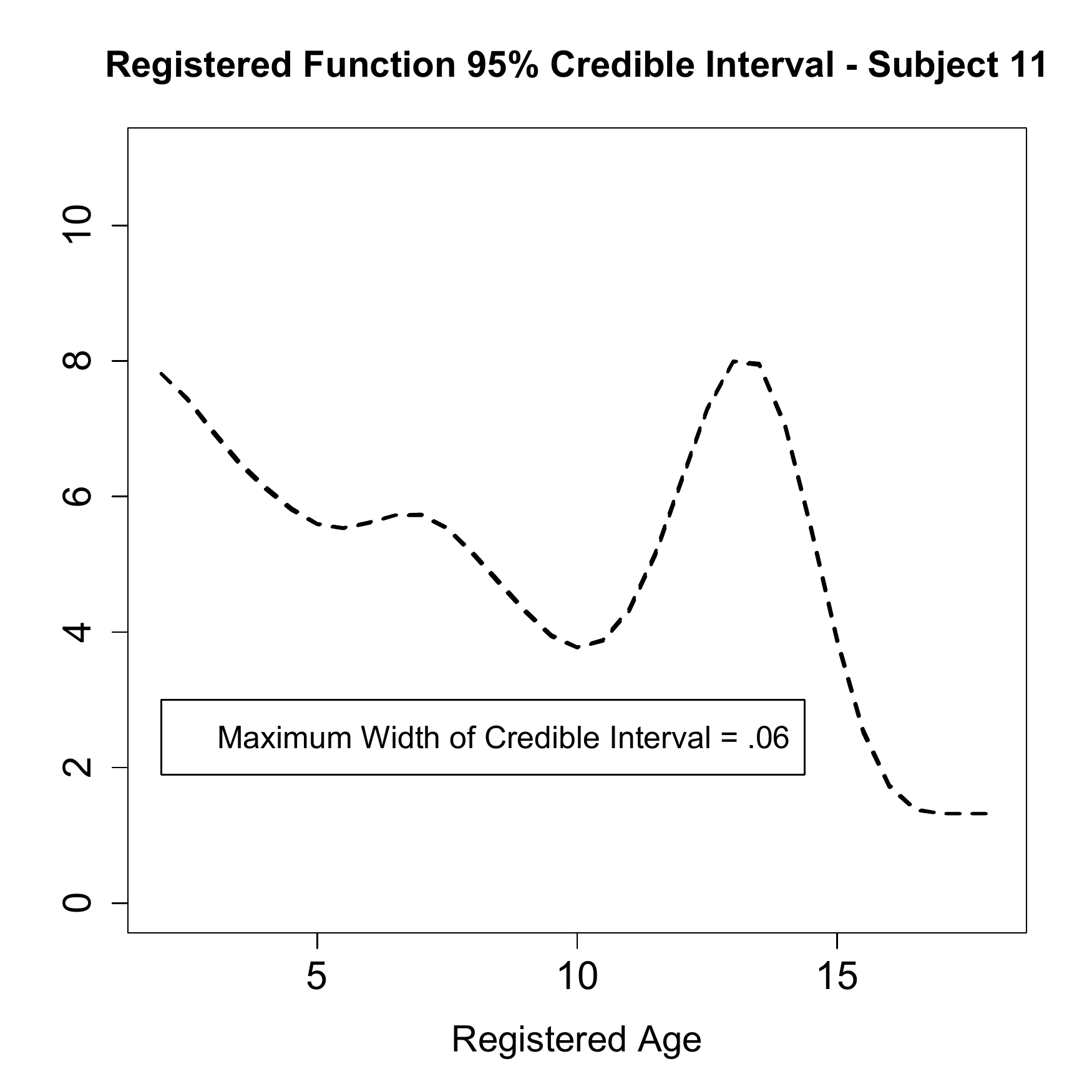} 
\end{tabular}
\caption{Examples of Credible Intervals for Noiseless Observations .  These are two examples from the Boys Growth Velocity Data of the tight credible bands that result from registering functions that are pre-smoothed.  In Figure \ref{fig:CIs}, the top and lower right illustrations contain the credible intervals for these same observations when the noise process is included in the model. }
\label{fig:cred811}
\end{figure}

 \begin{figure}
\begin{tabular}{cc}
\centering
\includegraphics[width=8cm]{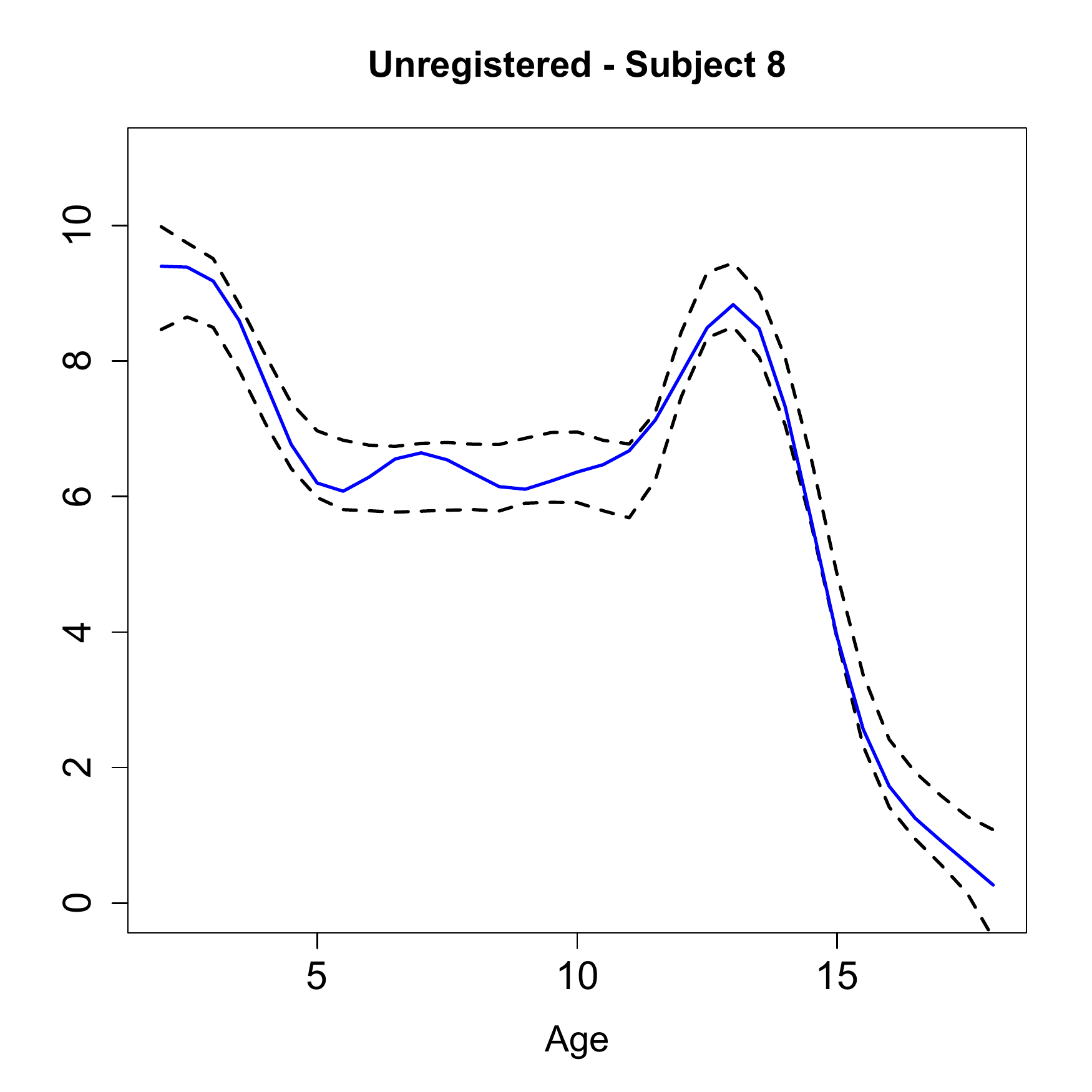} &
\includegraphics[width=8cm]{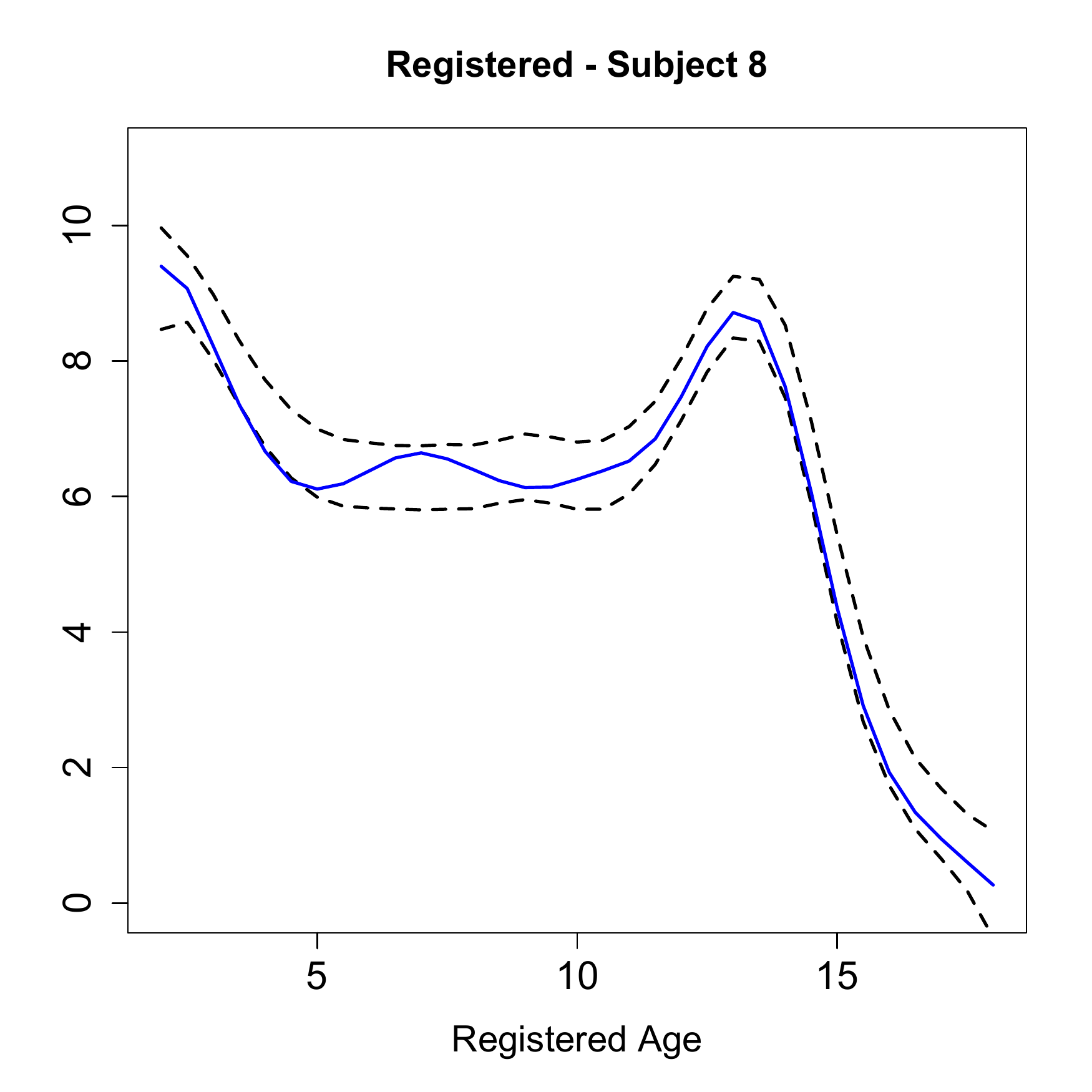}\\
\includegraphics[width=8cm]{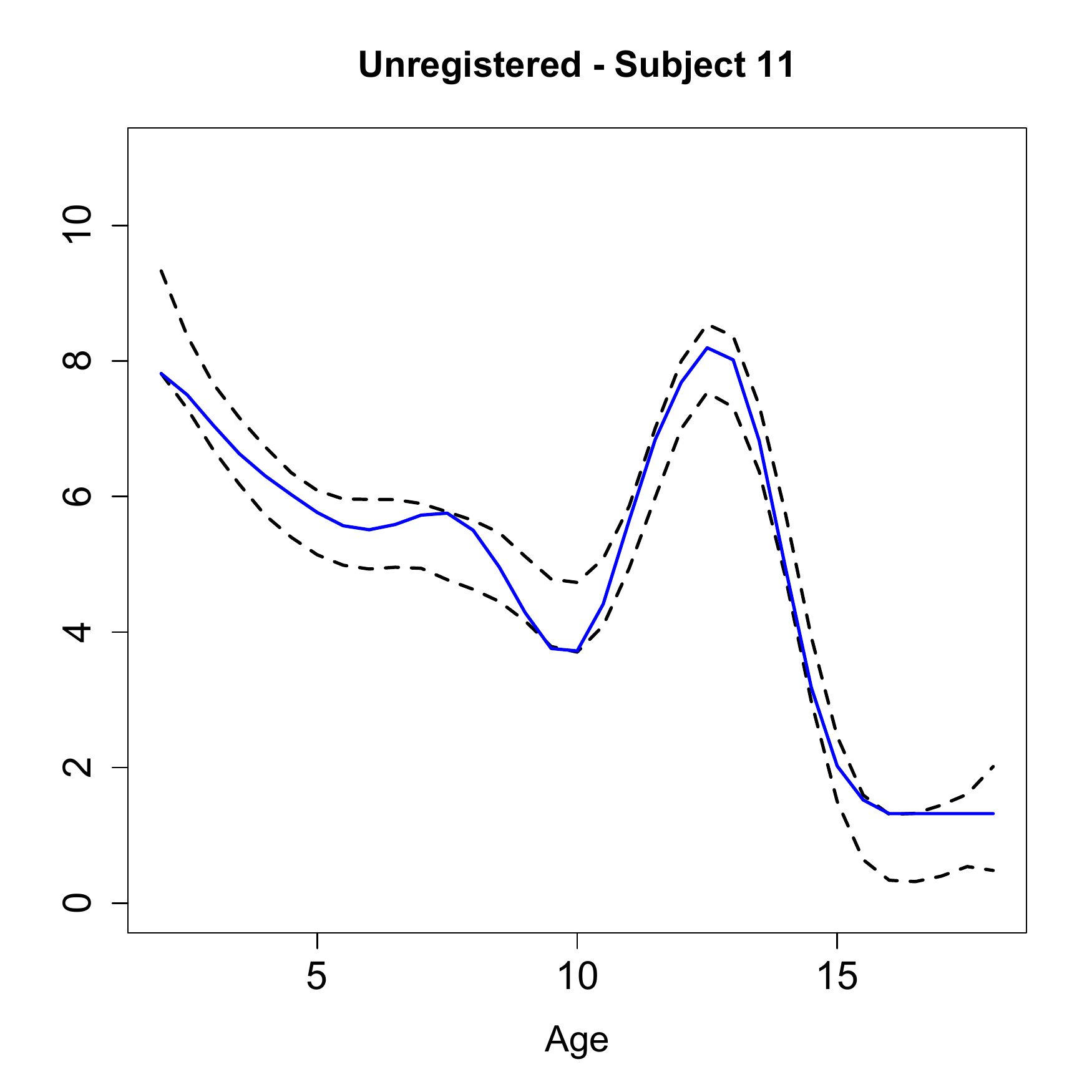}   &
\includegraphics[width=8cm]{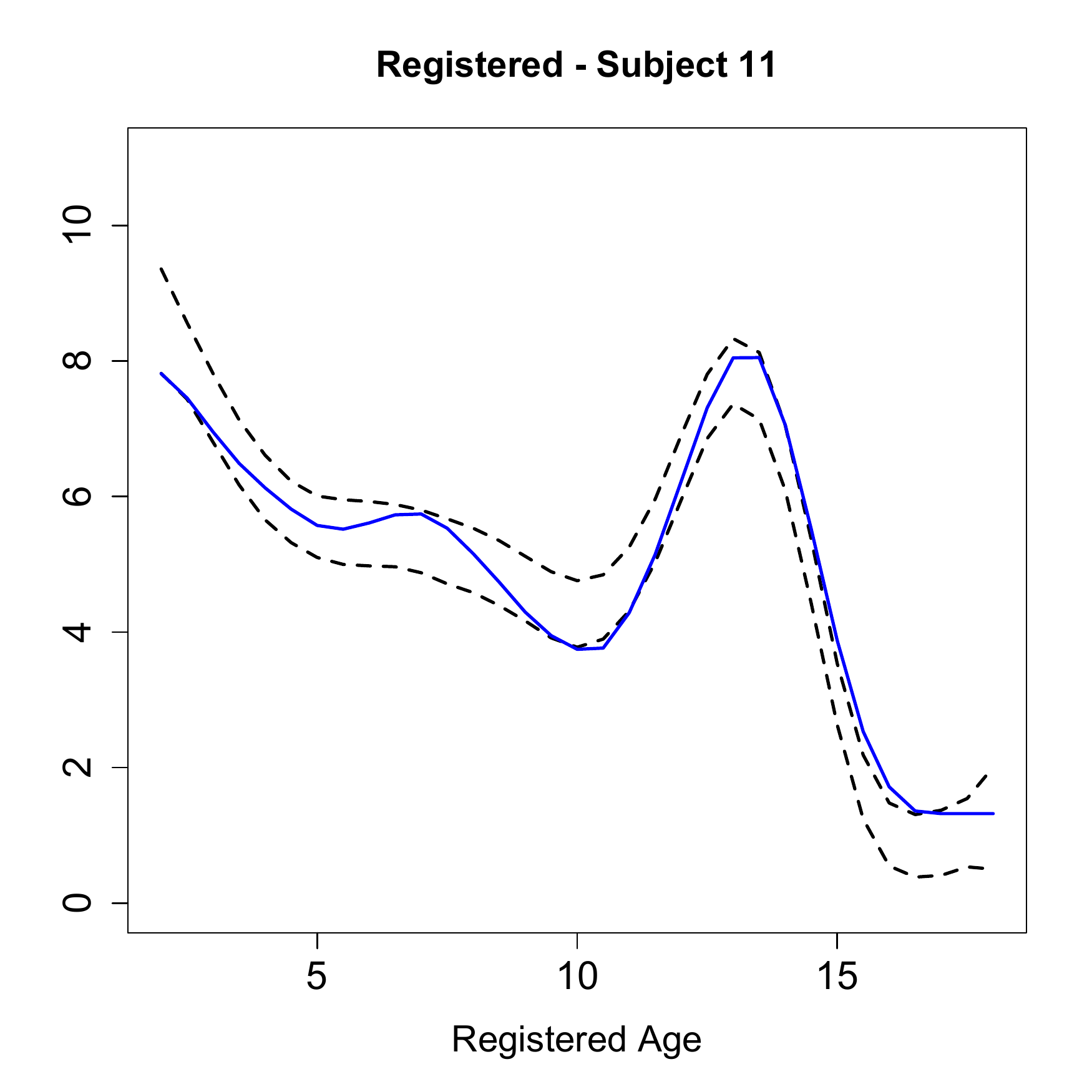}
\end{tabular}
\caption{Examples of Credible Bands for the Unregistered and Registered Functions when the Noise Process is Included in the Model.  \textbf{Top and Lower Left} 95\% credible bands for the unregistered functions are plotted with the original noiseless functions for subjects 8 and 11. \textbf{Top and Lower Right}  For subjects 8 and 11, 95\% credible bands for the registered functions are plotted with the estimate of the registered `true' functions.}
\label{fig:CIs}
\end{figure}

While it is common in statistical analysis to perform preprocessing steps before applying a particular inference procedure, failing to account for the variability in parameter estimates due to the preprocessing step leads to overly narrow confidence (or credible) regions.  In some cases, the effect may be fairly small, and not much is lost in this oversight.  However, as we show here, the underestimated variability can be substantial when uncertainty in the preprocessing steps is ignored.  

In Section 4.2 is an illustration of how closely AVB and MCMC estimates of the registered functions adhere to one another.  Not only do these estimates tend to be fairly similar when the functions are recorded without noise, but the uncertainty in these estimates is minimal.  Figure \ref{fig:cred811} contains the credible bands for two of the 39 pre-smoothed Boys Growth Velocity Functions.  These bands are so narrow the width between them cannot be seen.  Keep in mind the posterior distributions of the registered functions are certainly multi-modal. These credible bands result from imposing the restriction that the mean value of the warping functions at each time point over the sample must equal that time point.  Even with this restriction, the posterior distributions can be multi-modal.  However, these narrow credible bands reflect that our estimates are in a highly probable area of the posterior distribution with minimal local variance.  Figure \ref{fig:CIs} contains credible bands for both the unregistered and registered functions for the same two functions used in Figure \ref{fig:cred811} after noise has been added to the data and accounted for in the model.  The variability due to noise is substantial.  The solid line in all of the plots contains the noiseless version of these estimates (or observations in the case of the unregistered functions).

In addition to providing more accurate credible intervals, this model estimates the noise variance to be .258 (actual noise variance is .25).  This estimate is obtained using uninformative priors for both the noise variance, $\sigma_Y^2$ and the associated smoothing parameter $\lambda_X$.

This analysis illustrates how regularizing the data prior to statistical analysis for registration models severely limits inference for these models.  If significant noise is present in the data it is prudent to account for the variability in the registration process due to the noise.  Our proposed hierarchical model is one way to account for this variability.

{}

\end{document}